\pdfoutput=1
\documentclass[PAPER, atlasdraft=false, texlive=2023, cernpreprint, orcidlogo, UKenglish, texmf, orcidlogo]{atlasdoc}
 
\usepackage{atlaspackage}
\usepackage{atlasbiblatex}
 
\usepackage{atlasphysics}
 
\usepackage{siunitx}

\graphicspath{{logos/}{figures/}}
\usepackage{multirow}
 
\usepackage{ANA-HIGG-2018-30-PAPER-defs}

\AtlasTitle{Test of CP-invariance of the Higgs boson in vector-boson fusion production and in its decay into four leptons}
 
\AtlasVersion{2.0}
 
\AtlasAbstract{
A search for CP violation in the decay kinematics and vector-boson fusion production of the Higgs boson is performed in the $H \to ZZ^{*} \to 4\ell$ ($\ell = e,\mu$) decay channel.
The results are based on proton--proton collision data produced at the LHC at a centre-of-mass energy of 13~TeV and recorded by the ATLAS detector from 2015 to 2018, corresponding to an integrated luminosity of 139 $\mbox{fb\(^{-1}\)}$.
Matrix element-based optimal observables are used to constrain CP-odd couplings beyond the Standard Model in the framework of Standard Model effective field theory expressed in the Warsaw and Higgs bases.
Differential fiducial cross-section measurements of the optimal observables are also performed, and a new fiducial cross-section measurement for vector-boson-fusion production is provided.
All measurements are in agreement with the Standard Model prediction of a CP-even Higgs boson.
}

\AtlasRefCode{HIGG-2018-30}
 
\PreprintIdNumber{CERN-EP-2023-030}

\AtlasJournal{JHEP}
\AtlasJournalRef{JHEP 05 (2024) 105}
\AtlasDOI{10.1007/JHEP05(2024)105}

\hypersetup{pdftitle={ATLAS document},pdfauthor={The ATLAS Collaboration}}
 
\begin{document}
 
\maketitle
 
\tableofcontents
\clearpage
\section{Introduction}

The observed baryon asymmetry of the universe is one of the important open questions of physics today.
To explain this, the Sakharov criteria~\cite{Sakharov:1967dj} require sufficient violation of combined charge conjugation and parity (CP) invariance. The only known source of CP violation in the Standard Model (SM) is a complex phase in the Cabibbo--Kobayashi--Maskawa (CKM) matrix~\cite{Cabibbo:1963yz, Kobayashi:1973fv} for quark mixing. Other sources of CP violation may exist in the neutrino sector or the strong interaction.
The CKM complex phase does not lead to a sufficient amount of CP violation to explain the baryon asymmetry.
This motivates the exploration of potential new sources of CP violation, as performed in the present analysis of the production and decay of the Higgs boson~\cite{HIGG-2012-27, CMS-HIG-12-028}.
 
The spin-parity ($J^P$) of the Higgs boson is predicted to be $0^+$ with even C-parity by the Standard Model.
Any sign of CP-violation in the production or decay of the Higgs boson at the current level of sensitivity would therefore be an unambiguous indication of beyond the Standard Model phenomena (BSM).
With 25~\ifb of proton--proton collision data collected from the LHC with a centre-of-mass energy, $\sqrt{s}$,  of 7 and 8~\TeV, the ATLAS and CMS experiments excluded the pure spin-parity states $0^-$, $1^+$, $1^-$, $2^+$ and $2^-$ at more than \perc{99} confidence level (CL) based on the observed Higgs boson decays ($\gamma\gamma, ZZ, WW$)~\cite{CMS-HIG-12-041,HIGG-2013-01,HIGG-2013-17,CMS-HIG-14-018}.
These results also provided the first limits on a possible CP-odd contribution to Higgs boson to vector-boson couplings ($HVV$) in Higgs boson decays. With the same data set there was a first search for CP-odd $HVV$ couplings in vector-boson fusion (\VBF) production in the \htautau decay channel~\cite{HIGG-2015-06} and associated \VH production in the $bb$ decay channel~\cite{CMS-HIG-14-035}.
With an increased integrated luminosity of 139 \ifb of LHC collisions at $\sqrt{s}=13$ \TeV, an order of magnitude more Higgs boson candidates were collected.
Constraints on CP-invariance were tightened for the $HVV$ couplings and extended to Higgs Yukawa couplings to fermions, for example in \htautau decays and associated Higgs boson top pair (\ttH) production in both the diphoton and $b\bar{b}$ decay channels, and other channels~\cite{
CMS-HIG-17-011, 
CMS-HIG-18-002, 
CMS-HIG-17-034, 
HIGG-2018-14,   
HIGG-2019-01,   
CMS-HIG-19-013, 
CMS-HIG-19-009, 
CMS-HIG-20-006, 
HIGG-2017-13,   
CMS-HIG-21-013, 
CMS:2022uox,    
CMS:2022dbt,    
HIGG-2019-10,   
ATLAS:2023cbt,  
CMS:2023gjz,    
}.
 
For the \htollllbrief channel ($\ell=e,\mu$), constraints on CP-even and CP-odd BSM couplings have been set with 139 \ifb at $\sqrt{s}=13$ \TeV for Higgs boson production cross-section measurements, by comparing observations with SM expectations~\cite{HIGG-2018-28, CMS-HIG-19-009}.
However, potential deviations of the Higgs boson production cross-sections could be explained by either a CP-even or CP-odd BSM coupling, and would not distinguish between them.
The present analysis searches for a visible CP-odd effect in \VBF production and the \htollllbrief decay by employing \textit{optimal observables}~\cite{Atwood:1991ka, Davier:1992nw, Diehl:1996wm, Anderson:2013afp, CMS-HIG-14-018}, which are CP-odd by construction, with 139 \ifb of data collected at $\sqrt{s}=13$ \TeV by the ATLAS detector~\cite{PERF-2007-01}.
About 200 Higgs bosons decaying into four leptons are expected to be reconstructed in the detector for this sample, including about 10 events in a \VBF production phase space.
 
The optimal observables are constructed from matrix elements of the Standard Model effective field theory (SMEFT)~\cite{Weinberg:1978kz, Buchmuller:1985jz, Brivio:2017vri}, a specific formalism of an effective field theory (EFT).
An optimal-observable distribution is symmetric for the CP-even Higgs boson and becomes asymmetric when contributions from CP-odd BSM couplings are present.
There are three dimension-six operators with BSM CP-odd couplings in the SMEFT Lagrangian that can contribute to \VBF production or the \htollllbrief decay, each associated with a Wilson coefficient.
In this analysis, these three operators are considered in the Warsaw basis, and each operator has a different sensitivity to \VBF production and the \htollllbrief decay.
The optimal-observable distributions are used in two ways: the observed distributions are used directly to constrain CP-odd couplings, and they are unfolded to a fiducial phase space, complementing the \htollllbrief differential cross-section measurements in Ref.~\cite{HIGG-2018-29}, to allow reinterpretations in different models.
Since the present search is looking for a distinct CP-odd signature, the extraction of the CP-odd coupling constraints is based only on the shape of the optimal-observable distributions, ignoring the expected change in cross-section. Including cross-sections has only a small effect on the constraints, as shown in \Sect{\ref{sec:direct_results}}.
The fiducial phase space for the unfolded optimal-observable distributions is the same as for the \htollllbrief fiducial measurement~\cite{HIGG-2018-29}, and is dominated by gluon--gluon fusion (\ggF) production that allows constraints on CP-odd operators which are sensitive to the  \htollllbrief decay.
Extracting the differential distributions for the \VBF optimal observables in a \VBF-enriched fiducial phase space would improve the sensitivity to the CP-odd operator that is more sensitive to \VBF production, however this is not yet within statistical reach.
Instead this paper presents an inclusive cross-section measurement in a \VBF fiducial phase space. This measurement can be used to constrain BSM models affecting the Higgs boson to \zzstar vertex through their expected impact on event rates.


\FloatBarrier

\section{Theoretical framework and analysis methodology}
\label{sec:methodology}

While the analysis is designed to be sensitive to signatures of CP violation independently of a particular BSM physics model, the SMEFT framework is used to guide the analysis strategy and interpret the results. This framework also provides the coupling scenarios used to define the optimal observables at the centre of the analysis strategy. Both the SMEFT framework and the optimal-observable formalism are described in more detail in the following. In addition, the known dependence of cross-sections on the BSM couplings in the framework of SMEFT is exploited to parameterize the signal prediction in the direct BSM coupling measurement as described in \cref{sec:meth:morphing}.
 
\subsection{EFT framework}
In SMEFT, a complete set of
higher-order-in-mass operators invariant under the SM gauge group  $\mathrm{SU}(3)_C\times \mathrm{SU}(2)_L \times \mathrm{U}(1)_Y$ is built from the SM fields~\cite{Brivio:2017vri}.
If the dimension-five operator is ignored because it violates lepton number~\cite{Weinberg:1979sa}, the leading contributions to physical observables are expected to be from dimension-six operators. In this case the SMEFT Lagrangian becomes:
\begin{equation}
\label{eqn:LSMEFT_leading}
\mathcal{L}_{\mathrm{SMEFT}} =  \mathcal{L}_{\mathrm{SM}} + \sum_i \frac{c_i}{\Lambda^{2}} O_i^{(6)} \ ,
\end{equation}
where $O_i^{(6)}$ are the operators of mass-dimension six, which are invariant under the SM gauge group, the $c_{i}$ are the corresponding dimensionless real coupling constants, the Wilson coefficients, and $\Lambda$ is the energy cutoff scale of the effective field theory.
 
There are several complete sets of these dimension-six operators. One such set, used in many existing LHC EFT interpretations~\cite{HIGG-2020-08, HIGG-2020-16, ATL-PHYS-PUB-2022-037, CMS-TOP-21-004, CMS-HIG-19-009}, is the \textit{Warsaw basis}
~\cite{Buchmuller:1985jz, Grzadkowski:2010es}.
Only three CP-odd operators contributing to the $HVV$ vertex for both \VBF production and vector-boson decays
are considered, and are listed in the first three rows of Table~\ref{tab:CPcoup}\footnote{This analysis is sensitive to each CP-odd coupling normalised by $(1-\ensuremath{c_{\text{even}}})$ where $\ensuremath{c_{\text{even}}}$ is a linear combination of CP-even couplings that leave the shapes of the analysis observables invariant. Since these CP-even couplings are assumed to be zero, the measurements in the rest of this paper are simply denoted by the CP-odd coupling.}.

The same phenomena are described by an alternative complete set of operators, based on the mass eigenstates after the spontaneous symmetry breaking instead of the fields of the unbroken gauge symmetry.
This formulation is expressed in terms of the physical states of SM gauge bosons $W^+$, $W^-$, $Z$, and $\gamma$.
With these operators it is possible to build another basis~\cite{deFlorian:2016spz, Azatov:2022kbs} that can probe the effects of the Higgs boson couplings to other particles, and is more closely linked to experimental  physical observables than the Warsaw basis. It is called the \textit{Higgs basis} in which three independent CP-odd couplings, listed in Table~\ref{tab:CPcoup} together with their operators, mediate the same phenomena as the Warsaw-basis couplings.
Each coupling of one basis can be expressed as a linear combination of the three couplings in the other basis (see \App{\ref{sec:appconv}}).

Finally, a common parameterization used in existing experimental searches for CP-violation at the $HVV$ vertex~\cite{HIGG-2020-08,HIGG-2015-06} assumes a single BSM CP-odd Higgs boson coupling denoted by \dTilde.
In~\cite{HIGG-2015-06}, it was considered that the different contributions from the various electroweak gauge-boson fusion processes could not be distinguished experimentally, and assumed that the BSM CP-odd Higgs boson couplings \cHWtil and \cHBtil in the Warsaw basis are equal, leading to the remaining coupling \cHWBtil being zero.
The CP-odd parameter \dTilde is related to the Warsaw-basis couplings by the parameterization $\cHWtil = \cHBtil = \frac{\Lambda^2}{\varv^2} \dTilde$, where $\varv$ is the Higgs boson vacuum expectation value.
In the Higgs basis, it is equivalent to the direction $\tCza = 0$, $\tCaa = \sin^2 \theta_{W} \cos^2 \theta_{W} \tCzz $, where $\theta_{W}$ is the Weinberg angle.
 
It should be noted that, in this framework, the coupling strength associated to the SM-like coupling itself is not modified.
 
In the following, the focus is on interpretations with the three couplings in the Warsaw-basis framework, but results are also provided for the Higgs basis and \dTilde to facilitate comparisons.
Constraints are set on a set of benchmark scenarios where one or two of the aforementioned CP-odd coupling is none-zero at a given time, assuming that all other EFT couplings are zero.

\begin{table}[!htbp]
\renewcommand\arraystretch{1.2}
\centering
\caption{
SMEFT CP-odd dimension-six operators in the Warsaw and Higgs bases relevant for the measurement of CP-invariance in the \htollllbrief decay channel.
The gauge field strength tensors are denoted $W_{\mu\nu}^I$ and $B_{\mu\nu}$, while $\Phi$ is the scalar doublet Higgs boson field.
The photon and $Z$-boson field strength tensors after electroweak symmetry breaking are denoted as $A_{\mu\nu}$ and $Z_{\mu\nu}$, respectively, and $h$ denotes the massive Higgs field after electroweak symmetry breaking.
Combinations of couplings in one basis can be translated into equivalent combinations of couplings from the other basis that describe the same phenomena~\cite{deFlorian:2016spz, Azatov:2022kbs}.
}
\label{tab:CPcoup}
\hspace*{-1cm}
\begin{tabular}{ccc}
\hline
\hline
Operator & Structure & Coupling     \\
\hline
\multicolumn{3}{c}{Warsaw Basis} \\
\hline
$O_{\Phi \tilde{W}}$    &    $\Phi^{\dagger} \Phi  \tilde{W}_{\mu\nu}^IW^{\mu\nu I} $     & $\cHWtil$              \\
$O_{\Phi \tilde{W}B}$   &    $\Phi^{\dagger} \tau^I \Phi  \tilde{W}_{\mu\nu}^IB^{\mu\nu} $       & $\cHWBtil$   \\
$O_{\Phi \tilde{B}}$    &     $\Phi^{\dagger} \Phi  \tilde{B}_{\mu\nu}B^{\mu\nu} $      & $\cHBtil$                \\
\hline
\multicolumn{3}{c}{Higgs Basis} \\
\hline
$O_{h Z\tilde{Z}}$    &    $hZ_{\mu\nu}\tilde{Z}^{\mu\nu} $     & $\tCzz$   \\
$O_{h Z\tilde{A}}$    &    $hZ_{\mu\nu}\tilde{A}^{\mu\nu} $     & $\tCza$   \\
$O_{h A\tilde{A}}$    &    $hA_{\mu\nu}\tilde{A}^{\mu\nu} $     & $\tCaa$   \\
\hline
\hline
\end{tabular}
\end{table}

\subsection{Optimal observables}
\label{sec:meth:OO}
 
The cross-section for the combination of a set of CP-odd couplings, $c_{i}$, within the SM is proportional to the square of the matrix element composed of the sum of the individual matrix elements:
\begin{subequations}\label{eq:CouplingFormula}
\begin{align*}
\left|\mathcal{M}\right|^2 &= \left| \mathcal{M}_\mathrm{SM} + \sum_{i} \frac{c_{i}}{\Lambda^{2}} \mathcal{M}_{\mathrm{BSM},i}   \right|^2 \\
&= \left| \mathcal{M}_\mathrm{SM} \right|^2 + 2 \sum_{i} \frac{c_{i}}{\Lambda^{2}} \Re \left( \mathcal{M}_\mathrm{SM}^{\ast} \mathcal{M}_{\mathrm{BSM},i}   \right) + \sum_{i} \sum_{j} \frac{c_{i}c_{j}}{\Lambda^{4}} \Re \left( \mathcal{M}_{\mathrm{BSM},i}^{\ast} \mathcal{M}_{\mathrm{BSM},j}   \right)
\end{align*}
\end{subequations}
 
The expanded sum has a first term that is the squared SM matrix element, and a last term that represents the quadratic BSM contributions and the interference between BSM terms. These first and last terms are CP-even.
The middle term is composed of the cross terms between the SM and BSM couplings ($\Re$ is the real part) and represents the interference between the SM and BSM processes.\footnote{$ 2 \Re \left( \mathcal{M}_\mathrm{SM}^{\ast} \mathcal{M}_\mathrm{BSM}   \right) = \mathcal{M}_\mathrm{SM}^{\ast} \mathcal{M}_\mathrm{BSM} + \mathcal{M}_\mathrm{SM} \mathcal{M}_\mathrm{BSM}^{\ast} $}
This term is CP-odd and therefore suitable as a probe for CP violation.
 
Normalising the interference term by the SM term forms an `optimal observable' for each coupling, which can be used to detect a CP asymmetry:
\begin{align*}
\mathcal{OO} = \frac{2 \Re \left( \mathcal{M}_\mathrm{SM}^{\ast} \mathcal{M}_\mathrm{BSM}   \right)}{\left|\mathcal{M}_\mathrm{SM}\right|^2}.
\end{align*}
These observables are CP-odd by construction, implying a symmetric distribution with a vanishing mean in the absence of CP violation.
Any asymmetry observed in the distribution of the optimal observables would be direct evidence for CP-violation in the $HVV$ coupling.
This property is not shared by other CP-even optimal observables constructed from squared matrix elements, and motivates the choice of the interference-based observable definition in this search.
The LO matrix elements entering the optimal observable are calculated using \MGNLO~\cite{Alwall:2014hca} with the \NNPDF[2.3lo]~\cite{Ball:2012cx} parton distribution function (PDF) set.
Two types of optimal observables are considered.
\textit{Production-level} optimal observables reflect the VBF production vertex, not considering the Higgs boson decay.
They are constructed using VBF production matrix elements calculated from the reconstructed four-momentum of the Higgs boson candidate and the two leading jets assumed to form the VBF topology.
The \VBF production matrix element calculation is a PDF-weighted sum of matrix elements for several partonic processes.
In addition, \textit{decay-level} optimal observables are constructed using the matrix element describing the Higgs boson decay from the reconstructed four-momenta of the four decay leptons.
These decay-level optimal observables are agnostic to the production vertex, and only a single matrix element is needed for an optimal-observable evaluation.
The matrix element for decay-level optimal observables is evaluated separately for same flavour ($4\mu, 4e$) and different flavour ($2e2\mu$) decays to account for the differences of the CP-asymmetry from final state interference.
For each type of optimal observable, there is a choice of which BSM coupling to assume when calculating the interference term.
A dedicated observable is constructed for each of the seven EFT coupling parameters, three for each basis and one for \dTilde.
While each observable is sufficiently general to be sensitive to CP-violating effects from different couplings  to those assumed in its construction, the sensitivity to a CP-odd effect of a particular coupling is maximized using an observable made with the same coupling.
 
\subsection{Morphing method}
\label{sec:meth:morphing}

The distribution of the optimal observable for a signal process with an arbitrary combination of active EFT couplings is predicted as the weighted sum of a limited number of simulated predictions, described in \Sect{\ref{sec:simulation}}. The weights entering the sum are polynomial functions of the assumed EFT couplings, calculated using the \textit{morphing method}~\cite{ATL-PHYS-PUB-2015-047, CMS-HIG-14-018, CMS-HIG-19-009}.
This relies on the fact that, as shown in \Eqn{\ref{eq:CouplingFormula}}, the cross-section in each phase space bin, $\sigma$, is a polynomial function of the EFT coupling when excluding effects on the Higgs boson total width \footnote{These manifest in a rate change and can be factorised out of the consideration for the morphing method.},
\begin{equation}
\sigma = \sum_{k=0}^{4} c^k A_{k} = \vec{c} \cdot \vec{A}, \text{  with  } \vec{c} = \left(1, c, c^2, c^3, c^4\right) \text{ and } \vec{A} = \left(A_0, A_1, A_2, A_3, A_4 \right)
\label{eq:BasicMorphingFormula}
\end{equation}
where $c$ is the EFT coupling under study and $A_k$ denotes the component of the cross-section proportional to the $k^\mathrm{th}$ power of the BSM coupling. If up to two $HVV$ vertices are present, one for production and one for the decay, the maximum power in the BSM coupling is 4 as assumed above.
The same method applies to combinations of multiple active couplings and to different numbers of $HVV$ vertices. In this case, $\vec{c}$ contains all combinations of powers of the BSM couplings appearing in the amplitude, and $\vec{A}$ the corresponding cross-section contributions. The dimensionality of $\vec{A}$ and $\vec{c}$ then corresponds to the number of such unique coupling power combinations, including the case without any BSM couplings (SM-only amplitude). For example, for one vertex and three BSM couplings denoted $c_{1}$, $c_{2}$, and $c_{3}$, there would be ten elements $(1, c_{1}, c_{2}, c_{3}, c_{1}^2, c_{2}^2, c_{3}^2, c_{1} c_{2}, c_{1} c_{3}, c_{2} c_{3})$, and for two vertices and three BSM couplings, the dimensionality would increase to 35.
 
A set of simulated samples for different combinations of coupling values $\vec{c}_{\text{Sample i}}$ can be interpreted as a vector of cross-sections, which depends on the coupling values via a matrix $C$,
\begin{align*}
\vec{\sigma}_{\text{simulated}} =  C \times \vec{A},  \text{ with }  C = \left[\vec{c}_{\text{Sample 1}}, \vec{c}_{\text{Sample 2}}, \ldots\right]
\end{align*}
Inverting this relation, which is only possible if the number of simulated samples is equivalent to the dimensionality of $\vec{A}$, and inserting into \Eqn{\ref{eq:BasicMorphingFormula}} allows the cross-section for an arbitrary coupling set to be expressed as a linear combination of the existing, simulated cross-sections:
\begin{align*}
\sigma  = \vec{c} \cdot \vec{A} = \vec{c} \cdot \left(\vec{C}^{-1} \cdot \vec{\sigma}_{\text{simulated}}\right)  = \sum_j \underbrace{\left(\vec{C}^{-1}_{j} \cdot \vec{c} \right)}_{w_{j}\left(\vec{c}\right)} \sigma_{j}
\end{align*}
where the $w_{j}\left(\vec{c}\right)$ are the weights to be applied to each of the simulated samples to obtain the signal prediction.


\FloatBarrier

\section{ATLAS detector}

The ATLAS detector at the LHC is a multipurpose particle detector
with a forward--backward symmetric cylindrical geometry and a near \(4\pi\) coverage in
solid angle.\footnote{\AtlasCoordFootnote}
It consists of an inner tracking detector (ID) surrounded by a thin superconducting solenoid
providing a \SI{2}{\tesla} axial magnetic field, electromagnetic and hadron calorimeters, and a muon spectrometer (MS).
The ID covers the pseudorapidity range \(|\eta| < 2.5\).
It consists of silicon pixel, silicon microstrip, and transition radiation tracking detectors.
Lead/liquid-argon (LAr) sampling calorimeters provide electromagnetic (EM) energy measurements
with high granularity.
A steel/scintillator-tile hadron calorimeter covers the central pseudorapidity range (\(|\eta| < 1.7\)).
The endcap and forward regions are instrumented with LAr calorimeters
for both the EM and hadronic energy measurements up to \(|\eta| = 4.9\).
The MS surrounds the calorimeters and is based on
three large superconducting air-core toroidal magnets with eight coils each.
The field integral of the toroids ranges between \num{2.0} and \SI{6.0}{\tesla\metre}
across most of the detector.
The MS includes a system of precision tracking chambers and fast detectors for triggering.
A two-level trigger system is used to select events.
The first-level trigger is implemented in hardware and uses a subset of the detector information
to accept events at a rate below \SI{100}{\kHz}.
This is followed by a software-based trigger that
reduces the accepted event rate to \SI{1}{\kHz} on average,
depending on the data-taking conditions.
An extensive software suite~\cite{ATL-SOFT-PUB-2021-001} is used in data simulation, in the reconstruction and analysis of real
and simulated data, in detector operations, and in the trigger and data acquisition systems of the experiment.


\FloatBarrier

\section{Data set and event simulation}
\label{sec:simulation}

This measurement uses data from proton--proton collisions with a centre-of-mass
energy of $13~\TeV$ collected between 2015 and 2018 using single-lepton,
dilepton, and trilepton triggers~\cite{TRIG-2018-01,TRIG-2018-05} as detailed in Ref.~\cite{HIGG-2018-28}.
The combined efficiency of these triggers is approximately 98\%, 99\%, 97\%, and 99\% for the
$4\mu$, $2e2\mu$, $2\mu2e$, and $4e$ final states\footnote{The final state notation represents the mass pair closest to the $Z$ mass followed but the other pair.}, respectively, for the simulated \hfourl\ events
passing the event selection described in \Sect{\ref{sec:reco_selection}}\
(assuming \mH = \SI{125}{\GeV}).
After data-quality requirements are imposed,
the integrated luminosity of the data sample is \Lum~\cite{DAPR-2018-01}.
 
Details of the SM Higgs boson Monte Carlo (MC) samples used in this analysis can be found
in Refs.~\cite{HIGG-2018-28,HIGG-2018-29}.
Higgs boson production via the \ggF process was modelled at next-to-leading-order (NLO) accuracy
in the strong coupling constant \alphas\ using the \POWHEG\ NNLOPS generator~\cite{
Hamilton:2013fea,
Hamilton:2015nsa,
Alioli:2010xd,
Nason:2004rx,
Frixione:2007vw,
Hamilton:2012np,
Campbell:2012am,
Hamilton:2012rf,
Catani:2007vq}
with the \PDFforLHC[15nnlo] set of PDFs~\cite{Butterworth:2015oua}.
Higgs bosons produced via \VBF or in association with a vector boson (\VH) or in association with a top-quark pair (\ttH), were simulated at NLO accuracy with the \POWHEGBOX generator~\cite{Alioli:2010xd,Frixione:2007vw,Nason:2004rx},
using the \PDFforLHC[15nlo] PDF set.
Higgs boson production in association with a top quark (\tH)
was simulated at NLO accuracy using the \MGNLO{} generator~\cite{Alwall:2014hca,Wiesemann:2014ioa} with the
\NNPDF[3.0]\ PDF set~\cite{Ball:2014uwa}.
The production of a Higgs boson in association with a bottom quark pair (\bbH) was simulated at NLO with  \MGNLO{}~\cite{Alwall:2014hca,Wiesemann:2014ioa}, using the CT10 NLO PDF ~\cite{Lai:2010vv}.
For all signal processes, the \EVTGEN[1.6.0] generator~\cite{Lange:2001uf} was used for the simulation of the
bottom- and charm-hadron decays.
Correspondingly, the \PYTHIA[8] generator~\cite{Sjostrand:2007gs}\ was used for
the \htollllbrief decay and for parton showering, hadronization, and simulation of the underlying event.
These samples were normalized to cross-sections obtained from the most recent predictions provided by the LHC Higgs Working Group~\cite{deFlorian:2016spz}.
 
BSM Higgs boson samples with active CP-odd EFT couplings were generated for different coupling values at leading order in $\alpha_{\mathrm{s}}$ with \MGNLO and the \NNPDF[2.3] PDF set.
All diagrams contributing to the $HVV$ vertex for both \VBF/\VH production and vector-boson decays \zz into four lepton final state were generated.
The generation was performed using the effective Lagrangian of the SMEFT framework implemented via the \SMEFTmodel~\cite{Brivio:2017vri, Brivio:2017btx} model.
The CKKW-L method~\cite{Lonnblad:2001iq} was used for the jet merging in the \ggF~process.
The SMEFT v2 model with a U(3)$^5$ flavour symmetry and the $m_W$ input scheme was used.
The EFT cutoff scale was set to $\Lambda=1$~\TeV.
Separate samples were generated for the \ggF production mode and for the combination of the \VBF and \VH production modes.
Each sample was generated at a three-dimensional coupling point for values of the Warsaw basis of Table~\ref{tab:CPcoup}, which also corresponds to a single point in the three Higgs-basis couplings.
The choice of simulated coupling values was optimized to maintain high statistical power throughout the three-dimensional coupling space using the morphing method described in Section~\ref{sec:methodology}.
The optimal observable matrix elements were evaluated with \MGNLO with the \hcufo model and LHAPDF6~\cite{Buckley:2014ana}.
 
The non-resonant \zz background, hereafter denoted as \zzstar, is separated into quark-initiated \qqZZ/\gqZZ,
gluon-initiated \ggZZ, and vector-boson scattering (EW $ZZ$) production,
and was modelled using the \Sherpa[2.2.2] generator~\cite{Gleisberg:2008ta,Gleisberg:2008fv,Cascioli:2011va,Bothmann:2019yzt} with the \NNPDF[3.0nnlo]~\cite{Ball:2014uwa} PDF set.
The $\qqZZ/\gqZZ$ process was generated using matrix elements at NLO accuracy in QCD for up to one additional parton and at LO accuracy for up to three additional parton emissions. NLO EW corrections were applied as a function of the invariant mass of the \zzstar system, $m_{\zzstar}$~\cite{Biedermann:2016yvs,Biedermann:2016lvg}.
The samples for the gluon-induced processes (\ggZZ) were generated using LO-accurate matrix elements for up to one additional parton emission. A NLO QCD calculation~\cite{Caola:2015psa,Caola:2016trd} was used to correct the overall cross-section for this process. The EW $ZZ$ contribution was simulated at LO precision in QCD.

These samples were normalized to cross-sections obtained directly from the \Sherpa\ simulation.
Alternative \qqZZ samples,
produced with the \POWHEGBOX\ v2 and \MGNLO\ generators,
are used to study the theoretical modelling of the systematic uncertainties. For the alternative samples, the \PYTHIA[8] generator was used for parton showering, hadronization, and simulation of the underlying event.
 
Background events from $WZ$ and $t\bar{t}$ processes were generated with \POWHEGBOX[v2],
while contributions from $Z$ bosons produced in association with jets were simulated using the \Sherpa[2.2.1]
generator.
Minor contributions from processes with three electroweak bosons, denoted by $VVV$,
were modelled using \Sherpa[2.2.2]. Small backgrounds originating from top-quark production in
association with one or more electroweak bosons or additional top quarks, such as \ttZ, $tWZ$, \ttWW,
\ttWZ, \ttZg, \ttZZ, \ttt, \tttt, and $tZ$\ (denoted by $tXX$),
were simulated using the \MGNLO\ generator. The \PYTHIA[8] generator was used for parton showering, hadronization, and simulation of the underlying event.

Generated events were processed through the ATLAS detector simulation~\cite{SOFT-2010-01}
within the \GEANT framework~\cite{Agostinelli:2002hh} and
reconstructed in the same way as collision data.
Additional proton--proton interactions in the same or neighbouring bunch crossings, referred
to as \pileup, were included in the simulation.  The \pileup\ events were generated
using the \PYTHIA[8] generator with the A3 set of tuned parameters~\cite{ATL-PHYS-PUB-2016-017}
and the \NNPDF[2.3lo] PDF set.


\FloatBarrier
 
\section{Event reconstruction, selection and backgrounds}
\label{sec:reco_selection}

\subsection{Object reconstruction}
\label{sec:objects}
 
Muon candidates are reconstructed using
a combination of different algorithms~\cite{MUON-2018-03}.
The reconstruction of muon candidates within $|\eta|<2.5$ is primarily
performed by a global fit to reconstructed tracks in the ID and the MS\@.
In the central detector region ($|\eta|<0.1$),
where the MS has reduced geometrical coverage, muons are also identified
by matching a reconstructed ID track to either an MS track
segment (`segment-tagged muons') or a calorimetric energy deposit
consistent with a minimum-ionising particle (`calorimeter-tagged muons').
Calorimeter-tagged muons are required to have transverse momentum $\pt>15$~\GeV to avoid misidentifying leptons at low \pt.
For both the segment-tagged and calorimeter-tagged muons, the muon momentum
is measured from the ID track alone.
In the forward MS region ($2.5<|\eta|<2.7$)
outside the ID coverage, MS tracks with hits in three MS layers
are accepted as `stand-alone muons' and combined with track segments formed from
hits in the silicon tracker, if they exist.
Additionally, `loose' muon-identification criteria~\cite{MUON-2018-03} are applied to reject
low-quality tracks that have missing hits in the MS or have poor
agreement between the reconstructed MS and ID tracks.
Muons are required to be isolated by using
both calorimeter-based and track-based isolation variables and applying the `PflowLoose' criteria~\cite{MUON-2018-03}. They are matched to the hard-scatter vertex candidate by imposing a requirement $\left| z_{0} \sin \theta \right| < 0.5~\text{mm}$ on their longitudinal impact parameter $z_{0}$. Cosmic rays are removed  using a criterion $\left| d_{0} \right| < 1~\text{mm}$ on the transverse impact parameter $d_{0}$.

A reconstructed electron consists of a cluster of energy deposits in the calorimeter and a matched ID track~\cite{EGAM-2018-01}.
Variable-size clusters are created dynamically from
calorimeter-energy deposits, improving the invariant-mass resolution of the four-lepton system,
especially when bremsstrahlung photons are present.
Electron ID tracks are fitted using an
optimized Gaussian-sum filter~\cite{PERF-2017-01} that accounts
for non-linear effects arising from energy loss through bremsstrahlung.
Quality criteria are used to improve the purity of selected electron candidates.
The quality of an electron candidate is evaluated by using a likelihood method
that employs measurements from the tracking system and the calorimeter system,
and quantities that combine both tracking and calorimeter information~\cite{PERF-2017-01}.
The `loose' likelihood criteria, together with track hit requirements, are applied to electron candidates.
Electrons are required to be isolated using
both the calorimeter-based and track-based isolation variables as discussed in Ref.~\cite{HIGG-2018-28}. They are matched to the hard-scatter vertex using the same requirement as for the muons, $\left| z_{0} \sin \theta \right| < 0.5~\text{mm}$.

Jets are reconstructed using a particle flow algorithm~\cite{PERF-2015-09} from noise-suppressed positive-energy topological clusters~\cite{PERF-2014-07} in the calorimeter using the anti-$k_t$ algorithm~\cite{Cacciari:2008gp,Cacciari:2011ma} with a radius parameter $R$~=~0.4. Energy deposited in the calorimeter by charged particles is subtracted and replaced by the momenta of tracks that are matched to those topological clusters.
The jet four-momentum is corrected for the non-compensating calorimeter response, signal losses due to noise threshold effects, energy lost in non-instrumented regions, and contributions from pile-up~\cite{PERF-2016-04,PERF-2015-09,ATLAS-CONF-2016-035}. Jets are required to have $\pt>$~30~\GeV and $|\eta|<$~4.5.
Jets with \pT ranging between 20--60~\GeV\ for $|\eta|<$~2.4 and between 20--120~\GeV\ for $2.5 < |\eta| < 4.5$ are suppressed if tagged as pile-up using two jet-vertex-tagger multivariate discriminants~\cite{ATLAS-CONF-2014-018,PERF-2014-03,ATL-PHYS-PUB-2015-034,ATL-PHYS-PUB-2019-026}.
 
To avoid the double-counting of particles, a range of overlap removal criteria are applied to the electrons, muons, and jets. If any two leptons share an ID track or, in case of two electrons, have overlapping associated calorimeter energy clusters, only one is retained. Preference is given to higher-$\pt$ leptons and to muons over electrons, unless the muon is identified using only the calorimeter, in which case the electron is maintained.
Jets found within $\Delta R <0.2$ of leptons are removed.
 
\subsection{Event selection}
 
Events are required to contain at least four isolated leptons
emerging from a common vertex and forming two pairs of
oppositely charged same-flavour leptons.  Electrons are required to be
within the geometrical acceptance of the ID
($|\eta|<2.47$) and to have transverse energy $\et>7~\gev$, while muons
must be within the geometrical acceptance of the MS ($|\eta|<2.7$) and have $\pt>5~\gev$
(except for calorimeter-tagged muons, as explained in Section~\ref{sec:objects}).
At most one calorimeter-tagged or stand-alone muon is allowed per Higgs boson candidate.
The three higher-($\pt$ or $\et$) leptons in each quadruplet are required
to satisfy thresholds of 20, 15, and 10~\GeV, respectively. All lepton pairs within the quadruplet
must have an angular separation, $\Delta R(\ell, \ell^\prime)$, larger than 0.1, and opposite-sign same-flavour pairs must have a mass above 5~\GeV\ to avoid, for example, J/$\psi$ pairs.
Contributions from misidentified leptons are reduced by requiring the lepton tracks to have low significances of their
transverse impact parameters (less than 3 (5) for $e$ ($\mu$)) and to be compatible with originating from a common vertex.
A detailed description of the event selection can be found in Ref.~\cite{HIGG-2018-28}.
 
The lepton pair with an invariant mass closest to the $Z$~boson mass
in each quadruplet is referred to as the leading dilepton, while
the remaining pair is referred to as the subleading dilepton.
The selected quadruplets are separated into four subchannels,
according to the flavour of the leading and subleading pairs.
In order of decreasing expected selection
efficiency, they are $4\mu$, $2e2\mu$, $2\mu2e$, and $4e$.
Only one quadruplet is selected from each event, based on the mass of the
leading dilepton, the final state, and, for events with additional leptons, the largest leading-order squared matrix element for the SM Higgs boson decay,
as described in Ref.~\cite{HIGG-2018-28}.
Finally, final-state radiation photons are searched for in all  events following
the procedure described in Ref.~\cite{HIGG-2018-28}.
These photons are found in $4\%$ of the events and their energy is included in the mass
computation.

\subsection{Backgrounds}
\label{sec:backgrounds}
 
The dominant background contribution is non-resonant \zzstar\ production, accounting for approximately 30\% of events in the four-lepton invariant mass, \mfourl, signal region, 115--130~\GeV.
The evaluation of the non-resonant \zzstar\ is based on MC simulation, with a data-driven normalization obtained using the side-bands of the $\mfourl$ distribution.
The backgrounds from $Z+$jets, \ttbar and $WZ$, denoted by `reducible' in the following, are evaluated from data following the procedures described in~\cite{HIGG-2018-28, HIGG-2018-29}.
The tri-boson ($WWZ$, $WZZ$ and $ZZZ$) and $\ttbar  X$ backgrounds are evaluated from simulation and together with the reducible backgrounds amount to about 5\% of the events in the signal region.


\FloatBarrier

\section{Overview of analysis}

\subsection{Direct coupling measurement}
\label{sec:direct_coupling}

A binned maximum-likelihood fit to the observed distributions of optimal observables allows the EFT coupling parameters under consideration to be directly measured.
The measurement targets CP-violating effects on the optimal-observable shapes and is optimized to be insensitive to any potential CP-even signature of BSM physics, which would only affect event rates.
To achieve this, the impact of the EFT couplings on the event rates predicted by the signal model is not used in the fit, and only the change in the shapes of the optimal-observable distributions as a function of the BSM couplings is exploited. This is achieved by multiplying the predicted rates of the SM signal prediction and background by additional (a priori, unphysical) free parameters in the statistical model, and then allowing these to float when fitting the model to the data.

The optimal-observable distribution for the signal is predicted using the morphing method described in \Sect{\ref{sec:methodology}}.
In the case of production-level optimal-observable distributions, only the \VBF and \VH Higgs boson production modes are sensitive to BSM effects and considered as signal.
The observable shape for any combination of the three EFT couplings considered here is predicted using 35 simulated samples with different injected EFT coupling values covering both processes.
The number of samples is determined by the number of unique coupling combinations for three BSM couplings and two (production and decay) $HVV$ vertices as discussed for \Eqn{\ref{eq:BasicMorphingFormula}} in Section~\ref{sec:meth:morphing}.
 
All Higgs boson production modes represent a signal when considering decay-level optimal-observable distributions.
For  $\ggF$, the shape of the observable distributions is predicted using a set of 10 simulated samples.
Due to the negligible magnitude of the \ttH, \tH, and \bbH   contributions in the study of decay-level observable distributions, no dedicated morphing is performed for them, and their shape is taken from the $\ggF$ prediction, with the combined  yield accounted for by the free signal rate in the fit.
The shape of the decay observable distributions for $\VBF$ and $\VH$ events is predicted with the same 35 samples described above.
 
For most of the background processes, the optimal-observable shapes are derived directly from the SM MC samples described in \Sect{\ref{sec:simulation}}.
This is the case for the dominant \zzstar continuum background as well as the minor $t\bar{t}Z$/$t\bar{t}WW$ and triboson background contributions.
The background from events with non-prompt leptons has its optimal-observable shape derived from its data-driven analysis, but plays a negligible role in the measurement.
When the \ggF, \ttH, \tH and \bbH production modes are taken as background processes in the production-level optimal-observable measurement, the corresponding shape predictions are taken from the SM Higgs bosons samples described in \Sect{\ref{sec:simulation}}, rather than from the BSM Higgs boson samples, to benefit from the higher theoretical precision of the prediction.
 
Three different types of fits are performed: \textit{decay-only}, \textit{production-only}, and \textit{combined}.
In each case, events are categorized to maximize analysis sensitivity, and all categories are fitted simultaneously.
These categories are described in the following and illustrated in \Fig{\ref{fig:RecoAnalysisRegionSketch}}.

\begin{figure}
\centering
\includegraphics[width = 0.75 \textwidth]{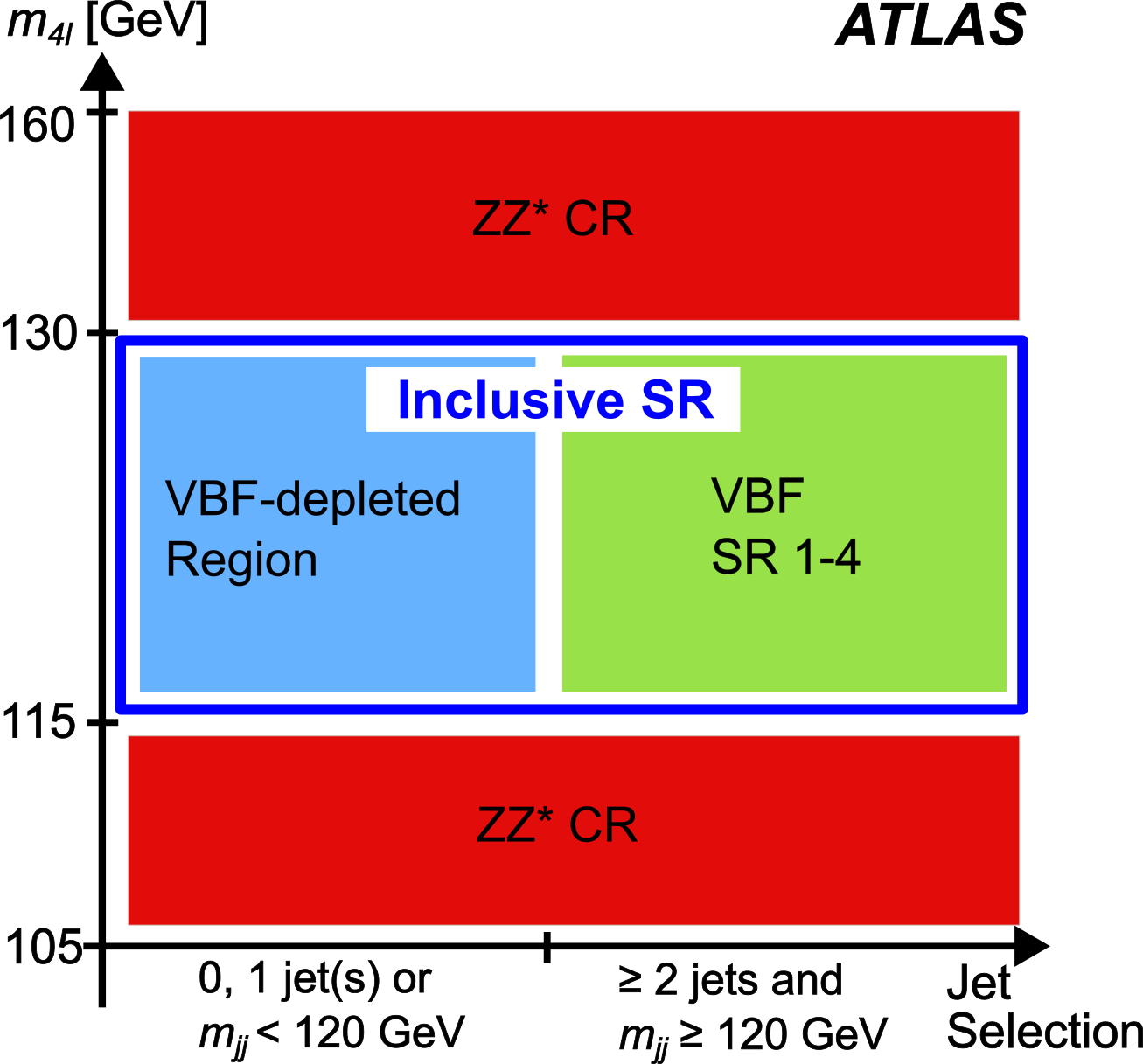}
\caption{Illustration of the regions used in the direct BSM coupling measurement. The horizontal axis represents the jet selection requirements ($\geq 2$ jets and $\mjj \geq 120 \GeV$) defining the \VBF-enriched signal regions (SR), while the vertical axis depicts the separation of the $\mfourl$ range into resonant and non-resonant regions. The `inclusive SR' comprizes events populating either the `VBF-depleted region' or any of the VBF signal regions. The side-band control region (CR), dominated by non-resonant \zzstar decays, is denoted by `\zzstar CR'. The multiple signal regions are used differently in the production-only, decay-only and combined fits.
\label{fig:RecoAnalysisRegionSketch}}
\end{figure}

A \textit{decay-only} fit exclusively targets CP-violating effects in the Higgs boson decay.
Two event categories are used.
The signal region, also referred to as `inclusive SR' in the following, comprises all events with $115\ \GeV < \mfourl < 130\ \GeV$. Events with invariant masses in the side-band of the resonance, $105\ \GeV < \mfourl < 115\ \GeV$ or $130\ \GeV < \mfourl < 160\ \GeV$, form the background-dominated control region.
This side-band, dominated by non-resonant \zzstar decays, is denoted by `\zzstar CR'.
The yield of the Higgs boson decay signal, independent of production mode, is parameterized with a normalization factor relative to the SM prediction.
A second normalization factor scales the yield of the non-resonant $\zzstar$ background relative to the SM prediction. In both cases, the relative distribution of the yield between the signal and control region is assumed to follow the SM prediction within uncertainties as described in \Sect{\ref{sec:systematics}}.
Both normalization factors are free parameters in the fit. In this way, the background rate in the signal region is determined by the control region data yield.
Decay-level optimal observables are considered in the signal region to test for signs of CP violation in the Higgs boson decay and to measure the  BSM couplings under consideration.
 
A \textit{production-only} fit searches for signs of CP-violation at the \VBF production vertex.
This fit uses events in the resonant $\mfourl$ region ($115\ \GeV < \mfourl < 130\ \GeV$) with at least two jets and $m_{jj} > 120\ \GeV$ for the signal extraction.
To further enhance the $\VBF$ signal over other Higgs boson production modes, four VBF signal regions are defined as ascending intervals of the \VBF score of a three-output neural network discriminant~\cite{HIGG-2018-28} trained to distinguish between $\VBF$, $\VH$ and $\ggF$ events.
The discriminant is trained on two-jet events using the di-jet mass, \Mjj, the transverse momentum of the full system, \pTHjj, and the transverse momentum and $\eta$ position of the individual leptons and jets, \ptl, \etal, \ptj and \etaj.
The discriminant interval boundaries are chosen to yield two regions with a large $\ggF$ admixture and two regions with a high $\VBF$ purity, while maintaining a steady variation in the event yield across categories. These are denoted by `VBF SR1' through `VBF SR4'.
The effect of \VH production on the production-only fit is small - after the \mjj $\geq$ 120 \GeV requirement it contributes less than 10\% of the events in the background-dominated \VBF SR1 bin. Although both \VBF and \VH production are affected by a non-zero $HVV$ CP-odd coupling, the \VBF optimal observable exhibits little asymmetry for \VH production.
 
In addition to the four signal regions, two control regions are used.
The first is the same \zzstar CR as used in the decay-only fit, which allows the non-resonant background to be constrained.
Events within the resonant $\mfourl$ region that do not enter the VBF SR regions form a second control region, denoted by the `VBF-depleted region'.
This is enriched in $\ggF$ events that represent the largest source of resonant background for this measurement.
Three separate normalization factors relative to the SM prediction are assigned to the $\VBF/\VH$ signal process, the \ggF/\ttH/\tH/\bbH resonant background and the non-resonant $\zzstar$ background. The factors are common across all categories, and the fraction of the total yield entering each of the analysis categories is taken from the SM prediction.
All three factors are free in the fit.
In addition, for the $\VBF$/$\VH$ signal process, the distribution of events among the four signal regions is allowed to vary in the fit, as BSM effects are expected to modify it.
The fraction of $\VBF$/$\VH$ events in the first two VBF regions is allowed to float in the fit while the last two regions are scaled accordingly to conserve the total yield across the four regions. The floating fraction is forced to be positive to ensure numerical stability of the fit.
Production-level optimal-observable distributions are used in the signal regions (VBF SR~1--4).
 
Finally, a \textit{combined} fit closely follows the production-only strategy. The VBF-depleted region takes the role of an additional signal region, in which decay-level optimal-observable distributions are fitted. Production-level observable distributions are fitted to events populating the four VBF signal regions. The same normalization strategy as for the production-only fit is used. This fit targets BSM signatures that have sensitivity in both the production and decay vertices.
 
In all fits, the reducible background is obtained directly from collision data and its normalization constrained to the measured value within its uncertainty.

\begin{figure}
\centering
\includegraphics[width = 0.8 \textwidth]{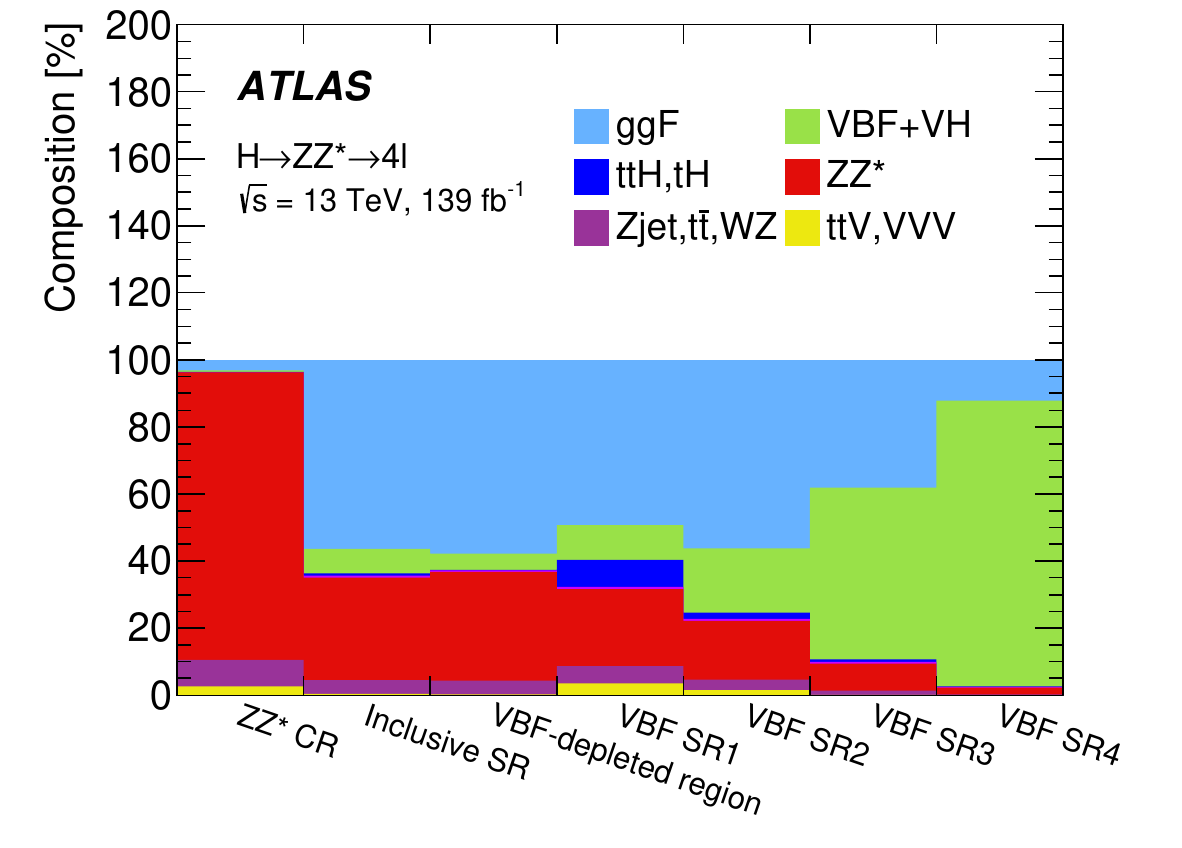}
\caption{The composition of the predicted event yield in the event categories used for the direct EFT coupling measurements.
The contribution from \bbH is negligible and not visible in the plot.
The MC statistical uncertainty in the fraction of each component is not shown but is below the 1\% level.
\label{fig:RegionComposition_analysisSummary}}
\end{figure}
 
\Fig{\ref{fig:RegionComposition_analysisSummary}} shows the composition of the predicted SM yield in each of the analysis categories. The \zzstar control region is highly pure in the non-resonant background.
The VBF-depleted region and inclusive signal region are both dominated by the resonant signal, with a non-resonant admixture of the order of $30\%$. The $\VBF$ signal regions are increasingly pure in the resonant process, with a $30\%$-level non-resonant admixture in the first region. The purity of the $\VBF$ process relative to the $\ggF$ process increases across the four regions, up to a value of $90\%$ in the final \VBF SR~4 category.
 
In the signal regions of all three fits, the optimal-observable distributions are binned in 48 bins for the decay-level observables and in 12 bins for the production-level ones.
The binning is optimized for each coupling assumption in the observable definition. The binning is constructed to yield an approximately flat distribution of expected events across the bins for a sample composed of 50\% SM events and 25\% events each simulated with a positive and a negative BSM coupling value close to the expected 68\% CL limit.

To measure a given BSM coupling, the optimal observable evaluated with a BSM matrix element for the same coupling is used, separately for production ($jj$) and decay ($4\ell$).
This provides optimal sensitivity, as discussed in \Sect{\ref{sec:meth:OO}}.
When simultaneously measuring pairs of BSM couplings using decay-level optimal observables, a two-dimensional binning of the observables targeting the $\cHWtil$ and $\cHBtil$ couplings yields the largest sensitivity.
In this case, six times six bins are used, optimized in the same way as for the one-dimensional binnings.
Finally, when simultaneously measuring the $\tCzz$ and $\tCaa$ couplings using the combined fit set-up, production-level observables targeting the $\tCzz$ coupling in the VBF signal regions are combined with decay-level observables targeting the $\tCaa$ coupling in the VBF-depleted region.
All control regions are treated as single-bin counting experiments.

\Fig{\ref{fig:CPoo1}} shows expected optimal-observable distributions for production-level \OOjjtCzz and decay-level \OOllllcHBtil for the SM and two BSM samples.
The optimal-observable distributions are symmetric in the SM case and asymmetric for CP-odd couplings.
The two distributions have different binning: \protect\subref{fig:CPoo1:jj:reco} has equal size binning, and \protect\subref{fig:CPoo1:llll:reco} has an equally  populated one, for a distribution composed of 50\% SM and 25\% for both $\cHBtil = \pm 0.75$.
Equally populated binning is used in the present analysis.
The distributions of both plots, and the ratios of BSM to SM, clearly demonstrate the importance of the tails for a CP-asymmetry measurement.

\begin{figure}[!h]
\centering
\subfloat[\OOjjtCzz ]{
\includegraphics[width=0.47\textwidth]{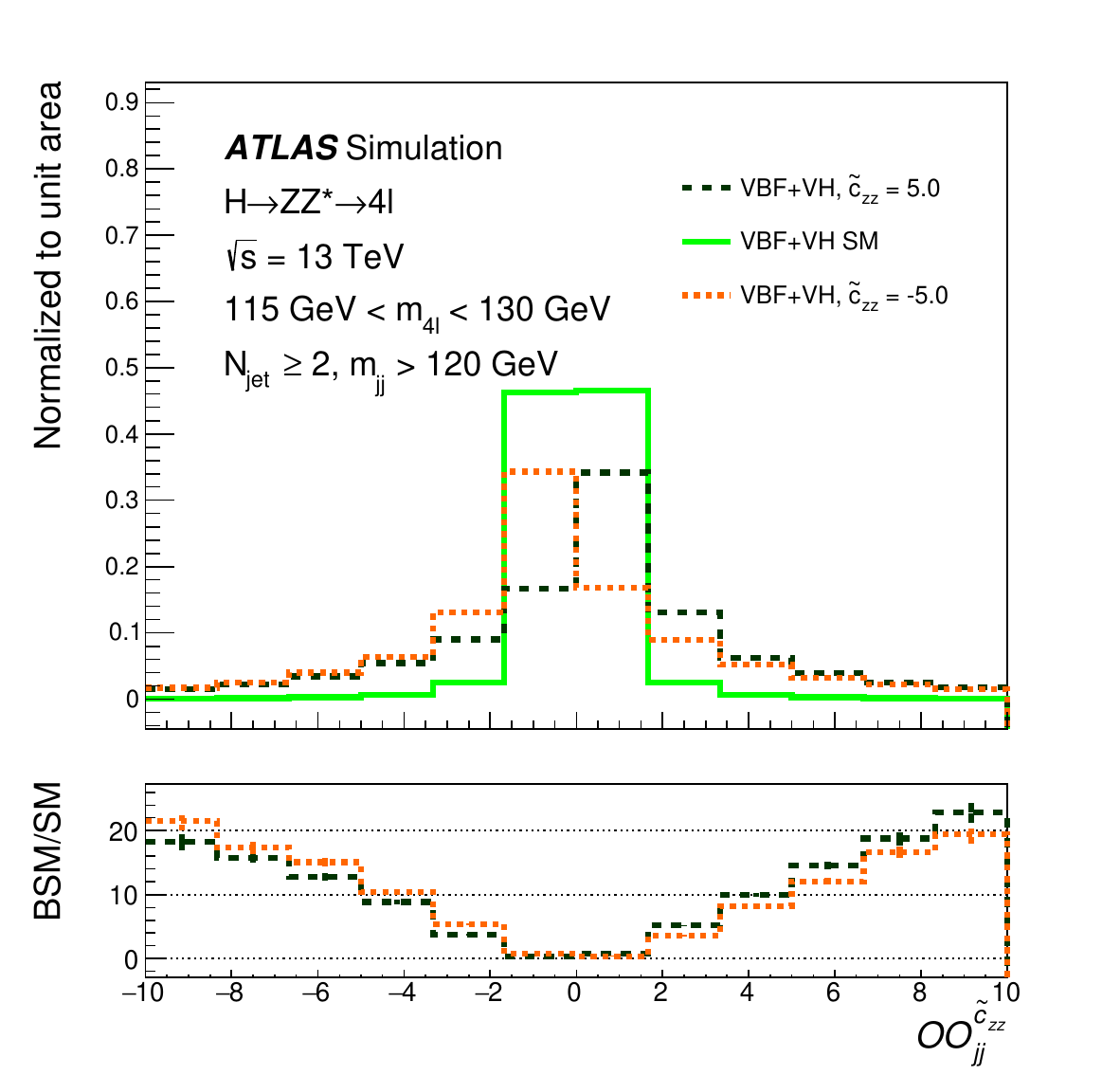} \label{fig:CPoo1:jj:reco}
}
\subfloat[\OOllllcHBtil]{
\includegraphics[width=0.47\textwidth]{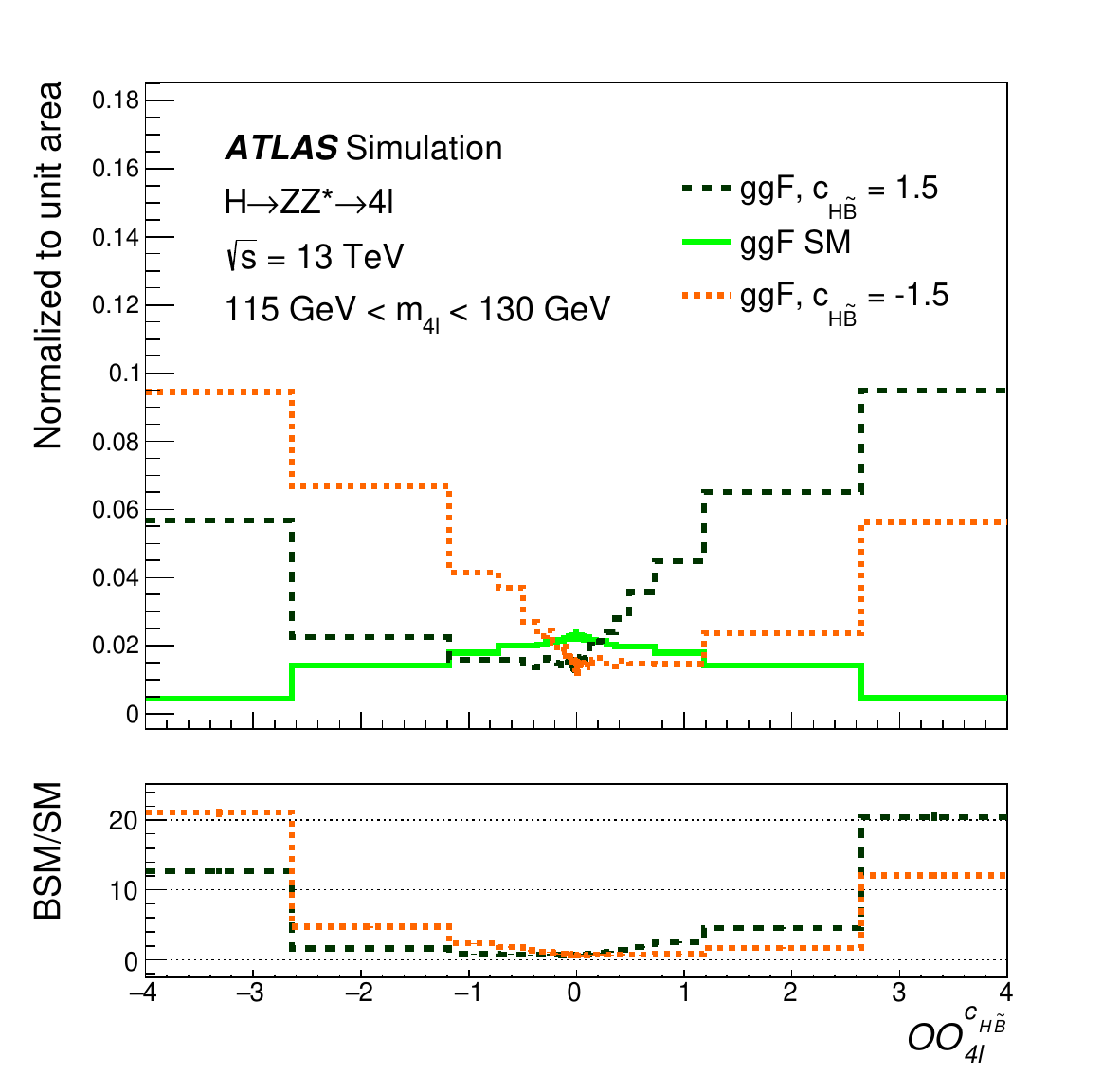}  \label{fig:CPoo1:llll:reco}
}
\caption{Expected distributions of
\protect\subref{fig:CPoo1:jj:reco} the \OOjjtCzz observable for \VBF-produced Higgs boson candidates combining \VBF SR1 through SR4 for the SM and for BSM coupling parameter values of \tCzz$= \pm5$, and \protect\subref{fig:CPoo1:llll:reco} the \OOllllcHBtil observable for the decay of \ggF~produced Higgs boson candidates in the inclusive SR for the SM and for BSM coupling parameter values of \cHBtil$= \pm1.5$.
\protect\subref{fig:CPoo1:jj:reco} has 12 equal-size bins, and \protect\subref{fig:CPoo1:llll:reco} has 48 equal population bins.
In \protect\subref{fig:CPoo1:llll:reco} the first and last bins extend up to $\pm$40, but the bin range is truncated to $\pm$4 in the plot for clarity.
The lower panels show the ratio of the BSM to the SM optimal-observable distributions.
}
\label{fig:CPoo1}
\end{figure}

\subsection{Differential cross-section measurement}
\label{sec:diff_xsec}
 
The main motivation for performing differential cross-section measurements is to have a model-independent result sensitive to possible deviations from the SM\@. The differential cross-section measurement of the optimal observables allows any CP-odd effects affecting the Higgs boson coupling to be probed. Unlike the direct coupling measurement, a differential cross-section measurement is sensitive to possible BSM effects on both the yields and the shape of the observable. The measurement is performed by unfolding the corresponding observable distribution within a fiducial phase space.
 
The fiducial phase space is defined using simulation at particle level and applying selection requirements to minimize model-dependent acceptance extrapolations. These are chosen to closely match the selection requirements of the detector-level analysis after the event reconstruction. The fiducial phase space definition is the same as described in Ref.~\cite{HIGG-2018-28} and summarized in \Tab{\ref{tab:FiducialRegion}}.
 
\begin{table}[th]
\centering
\caption{List of event selection requirements that define the fiducial phase space for the cross-section measurement. SFOC lepton pairs are same-flavour
opposite-charge lepton pairs.
For the mass requirement of the subleading lepton pair, $m_{\textrm{threshold}}$ is 12~\GeV for $m_{4\ell}<140$~\GeV and rises linearly until reaching 50~\GeV for $m_{4\ell} = 190$~\GeV.
\label{tab:FiducialRegion}}
\vspace{0.2cm}
\resizebox{\columnwidth}{!}{\begin{tabular}{ll}
\hline\hline
\multicolumn{2}{c}{\textbf{Leptons and jets}} \\
\hline
Leptons & $\pt>5$ \GeV, $|\eta|<2.7$  \\
Jets &  \pt~$>$ 30 \gev, $|y|<$ \,4.4 \\
\hline
\multicolumn{2}{c}{\textbf{Lepton selection and pairing}} \\
\hline
Lepton kinematics & $\pt > 20, 15, 10 $ \GeV  \\
Leading pair ($m_{12}$) & SFOC lepton pair with smallest $|m_Z-m_{\ell\ell}|$  \\
Subleading pair ($m_{34}$) & Remaining SFOC lepton pair with  smallest $|m_Z-m_{\ell\ell}|$  \\
\hline
\multicolumn{2}{c}{\textbf{Event selection (at most one quadruplet per event)}} \\
\hline
Mass requirements & 50 \GeV$ < m_{12} < 106$ \GeV and $m_{\textrm{threshold}}$ $< m_{34} < 115$ \GeV  \\
Lepton separation & $\Delta R(\ell_{i},\ell_{j}) > 0.1$ \\
Lepton/Jet separation & $\Delta R(\ell_{i}, \textrm{jet}) > 0.1$ \\
$J/\psi$~veto & $m(\ell_{i},\ell_{j}) > 5$ \GeV  for all SFOC lepton pairs  \\
Mass window &  105~\GeV $< m_{4\ell} < 160 $ \GeV  \\
\multicolumn{2}{l}{If an extra lepton with $\pt>12$~\GeV\ is found, the quadruplet with the largest squared matrix element value is kept} \\
\hline\hline
\end{tabular}}
 
\end{table}

Furthermore, as mentioned in Section~\ref{sec:direct_coupling}, it is also interesting to investigate CP-odd effects using the $\VBF$ Higgs boson production vertex. For this reason, a fiducial cross-section is measured in a $\VBF$-enriched phase space.
In addition to the selections in \Tab{\ref{tab:FiducialRegion}}, the \VBF-enriched fiducial region requires the presence of at least two jets with an invariant mass $m_{jj} \ge 400$ \GeV, and a difference between the $\eta$ of the leading and of the subleading jet of $|\Delta \eta_{jj}| \ge 3.0$.
These selections replace the neural network used to define the \VBF SR regions of the direct coupling measurement.
Two separate fiducial cross-sections are extracted from the \VBF-enriched fiducial region, resulting in different levels of \VBF purity.
For the first, a `combined production mode' Higgs boson signal is extracted with \ggF, \VBF, \VH and \ttH considered as signal.
For the second the Higgs boson signal is extracted while additionally removing the $\ggF$ production mode signal by treating it as background using the detector-level MC expectation.
This \VBF-enriched fiducial region has a signal composition of $59.3 \%$ VBF, $37.5\%$ \ggF, $1.7\%$ VH and $1.5\%$ \ttH. The VBF purity increases to 95\% if the \ggF is removed as a background.
Although the first fiducial cross-section is less pure in \VBF signal, it is less model dependent because it has no assumption on the signal composition.
The second fiducial cross-section relies on the assumption of SM \ggF production.

To extract the number of signal events in each bin of a differential distribution, invariant mass templates for the Higgs boson signal and the background processes are fitted to the $m_{4\ell}$ distribution in data, following the same strategy as used in Ref.~\cite{HIGG-2018-28}.
The non-resonant \zzstar background is fitted simultaneously with the signal and constrained by extending the $m_{4\ell}$ fit range from 115--130~\GeV to 105--160~\GeV, similar to the control region used in the direct coupling measurement.
Furthermore, the SM $\zzstar \rightarrow 4\ell$ decay fractions are assumed.
To reduce the effects of bin-to-bin statistical fluctuations of the non-resonant \zzstar background, neighbouring bins are combined for this background estimate.
The overall normalization and shape of the reducible background from a data-driven evaluation (\Sect{\ref{sec:backgrounds}}) are allowed to vary within their associated systematic uncertainties (Section~\ref{sec:systematics}).
 
The expected number of events $N_{i}$ in each detector-level observable bin $i$, expressed as a function of  $m_{4\ell}$, is given by
\begin{align}
\label{eq:fidXS_matr}
N_{i}(m_{4\ell})=\sum_{j} r_{ij}\cdot(1+f_{i}^{\mathrm{non fid}})\cdot\sigma^{\mathrm{fid}}_{j}\cdot\mathcal{P}_i(m_{4\ell})\cdot\mathcal{L}+N^{\mathrm{bkg}}_{i}(m_{4\ell})
\end{align}
with the fiducial cross-section in each fiducial bin $j$ given by
\begin{align*}
\sigma_{j}^{\mathrm{fid}}= \sigma_j \cdot A_j \cdot \mathcal{B}
\end{align*}
 
where $A_j$ is the acceptance in the fiducial phase space for a total cross-section $\sigma_j$ in fiducial bin $j$, $\mathcal{L}$ is the integrated luminosity, $\mathcal{B}$ is the decay branching ratio and $N^{\mathrm{bkg}}_{i}(m_{4\ell})$ is the background contribution. The index $j$ runs over all observable bins in the fiducial phase space.
The term $\mathcal{P}_i(m_{4\ell})$ is the $m_{4\ell}$ signal shape containing the fraction of events as a function of $m_{4\ell}$ expected in each reconstruction bin, taken from SM MC simulation. The term  $r_{ij}$ represents the detector response matrix, created with SM simulated signal samples and averaged across the different production modes using the expected SM cross-sections~\cite{deFlorian:2016spz}. These factors correspond to the probability that an event generated within the fiducial volume in the observable bin $j$ is reconstructed in bin $i$.
The normalization, $f_{i}^{\mathrm{non fid}}$, represents the fraction of events that are outside of the fiducial region but are reconstructed within the SR\@. This ranges from 1.1\% to 1.7\% depending on the bin of the unfolded observable or the final state. The detector response matrix accounts for bin-to-bin migrations in the unfolding of the signal. The binning choice made for all observables ensures a statistical significance of more than two standard deviations for the SM signal process. The binning is also chosen to minimize migrations between bins. In most cases, the bin width is more than twice the experimental  resolution.
 
For the \VBF-enriched fiducial cross-section measurements, the signal is extracted by unfolding the two-dimensional distribution $m_{jj}$ versus $|\Delta \eta_{jj}|$ that is divided into three bins (see \Fig{\ref{fig:VBF_mjj_detajj}}),
following the strategies explained previously. The difference for the second measurement is that the \ggF is considered to be a background and the expected \ggF signal is moved from $\mathcal{P}_i(m_{4\ell})$ to $N^{\mathrm{bkg}}_{i}(m_{4\ell})$ of \Eqn{\ref{eq:fidXS_matr}}.


\FloatBarrier

\section{Systematic uncertainties}
\label{sec:systematics}

 
The statistical uncertainty in the data is generally dominant in all aspects of the analysis, and systematic uncertainties play a minor role. Systematic uncertainties arise from experimental sources, such as lepton and jet momentum scales, resolutions and reconstruction efficiencies, and theoretical uncertainties related to the modelling of the signal and background processes. In addition, a set of uncertainties specific to the cross-section extraction method affect the differential cross-section measurement.
 
As the rates of all key contributing processes are determined using the collision data, normalization uncertainties affecting their total calculated cross-sections have no impact.
Uncertainties in the determination of the kinematic observables characterising the final state, in particular jet and lepton momentum scale and resolutions, affect the analysis in the most direct way through the reconstructed value of the optimal observables.
Uncertainties in the relative rates between the various categories, arising for example from theoretical sources, affect the results indirectly via changes in the process composition in the signal regions and migration effects in the cross-section measurement. While still small compared to statistical uncertainties, these composition uncertanties are the leading source of systematic uncertainty in the direct coupling measurement.
The measured differential cross-sections are also sensitive to luminosity and lepton reconstruction and isolation efficiency uncertainties, which play a negligible role for the direct coupling measurement.
 
When predicting BSM effects in the direct coupling measurement, theoretical uncertainties due to the PDF, parton shower, and QCD scale variations are assigned to the signal predictions based on SM signal samples, which are simulated with higher-order precision. The same uncertainties are assigned to all corresponding BSM signal predictions, since it is observed using the MC signal samples simulated at leading-order accuracy that the uncertainties change negligibly as a function of the Wilson coefficients.
 
An overview of the main sources and their impact on the analysis is provided below. When quantifying the impact of uncertainties on the direct coupling measurement, where no evidence for a non-zero coupling value is obtained in the fit to collision data, the impacts on a hypothetical BSM signal obtained using simulation are provided instead.
 
\subsection{Experimental uncertainties}
 
The uncertainty in the measured differential cross-sections due to pile-up modelling is of the order of 1\%.
The uncertainty in the combined 2015--2018 integrated luminosity is 1.7\%~\cite{ATLAS-CONF-2019-021}, obtained using the LUCID-2 detector~\cite{LUCID2} for the primary luminosity measurements. The luminosity uncertainty directly propagates to the measured cross-sections. Neither source affects the direct coupling measurement.
 
The electron (muon) reconstruction and identification efficiency uncertainties in the extracted cross-sections are of the order of  1--2\% ($\leq 1\%$). Isolation efficiency uncertainties affect the results at a similar magnitude. Lepton efficiencies have a negligible impact on the direct coupling measurement since they do not induce asymmetries in the CP-sensitive observables.
 
Lepton energy, momentum scale and resolution uncertainties affect both the direct coupling measurement and the extracted cross-sections at the order of 1--2\% level due to their relation to the reconstructed rest-frame kinematics determining the optimal observables.
 
The impact of uncertainties in the jet energy scale and resolution is only relevant for production-level optimal observables. There, they affect the differential cross-sections at the 1--5\% level and the directly extracted coupling values at the 1--3\% level.
 
The impact of the precision of the Higgs boson mass measurement~\cite{HIGG-2014-14} on the signal acceptance due to the signal region mass-window requirement is considered for the cross-section extraction, but found to have a negligible impact ($<1\%$).
 
For the data-driven measurement of the reducible background, three sources of uncertainty are considered: statistical uncertainty, overall systematic uncertainty for muon and electron background fake estimates~\cite{HIGG-2018-28}, and a shape systematic uncertainty that varies with the differential variable~\cite{HIGG-2018-29}. All of these have a negligible impact on the analysis.
 
\subsection{Theoretical uncertainties}
\label{sec:systematics_theory}
 
The theoretical modelling of the signal and background processes is affected by uncertainties due to missing higher-order corrections, modelling of parton showers and the underlying event, and PDF plus $\alphas$ uncertainties.
 
One of the dominant sources of theoretical uncertainty is the prediction of the \ggF process, in particular its contribution to the dijet signal regions targeting production observables.
To estimate the variations due to the impact of higher-order contributions not included in the calculations, the approach described in Refs.~\cite{deFlorian:2016spz,Stewart:2011cf} is used, which exploits the latest predictions for the inclusive jet cross-sections. In particular, the uncertainty from the choice of factorization and renormalization scales, the choice of resummation scales, and the migrations between the 0-jet and 1-jet phase space bins or between the 1-jet and $\geq 2$-jet bins are considered.
The detailed procedure is described in Ref.~\cite{HIGG-2018-28}. The impact of these uncertainties can reach $5\%$ of the directly measured coupling value and 1--2\% of the measured differential cross-sections.
 
The uncertainties in the acceptance and distribution of yields among the analysis regions due to the modelling of parton showers and the underlying event are estimated with \AZNLO tune~\cite{STDM-2012-23} eigenvector variations and by comparing the results using the parton showering algorithm from \PYTHIA[8] with that from \HERWIG[7]~\cite{Bellm:2015jjp} for all signal processes. The eigenvector variations are correlated between the Higgs boson production modes, while the algorithm uncertainty is treated as uncorrelated. This is the largest individual systematic uncertainty source when considering production-level optimal observables, with a magnitude in the range of 2--10\% of the measured cross-sections in the differential measurement and 1--5\% of the directly measured BSM coupling values.
 
For the \VBF, \VH and \ttH production modes, the uncertainty due to missing higher orders in QCD is parameterized using the scheme outlined in Ref.~\cite{bendavid2018les}.
Compared to the uncertainties in the $\ggF$ contribution, both the scale and the shower uncertainties in these predictions are negligible.
 
The impact of the PDF uncertainty is estimated with the
30 eigenvector variations of the \PDFforLHC[nlo30] Hessian PDF set following the PDF4LHC recommendations~\cite{Butterworth:2015oua}. The modification of the predictions
originating from each eigenvector variation is added as a separate source of uncertainty in the model. The same procedure is applied for the \ggF, \VBF, \VH and \ttH processes.
PDF uncertainties have a negligible impact.
 
Beside the resonant process, uncertainties in the prediction of the main non-resonant \qqZZ background process affect the analysis. As the normalization of this background is obtained from collision data, the main impact is from migration effects between various regions.
The uncertainties due to missing higher-order effects in QCD are estimated by varying the
factorization and renormalization QCD scales by a factor of two; the impact of the PDF uncertainty
is estimated by using the MC replicas of the \progname{NNPDF3.0} PDF set. Uncertainties due to parton shower modelling for the \zzstar\ process are considered as well.
For the differential cross-section measurement, a comparison of the $\zzstar$ predictions of the $\mfourl$ spectrum used to normalize this background to the ones obtained using the alternative generators described in Section~\ref{sec:simulation} is used to define an additional systematic uncertainty.
The impact of these uncertainties is very small, less than 2\% on the directly measured couplings and negligible for the differential cross-sections.
 
\subsection{Uncertainties specific to the differential cross-section extraction}
Unfolding-related uncertainties arise from the production mode composition that determines the average response matrices, and from residual biases introduced by the unfolding method~\cite{HIGG-2018-29}. For the former, an uncertainty is assessed by varying the production cross-sections within their measured uncertainties taken from Ref.~\cite{HIGG-2016-22}, with an impact of less than 1\%.
The uncertainty arising from unfolding biases is obtained independently for each bin by comparing the unfolded cross-section from simulation with that expected when varying the underlying particle-level cross-sections of the simulated data sample within the expected statistical error. In the same way, the uncertainty is extended to also account for differences that arise from using an SM response matrix to measure observable distributions possibly affected by BSM behaviour by means of signal injection (up to 95\% CL of a possible CP-odd coupling).
The impact of this uncertainty is typically around 1--2\% in the decay optimal observable, where the response matrix is largely diagonal, and increases up to 7\% in the production optimal-observable distributions, which have larger bin migrations.
 
\subsection{Monte Carlo statistics}
The effect of MC statistical uncertainties is included in the statistical model used for the measurement, as the direct coupling measurement relies on precise shape predictions for the optimal observables and other systematic uncertainties are very small.
The signal modelling technique results in a negligible impact of statistical uncertainties on the prediction of BSM effects. The largest contribution to the MC statistical uncertainty is linked to the non-resonant \qqZZ background prediction, affecting the measured coupling values at the level of few percent. This is negligible compared to statistical uncertainties in the data, but comparable (1--2\% impact) to the leading systematic uncertainty sources for coupling measurements using decay-level observables, due to the small size of the systematic uncertainties in these measurements.


\FloatBarrier

\section{Results}


\subsection{Direct coupling measurement results}
\label{sec:direct_results}

The number of events observed and the expected (pre-fit) contributions in each of the categories used in the coupling measurements,  described in \Sect{\ref{sec:direct_coupling}} and \Fig{\ref{fig:RecoAnalysisRegionSketch}}, are shown in \Fig{\ref{fig:cat_data}} and \Tab{\ref{tab:recoyields}}.
Combined statistical and systematic uncertainties are included for the predictions (see \Sect{\ref{sec:systematics}}).
As discussed in \Sect{\ref{sec:direct_coupling}}, the \zzstar CR and Inclusive SR contribute to the decay observable fits, and the rest contribute to the production-only and combined observable fits.
In all categories, the  number of events observed and the expectation agree within one to two standard deviations. The largest difference occurs in the most sensitive \VBF category, \VBF SR4, where the small number of observed events exceeds the expectation by about a factor of two.
\newenvironment{DIFnomarkup}{}{}

\begin{DIFnomarkup}
\begin{table*}[htbb!]
\centering
\caption{The numbers of expected and observed events in the event categories for the direct CP measurement, in the mass range \hbox{115~\gev$< \mfourl <$ 130~\gev}.
The sum of the number of expected SM Higgs boson events and the estimated background yields is compared with the data. Combined statistical and systematic uncertainties are included for the predictions.
The Inclusive SR, used in the decay-only fit, is composed of the \VBF-depleted Region and \VBF SR1 to SR4.
\label{tab:recoyields}}
\vspace{0.4cm}
\resizebox{\columnwidth}{!}{\begin{tabular}{c|*{2}{r@{\,}c@{\,}l|}*{5}{r@{\,}c@{\,}l}}
 
\hline
& \multicolumn{3}{c|}{\multirow{2}{*}{\zzstar CR}}
& \multicolumn{3}{c|}{\multirow{2}{*}{Inclusive SR}}
& \multicolumn{3}{c}{VBF-depleted}
& \multicolumn{12}{c}{VBF}  \rule{0pt}{2.6ex} \\
& \multicolumn{3}{c|}{}
& \multicolumn{3}{c|}{}
& \multicolumn{3}{c}{Region}
& \multicolumn{3}{c}{SR1}
& \multicolumn{3}{c}{SR2}
& \multicolumn{3}{c}{SR3}
& \multicolumn{3}{c}{SR4} \rule[-1.2ex]{0pt}{0pt} \\
\hline
\ggF       & $8.2$ & $\pm$ & $1.3$
& $181$ & $\pm$ & $12$
& $165$ & $\pm$ & $12$
& \multicolumn{3}{c}{$7.5^{+3.0}_{\text{--}2.4}$}
& \multicolumn{3}{c}{$5.6^{+1.8}_{\text{--}1.5}$}
& $2.2 $ & $\pm$ & $0.6$
& $0.49$ & $\pm$ & $0.17$ \rule[-1ex]{0pt}{4ex}  \rule{0pt}{2.6ex} \\
\bbH       & \multicolumn{3}{c|}{$0.087^{+0.016}_{\text{--}0.015}$}
& $1.85$ & $\pm$ & $0.05$
& $1.65$ & $\pm$ & $0.05$
& $0.11$ & $\pm$ & $0.01$
& \multicolumn{3}{c}{$0.072^{+0.010}_{\text{--}0.009}$}
& \multicolumn{3}{c}{$0.020^{+0.005}_{\text{--}0.003}$}
& \multicolumn{3}{c}{$<0.01$}  \rule[-1ex]{0pt}{4ex}  \\
\VBF/\VH   & $1.39$ & $\pm$ & $0.16$
& $23.8$ & $\pm$ & $0.7$
& $13.8$ & $\pm$ & $0.6$
& \multicolumn{3}{c}{$1.60^{+0.09}_{\text{--}0.08}$}
& $1.89$ & $\pm$ & $0.11$
& $3.01$ & $\pm$ & $0.18$
& $3.5$  & $\pm$ & $0.4$ \rule[-1ex]{0pt}{4ex}\\
\ttH,\tH   & \multicolumn{3}{c|}{$0.22^{+0.03}_{\text{--}0.04}$}
& \multicolumn{3}{c|}{$1.89^{+0.21}_{\text{--}0.22}$}
& $0.44$  & $\pm$ & 0.05
& $1.22$  & $\pm$ & $0.14$
& $0.179$ & $\pm$ & $0.023$
& \multicolumn{3}{c}{$0.046^{+0.009}_{\text{--}0.010}$}
& \multicolumn{3}{c}{$<0.01$}  \rule[-1ex]{0pt}{4ex}\\
\ttV,$VVV$ & $6.79$  & $\pm$ & $0.13$
& $1.31$  & $\pm$ & $0.06$
& $0.62$  & $\pm$ & $0.04$
& $0.53$  & $\pm$ & $0.04$
& $0.150$ & $\pm$ & $0.020$
& \multicolumn{3}{c}{$<0.01$}
& \multicolumn{3}{c}{$<0.01$} \rule[-1ex]{0pt}{4ex}\\
\zzstar    & \multicolumn{3}{c|}{$229^{+20}_{\text{--}25}$}
& \multicolumn{3}{c|}{$98^{+6}_{\text{--}9}$}
& \multicolumn{3}{c}{$92^{+6}_{\text{--}8}$}
& \multicolumn{3}{c}{$3.5^{+1.3}_{\text{--}1.7}$}
& $1.7$ & $\pm$ & $0.6$
& \multicolumn{3}{c}{$0.48^{+0.16}_{\text{--}0.15}$}
& \multicolumn{3}{c}{$0.086^{+0.025}_{\text{--}0.028}$} \rule[-1ex]{0pt}{4ex}\\
$Z$jet,$t\bar{t}$,$WZ$
& $21$   & $\pm$ & $5$
& $13$   & $\pm$ & $4$
& $12$   & $\pm$ & $3$
& $0.8$  & $\pm$ & $0.9$
& $0.3$  & $\pm$ & $0.6$
& $0.07$ & $\pm$ & $0.26$
& $0.01$ & $\pm$ & $0.09$ \rule[-1ex]{0pt}{4ex} \rule[-1.2ex]{0pt}{0pt} \\
\hline
Total SM   & \multicolumn{3}{c|}{$267^{+21}_{\text{--}26}$}
& \multicolumn{3}{c|}{$321^{+14}_{\text{--}15}$}
& \multicolumn{3}{c}{$286^{+14}_{\text{--}15}$}
& $15$  & $\pm$ & $3$
& \multicolumn{3}{c}{$9.9^{+2.0}_{\text{--}1.7}$}
& $5.9$ & $\pm$ & $0.7$
& $4.1$ & $\pm$ & $0.5$ \rule[-1ex]{0pt}{4ex} \rule{0pt}{2.6ex}\rule[-1.2ex]{0pt}{0pt} \\
[3pt] \hline
Data       & \multicolumn{3}{c|}{$294$}
& \multicolumn{3}{c|}{$311$}
& \multicolumn{3}{c}{$276$}
& \multicolumn{3}{c}{$14$}
& \multicolumn{3}{c}{$9$}
& \multicolumn{3}{c}{$4$}
& \multicolumn{3}{c}{$8$} \rule{0pt}{2.6ex}\rule[-1.2ex]{0pt}{0pt} \\
\hline
\end{tabular}}
 
\end{table*}
\end{DIFnomarkup}

\begin{figure}[!htbp]
\centering
\includegraphics[width=0.8\textwidth]{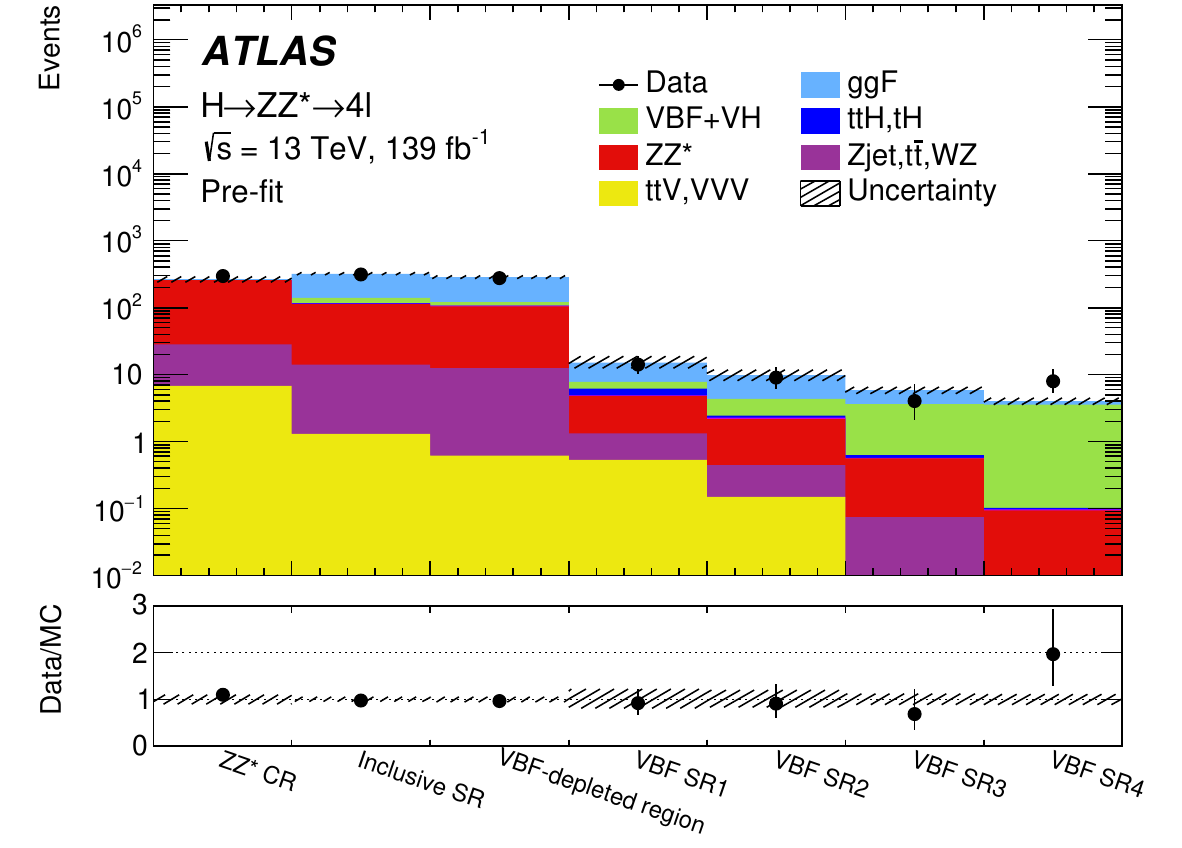}
\caption{
The number of observed events and the expected contributions in each of the categories used in the decay-only, production-only, and combined observable fits for the coupling measurements.
The contribution from \bbH is negligible and not visible in the plot.
The lower panel shows the ratio of data to expectation.
The shaded bands in both the upper and lower panels correspond to the combined statistical and systematic uncertainties in the predictions.
}
\label{fig:cat_data}
\end{figure}

The observed and expected distributions for the production-level observable \OOjjtCzz  in each of the four \VBF signal regions are shown in
\Fig{\ref{fig:tCzz_obs_data}}.
The corresponding observed and expected distributions for the decay level observables \OOllllcHBtil, \OOllllcHWBtil, \OOllllcHWtil, and \OOlllldTilde are shown in \Fig{\ref{fig:oollll_obs_data}}.
The expected distributions for the SM are shown as stacked histograms, and the two lines show the expectation for an additional positive or negative CP-odd BSM coupling that has a clear excess in the tail.
The combined statistical and systematic uncertainties are shown in the ratio plot below each distribution.
Overall there is good agreement between data and the SM expectation.
The mean value for the data is given for each distribution, calculated from the observable value for each event.
All mean values are compatible with zero, indicating that the data exhibit no measurable asymmetry.

\begin{figure}[!htbp]
\centering
\subfloat[\VBF SR1 ]{
\includegraphics[width=0.47\textwidth]{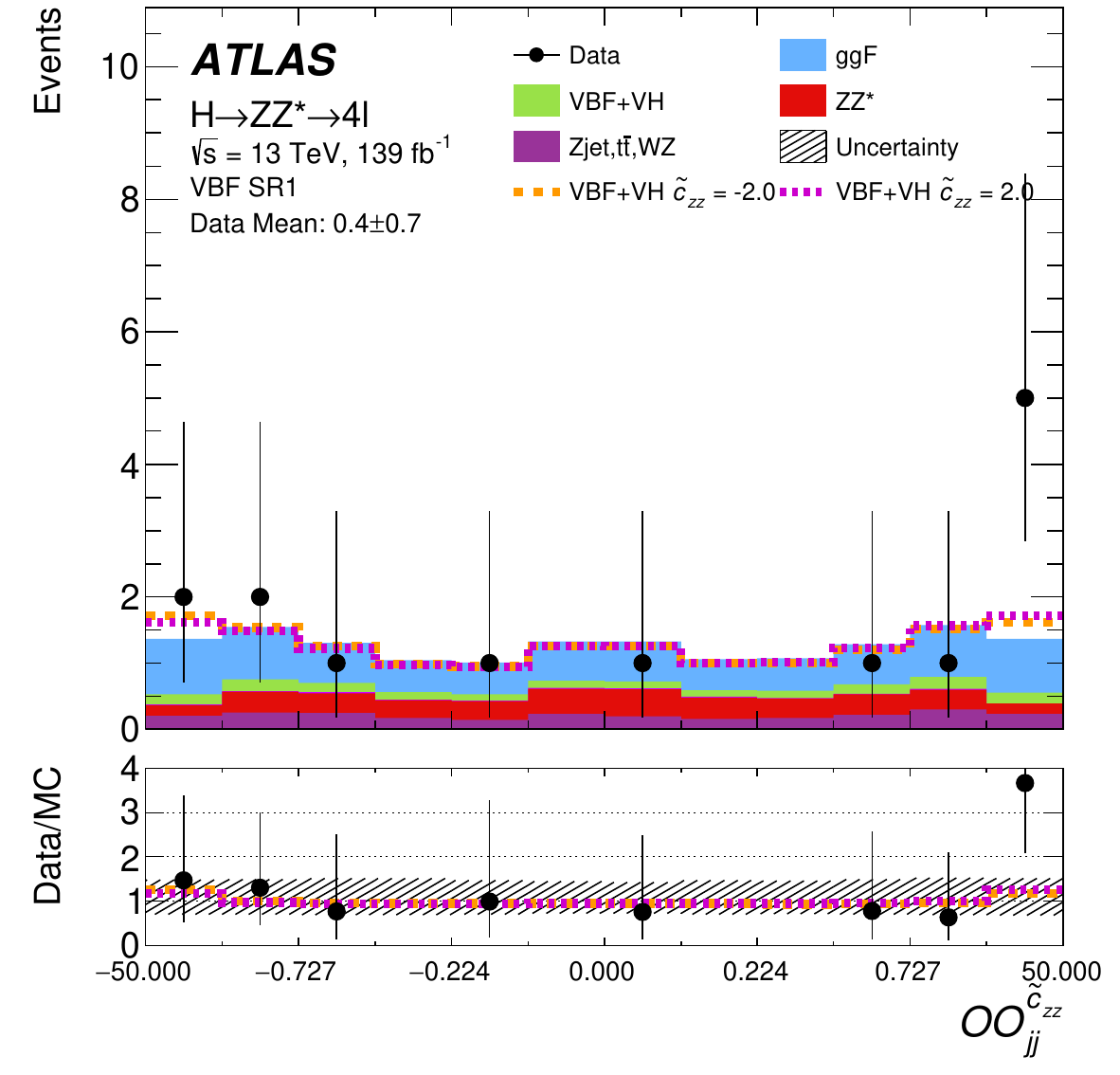} \label{fig:tCzz_obs_data_sr1}
}
\subfloat[\VBF SR2 ]{
\includegraphics[width=0.47\textwidth]{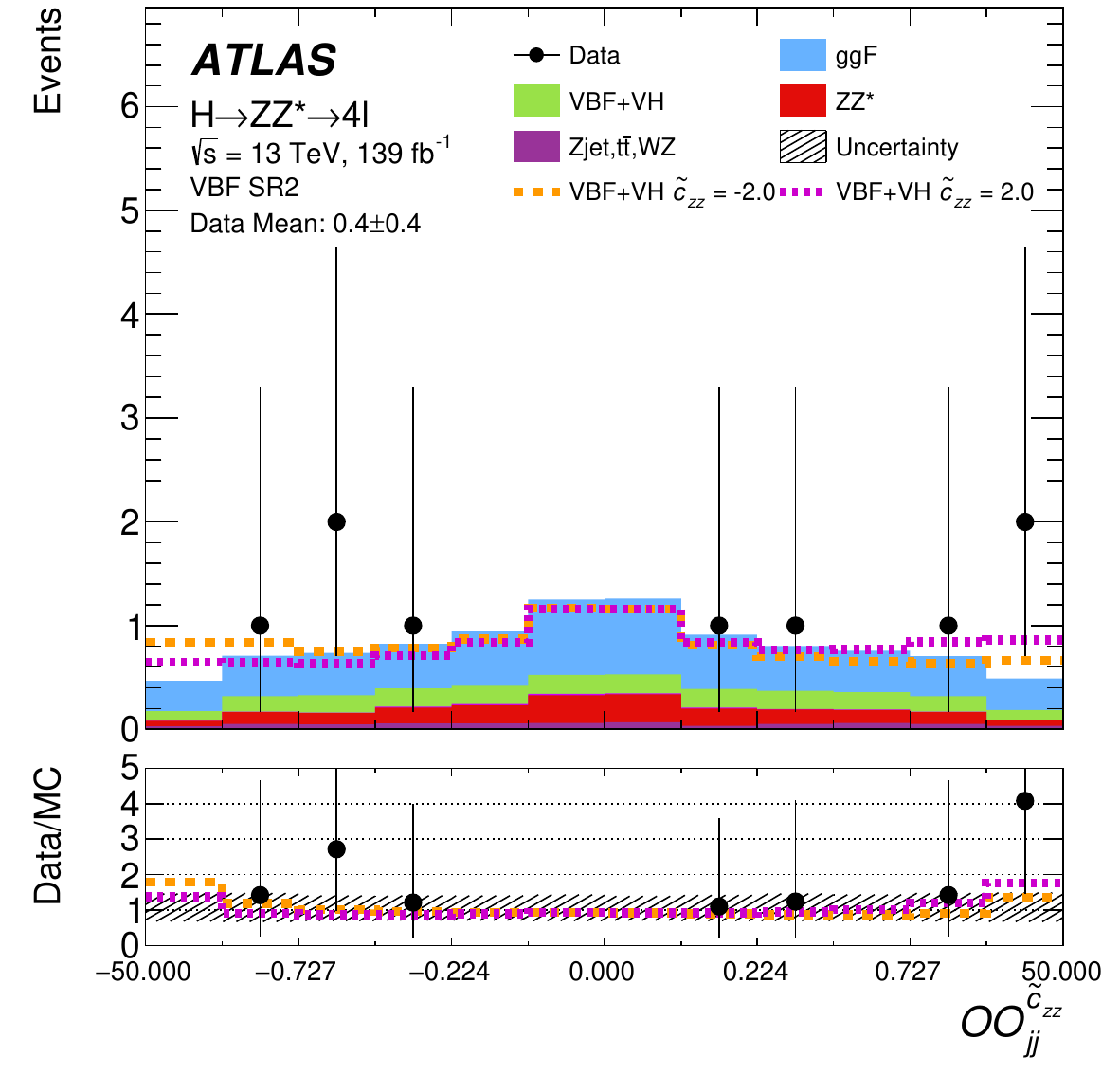} \label{fig:tCzz_obs_data_sr2}
}
\quad
\subfloat[\VBF SR3 ]{
\includegraphics[width=0.47\textwidth]{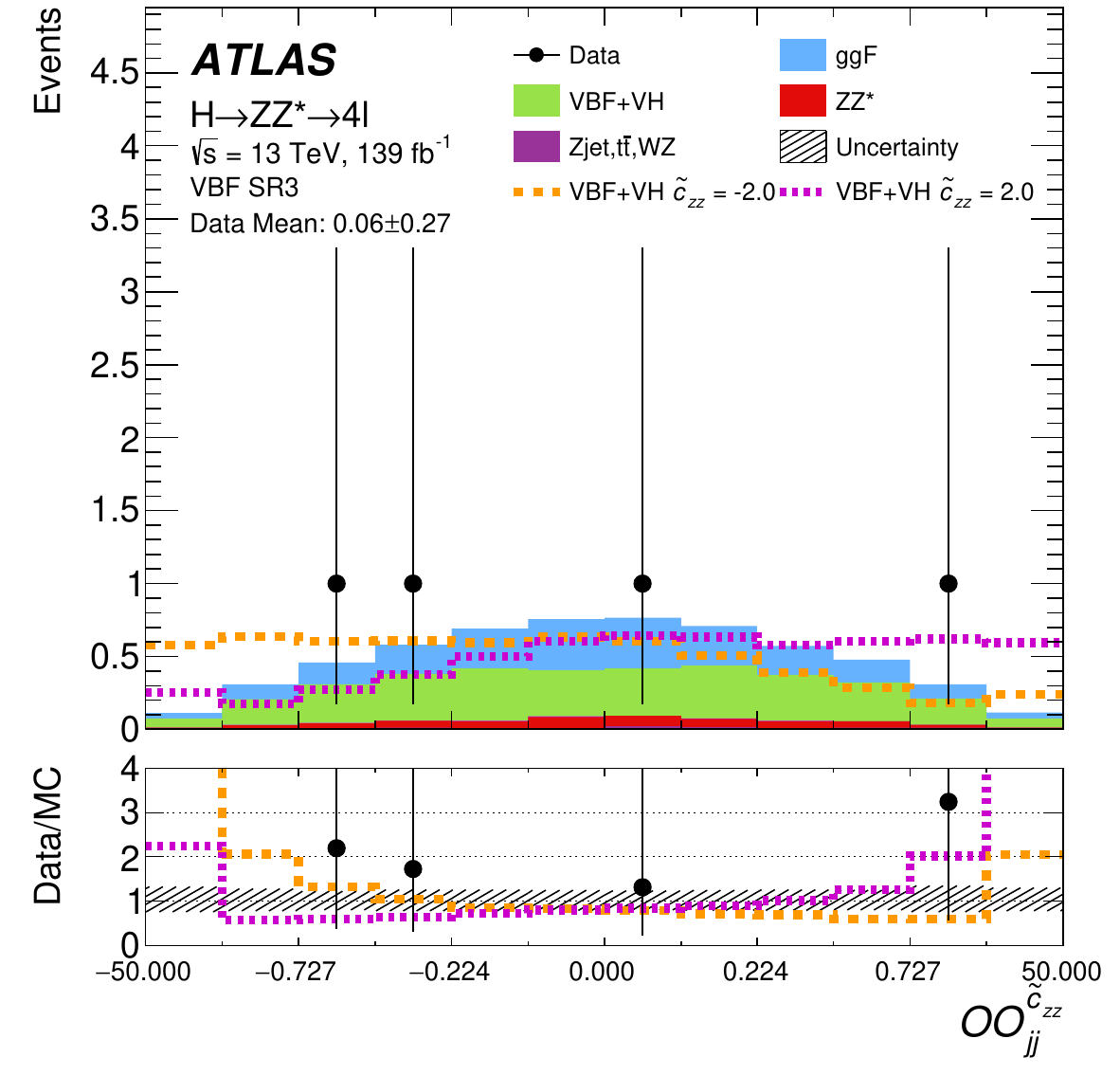} \label{fig:tCzz_obs_data_sr3}
}
\subfloat[\VBF SR4 ]{
\includegraphics[width=0.47\textwidth]{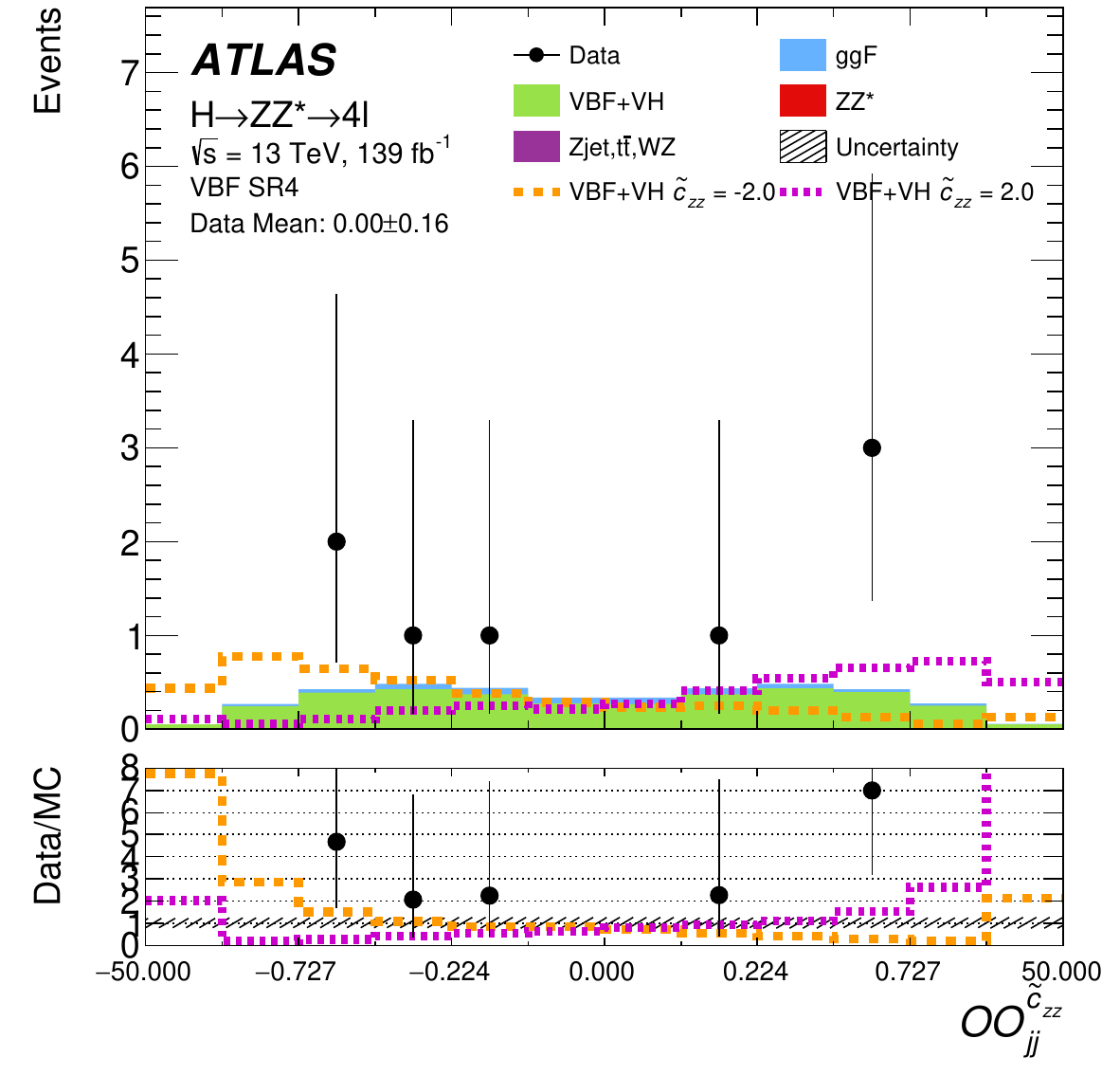} \label{fig:tCzz_obs_data_sr4}
}
\caption{
The observed and expected distributions for the production-level observable \OOjjtCzz  in each of the four \VBF signal regions:
\protect\subref{fig:tCzz_obs_data_sr1} SR1, \protect\subref{fig:tCzz_obs_data_sr2} SR2, \protect\subref{fig:tCzz_obs_data_sr3} SR3, and
\protect\subref{fig:tCzz_obs_data_sr4} SR4.
The expected distribution for the \VBF signal and the backgrounds from \ggF, \zzstar, and reducible backgrounds are shown as a stacked histogram.
Also shown are the expected distributions in presence of CP-odd contributions for $\tCzz = \pm2.0$, assuming the $\VBF/\VH$ event rate remains that of the SM\@.
The bin size is variable, optimized for an approximately flat distribution for the SM combined with a positive and negative BSM scenario.
All events have optimal observable values within the limits of the plots.
The mean value of the optimal observable across the data events is given with its standard deviation.
The lower panels show the ratio of data to expectation and the shaded band corresponds to the combined statistical and systematic uncertainties in the predictions.
}
\label{fig:tCzz_obs_data}
\end{figure}

\begin{figure}[!htbp]
\centering
\subfloat[\OOllllcHBtil ]{
\includegraphics[width=0.47\textwidth]{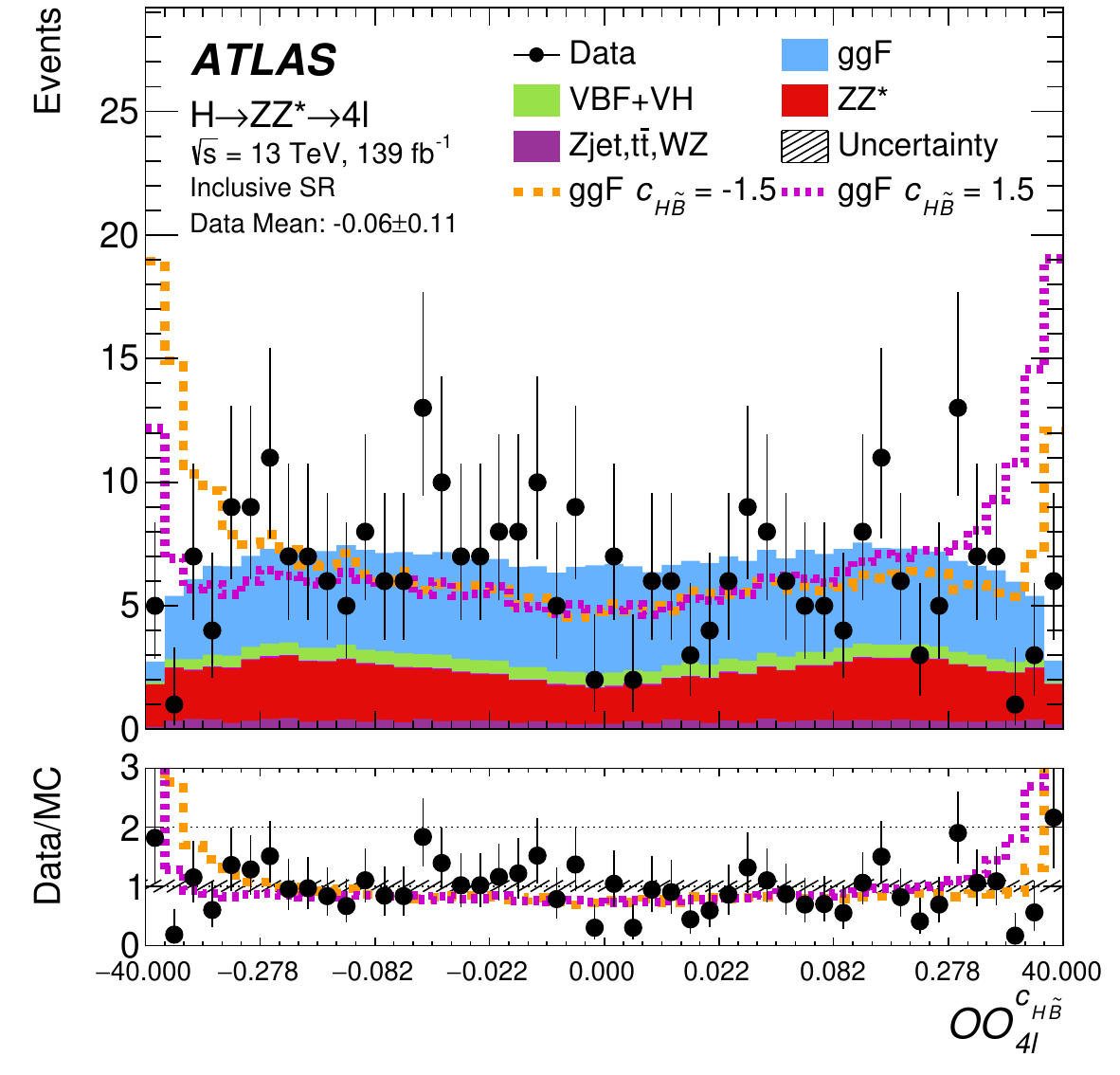} \label{fig:cHB_obs_data}
}
\subfloat[\OOllllcHWBtil ]{
\includegraphics[width=0.47\textwidth]{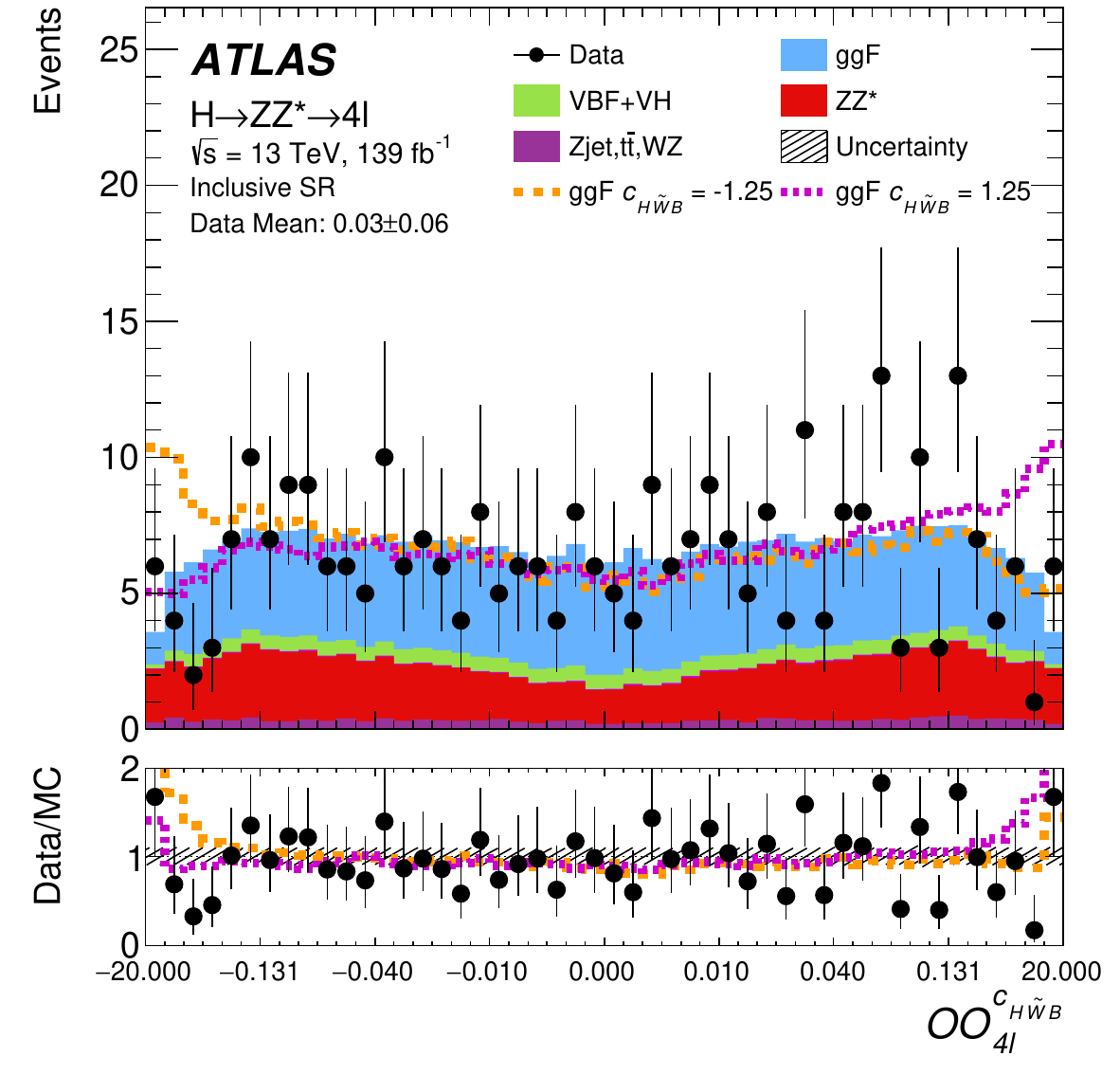} \label{fig:cHWB_obs_data}
}
\quad
\subfloat[\OOllllcHWtil ]{
\includegraphics[width=0.47\textwidth]{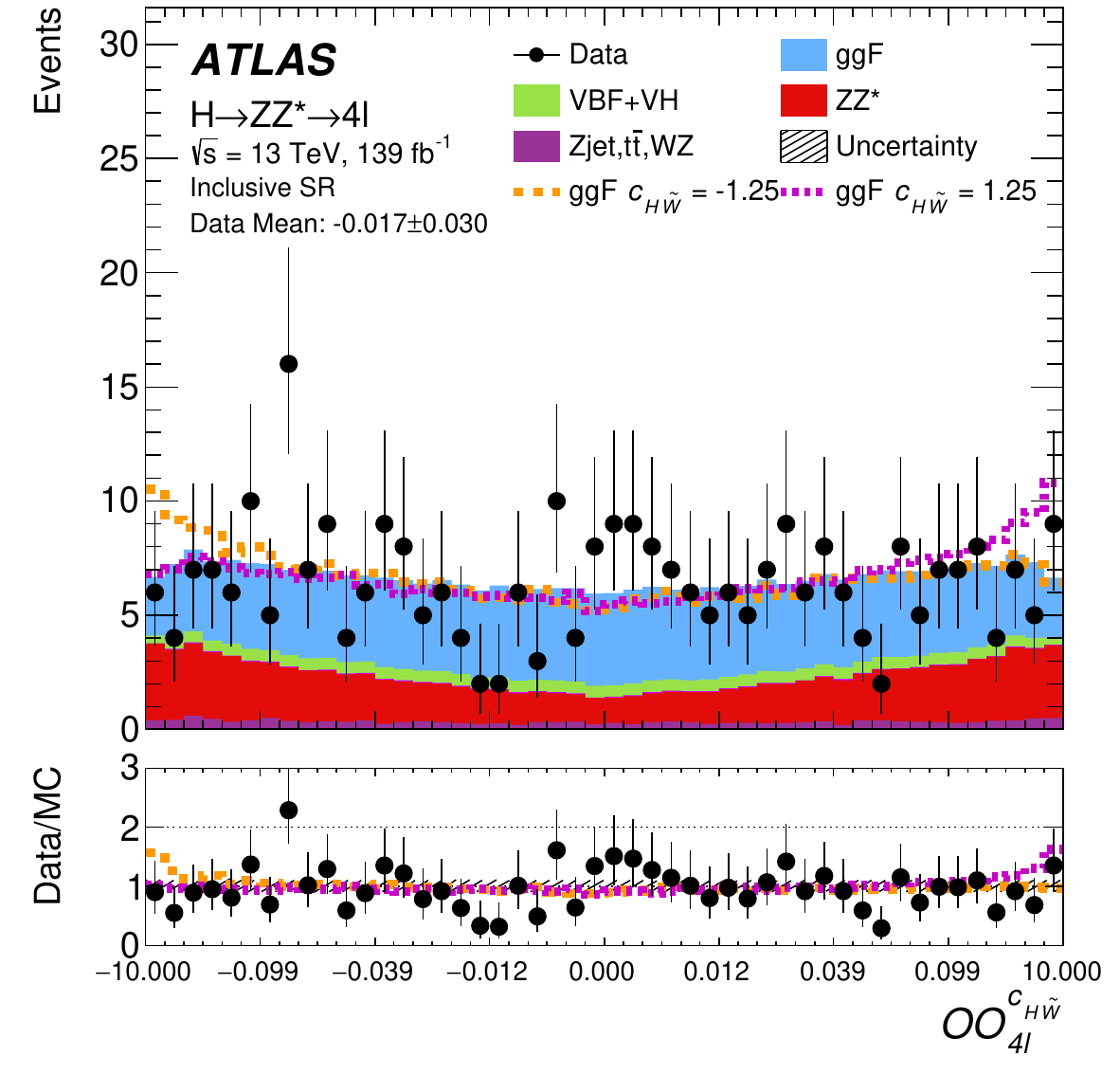} \label{fig:cHW_obs_data}
}
\subfloat[\OOlllldTilde ]{
\includegraphics[width=0.47\textwidth]{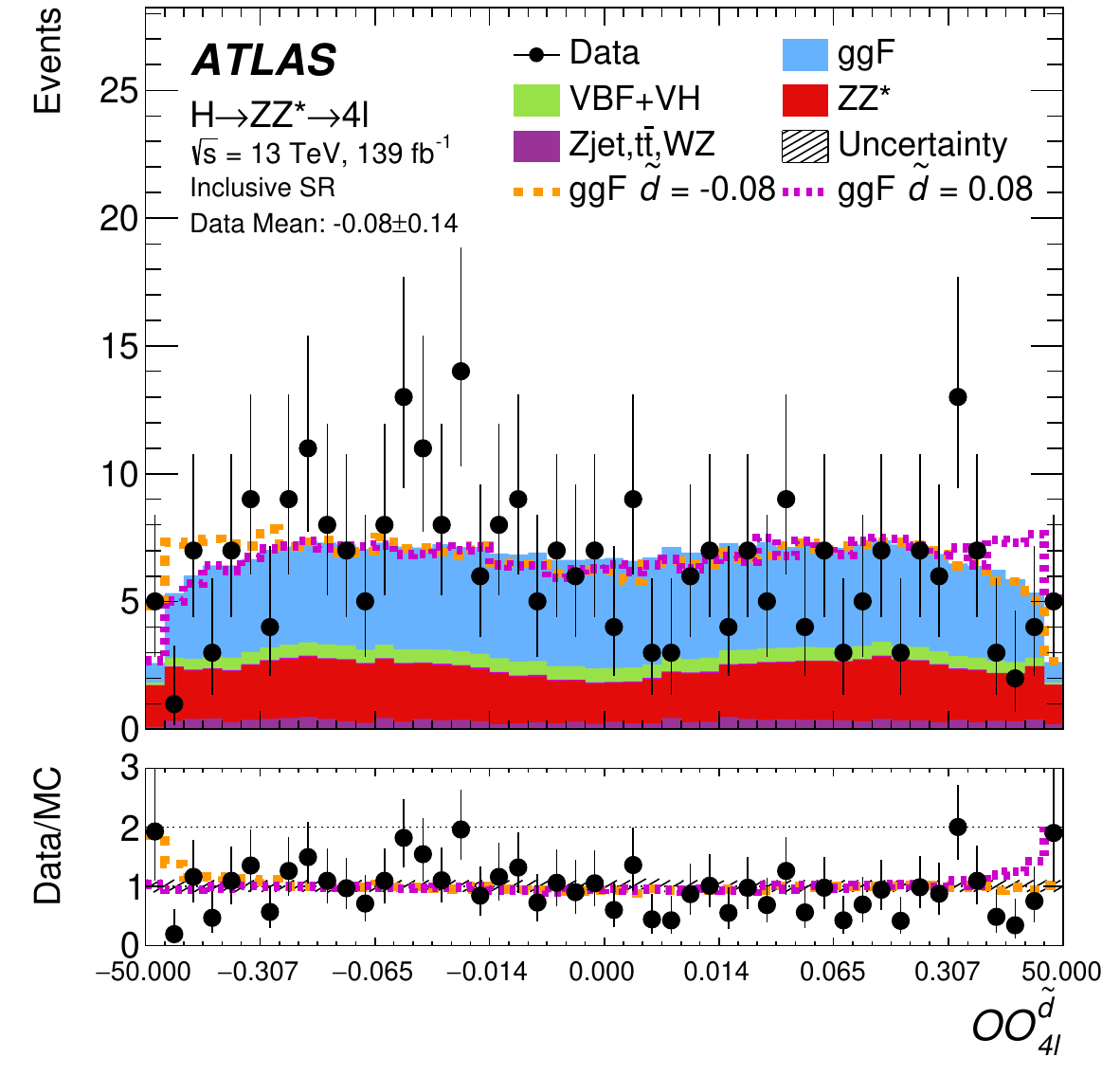} \label{fig:dTilde_obs_data}
}
\caption{
The observed and expected distributions for the decay level observables: \protect\subref{fig:cHB_obs_data} \OOllllcHBtil, \protect\subref{fig:cHWB_obs_data} \OOllllcHWBtil, \protect\subref{fig:cHW_obs_data} \OOllllcHWtil and \protect\subref{fig:dTilde_obs_data} \OOlllldTilde.
The expected distribution for the \ggF and \VBF signals and the backgrounds from \zzstar and reducible backgrounds are shown as a stacked histogram.
Also shown are the expected distributions in presence of CP-odd contributions for \protect\subref{fig:cHB_obs_data} $\cHBtil = \pm 1.5$, \protect\subref{fig:cHWB_obs_data} $\cHWBtil = \pm 1.25$, \protect\subref{fig:cHW_obs_data} $\cHWtil = \pm 1.5$ and \protect\subref{fig:dTilde_obs_data} $\dTilde = \pm 1.25$, assuming the signal event rate remains that of the SM\@.
The bin size is variable, optimized for an approximately flat distribution for the SM combined with a positive and negative BSM scenario.
There are no observed events that fall outside the optimal-observable distribution range, with rate expectations below 0.1\%.
The mean value of the optimal observable across the data events is given with its standard deviation.
The lower panels show the ratio of data to expectation and the shaded band corresponds to the combined statistical and systematic uncertainties in the predictions.
}
\label{fig:oollll_obs_data}
\end{figure}

In absence of positive evidence for BSM physics,  68\% and 95\% confidence intervals on the CP-odd couplings ($\boldsymbol{c}_{\mathrm{BSM}}$) are extracted with the profile likelihood ratio~\cite{Cowan:2010js} statistical test, using the negative-log likelihood (NLL) test statistic:
\begin{equation}
-2\ln\lambda (x|\boldsymbol{c}_{\mathrm{BSM}}) = -2 \ln \frac{\mathcal{L}(x|\boldsymbol{c}_{\mathrm{BSM}},\hat{\hat{\theta}})}{\mathcal{L}(x|\hat{\boldsymbol{c}}_{\mathrm{BSM}},\hat{\theta})}
\end{equation}
where $\theta$ corresponds to the set of systematic uncertainties, or nuisance parameters. The numerator denotes the conditional likelihood estimator of $\theta$, ({\em i.e.}, $\hat{\hat{\theta}}$ is the value of $\theta$ that maximizes the likelihood function for a given $\boldsymbol{c}_{\mathrm{BSM}}$), and the denominator denotes the maximized likelihood estimator ({\em i.e.}, $\hat{\boldsymbol{c}}_{\mathrm{BSM}}$ and $\hat{\theta}$ are the values of $\boldsymbol{c}_{\mathrm{BSM}}$ and $\theta$ that maximize the unconditional likelihood function). The effect of the nuisance parameters is to broaden the profile likelihood ratio reflecting the loss of information originating from the inclusion of systematic uncertainties.
For each BSM coupling, a scan across the values of $\boldsymbol{c}_{\mathrm{BSM}}$ of the NLL is performed by fitting all other parameters (normalization parameters, other BSM couplings, and nuisance parameters) to obtain the values that minimize the NLL for the current value of the coupling being scanned, or couplings for two-dimensional scans.
This result relies on the assumption that for any value of the coupling being scanned, the NLL behaves as a $\chi^2$ distribution with one degree of freedom (\emph{asymptotic approximation}~\cite{Bohm:2014vmk}). Based on this, the confidence intervals are constructed from the scan parameter values for which the NLL crosses the $68\%$ and $95\%$ quantile locations of a $\chi^2$ distribution with one degree of freedom. The validity of the asymptotic approximation for this measurement was confirmed by sampling the distribution of the test statistic observed on toy datasets obtained by random sampling from the predicted likelihood model.
 
The expected sensitivity of the seven couplings explored in this analysis (\Sect{\ref{sec:meth:morphing}}) are given in \Tab{\ref{tab:coupling_expected}}. The expected 95\% confidence intervals estimated separately for either \VBF production or Higgs boson to four-lepton decay and as well for their combination are shown for each coupling.
There are about ten \VBF events expected in the signal region and over 200 expected Higgs boson decays from all production modes in the inclusive signal region, giving a significant statistical advantage to the decay observable fits.
For the Higgs basis that has couplings in the mass Lagrangian, production is predominantly sensitive to the \tCzz coupling with a limited sensitivity to \tCza, and decay is sensitive to both \tCza and \tCaa.
This can be understood from the fact that the \VBF production includes also $W$-boson diagrams and occurs at larger energy scales compared to the decay vertex, which happens at the scale of the Higgs mass and exclusively via neutral gauge bosons, enhancing the role of virtual photon diagrams.
For the Warsaw basis that has couplings in the interaction Lagrangian before SU(2) symmetry breaking, the physical interpretation is less intuitive, as all operators mix the effects of $Z^{\left(\ast\right)}$ and $\gamma^\ast$ propagators on both legs of the HVV vertex. \cHWtil has sensitivity to both production and decay with the latter having improved sensitivity due to the larger number of events, \cHBtil and \cHWBtil are both only sensitive to the decay.
Finally \dTilde, which is proportional to \cHBtil with the constraints \cHBtil = \cHWtil and \cHWBtil = 0, is mostly sensitive to the decay.
The combined observable fits use the \VBF SR1 to \VBF SR4 for the production observable and the \VBF-depleted signal region for the decay observable.
This reduces by about 10\% the number of events entering the decay observable fit, i.e., those in  \VBF SR1 to \VBF SR4, and can lead to a degraded sensitivity of the combined fit compared to the decay fit in the inclusive signal region.
This is the case for \tCza and \dTilde, since their production sensitivity is poor.
Whereas for \cHWtil, which has a more balanced sensitivity for production and decay, the combination limits improve over either production or decay only.
 
\begin{DIFnomarkup}
\begin{table}[!htbp]
\centering
\renewcommand{\arraystretch}{1.2}
\caption{The expected 95\% confidence intervals of production-only, decay-only and combined production and decay likelihood scans for the CP-odd Wilson coefficients in the Warsaw and Higgs bases for an integrated luminosity of \Lum at $\sqrt{s}=13 \ \TeV$. Only one Wilson coefficient is fitted at a time while all others are set to zero.
Limits denoted by `$-$' indicates no sensitivity or no observable gain for the combined fits.
All couplings scale as $1/\Lambda^2$ with the assumed value of $\Lambda =1$~\TeV.
}
\vspace{0.1cm}
\begin{tabular}{c
S[table-format=1.3]
S[table-format=1.3]
S[table-format=1.3]
}
\hline\hline
\noalign{\vspace{0.05cm}}
EFT coupling    &
\multicolumn{3}{c}{Expected 95\% CL} \\
&
\multicolumn{1}{c}{production-only} &
\multicolumn{1}{c}{decay-only}      &
\multicolumn{1}{c}{combined}        \\
\hline
$\cHBtil$  & \multicolumn{1}{c}{$-$} & \pm 0.37 & \multicolumn{1}{c}{$-$} \\
$\cHWBtil$ & \multicolumn{1}{c}{$-$} & \pm 0.72 & \multicolumn{1}{c}{$-$} \\
$\cHWtil$ & \pm 4.8 & \pm 1.34 & \pm 1.27 \\
$\dTilde$ & \pm 0.63 & \pm 0.018 & \pm 0.019 \\
$\tCzz$ & \pm 2.4  &  \multicolumn{1}{c}{$-$} & \multicolumn{1}{c}{$-$} \\
$\tCza$ & \pm 6.6  &  \pm 0.76 & \pm 0.80 \\
$\tCaa$ & \multicolumn{1}{c}{$-$}  &  \pm 0.76 & \multicolumn{1}{c}{$-$} \\
\noalign{\vspace{0.05cm}}
\hline\hline
\label{tab:coupling_expected}
\end{tabular}
\end{table}
\end{DIFnomarkup}
 
\FloatBarrier
 
The expected and observed distributions for the \tCzz coupling NLL production scans are shown in \Fig{\ref{fig:tCzz_prod_scan}}.
The observed central value for \tCzz at 0.78 and the non-parabolic shape of the NLL is due to the small excess of \VBF candidates at positive \tCzz in SR1 (\Fig{\ref{fig:tCzz_obs_data_sr1}}).
The effect of systematic uncertainties, primarily jet-related, is about 15\% for the expected 95\% CL\@.
In the observed NLL scan, a local enhancement of the impact of systematic uncertainties for small absolute values of the BSM coupling is driven by the $\ggF$ parton-shower uncertainty, which primarily affects the migration of $\ggF$ events between the control region and the VBF-enriched signal regions, as discussed in \Sect{\ref{sec:systematics_theory}}. For small couplings, where the $\VBF$ optimal-observable distribution is like the SM, this uncertainty allows the fit model to adapt to the observed local downward fluctuation in the central bins of the VBF SR1, leading to a comparably large impact of this uncertainty.
In \Fig{\ref{fig:cHW_scans}}, the observed NLL distributions are shown for \cHWtil for production, decay and their combination,
as well as the expected distribution for the combination.
The observed central value at 1.5 of the production-only \cHWtil scan arises from the same excess of \VBF candidates as for \tCzz.
Combining with the decay-only scan (central value about 0.1), reduces the combined central value to 0.6.
There is a small effect of systematic uncertainties on the production-only limits, with a local increase of their impact around zero originating from the same effect described above for the $\tCzz$ coupling.
The \cHWtil decay-only scan has a better statistical precision than the production-only scan and relies only on the well-measured lepton kinematics, so the effect of systematic uncertainties is negligible.
 
\begin{figure}[!htbp]
\centering
\subfloat[\tCzz production ]{
\includegraphics[width=0.47\textwidth]{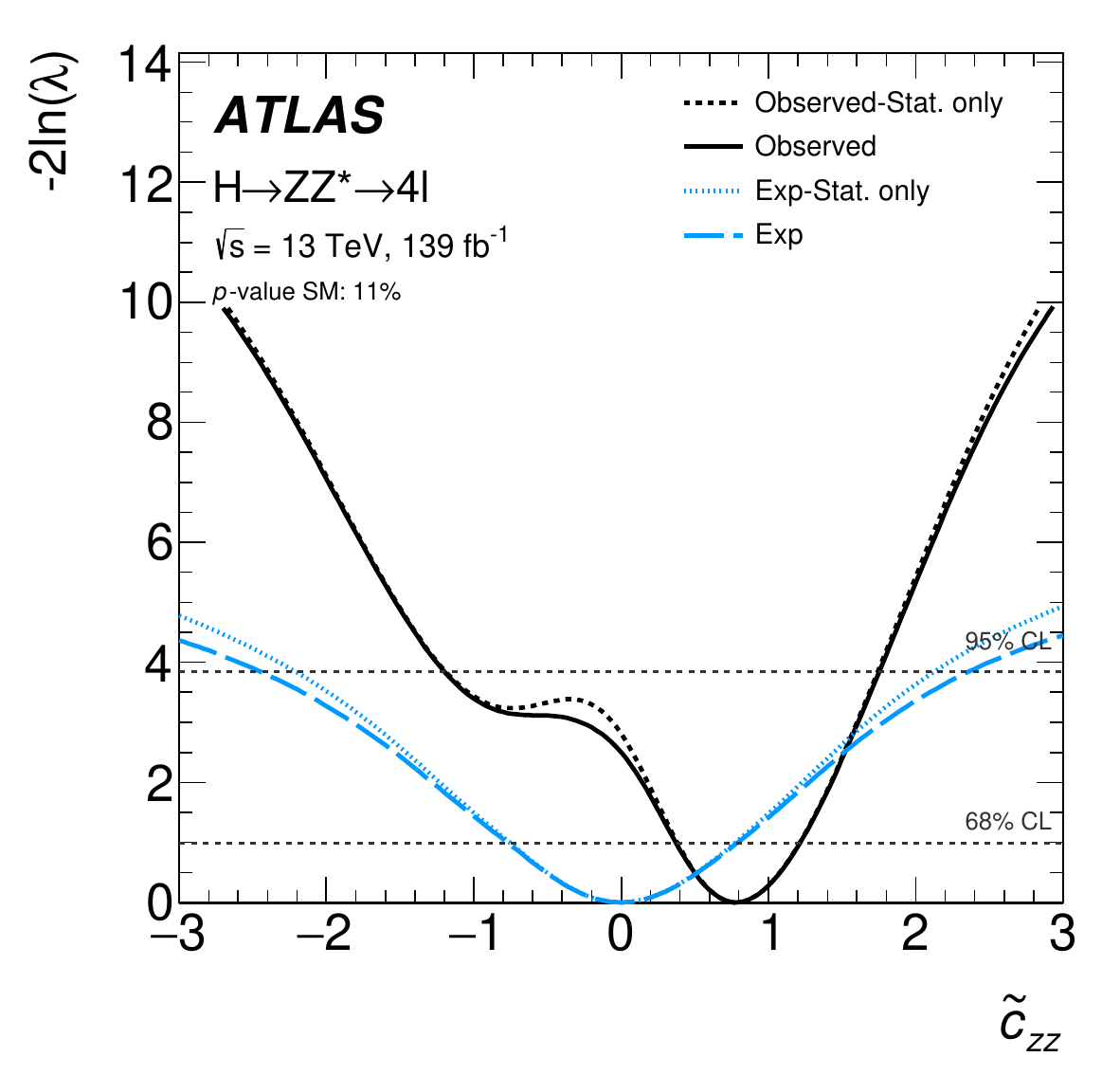}  \label{fig:tCzz_prod_scan}
}
\subfloat[\cHWtil ]{
\includegraphics[width=0.47\textwidth]{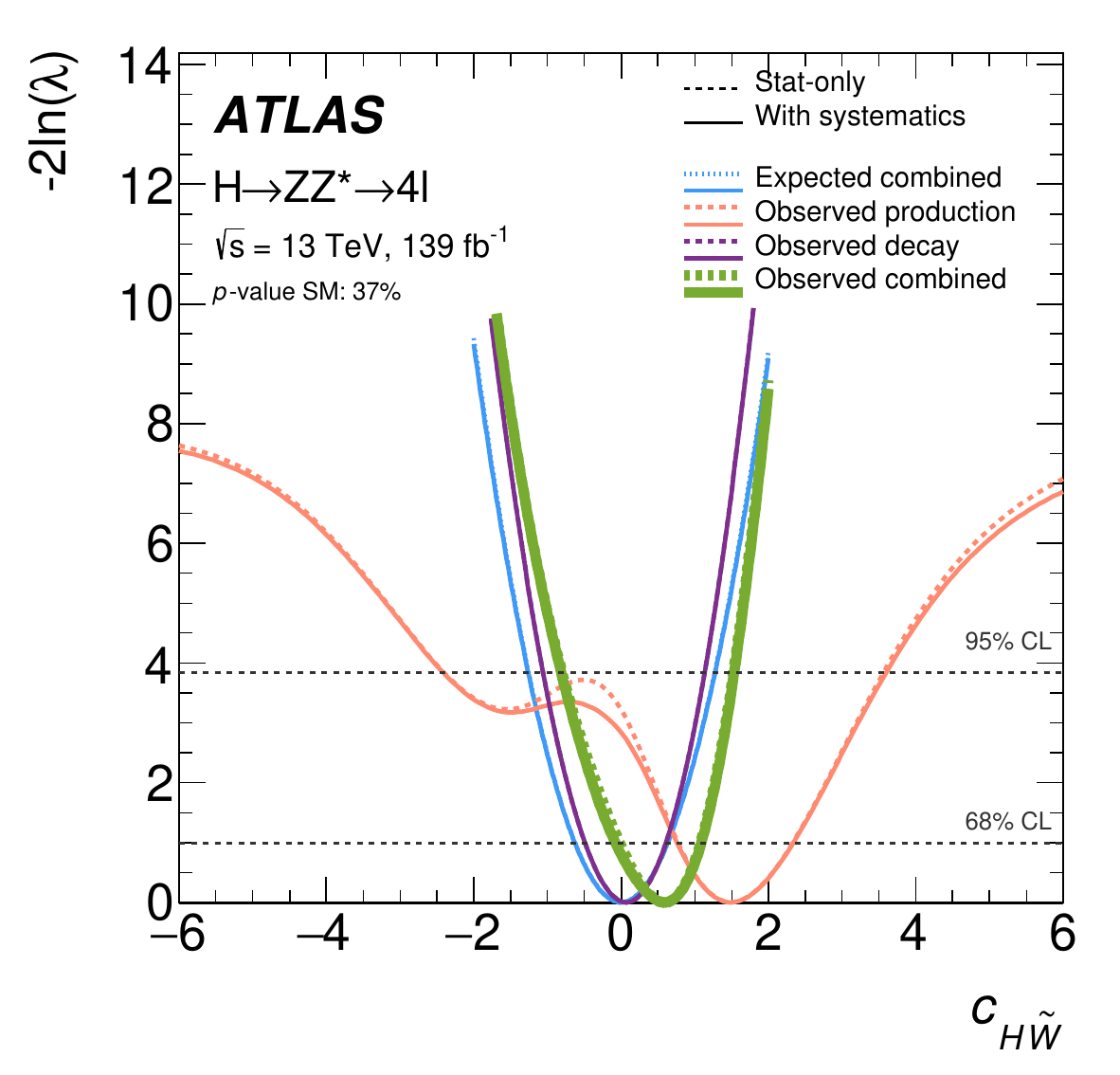}  \label{fig:cHW_scans}
}
\caption{
The NLL observable scans as a function of \protect\subref{fig:tCzz_prod_scan} \tCzz for the production-only \OOjjtCzz fit, and \protect\subref{fig:cHW_scans} \cHWtil for production-only \OOjjcHWtil, decay-only \OOllllcHWtil, and combined fits.
The \tCzz coupling NLL scan shows the expected and observed distributions with (long-dashed/solid lines) and without (dashed line) systematic uncertainties.
For the \cHWtil coupling, sensitive to both production and decay, the expected NLL scan is shown only for the combined fit (blue).
The observed NLL scans for \cHWtil are shown for production-only , decay-only, and their combined (thicker line) fit.
The $p$-value for agreement with the SM is 0.11 (0.37) for the \tCzz (\cHWtil) coupling NLL scan.
All couplings scale as $1/\Lambda^2$ with the assumed value of $\Lambda =1$~\TeV.
}
\label{fig:prod_scans}
\end{figure}
 
The observed and expected NLL observable scans for the decay level couplings for the Warsaw basis \cHBtil, \cHWBtil, \cHWtil and \dTilde are shown in \Fig{\ref{fig:wb_decay_scans}}, and for the Higgs basis \tCza and \tCaa are shown in \Fig{\ref{fig:hb_decay_scans}}.
Similarly to the combined scan of \cHWtil in \Fig{\ref{fig:cHW_scans}}, the decay-only optimal-observable fits have a negligble contribution from systematic uncertainties.
Due to the small upward fluctuation of the data in the first and last observable bins in \Fig{\ref{fig:oollll_obs_data}} the observed NLL scans for \OOllllcHBtil, \OOllllcHWBtil and \OOlllldTilde are less constraining than expected.
The small upward fluctuation around zero for \OOllllcHWtil leads to tighter limits than expected.
In the Higgs basis, the observed decay-only scans for \OOlllltCza and \OOlllltCaa are also less constrained.

\begin{figure}[!htbp]
\centering
\subfloat[\cHBtil ]{
\includegraphics[width=0.47\textwidth]{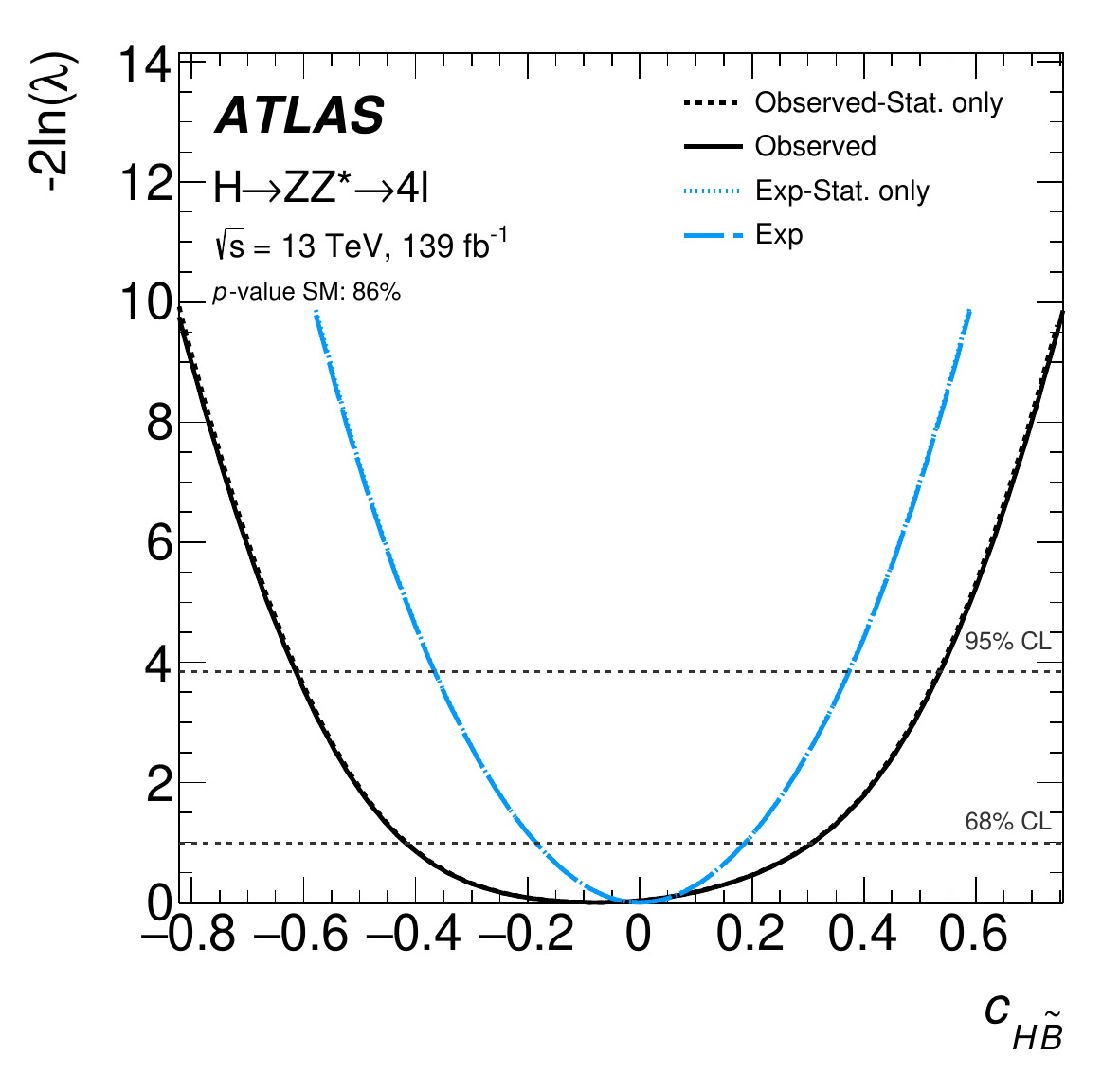}  \label{fig:cHB_decay_scan}
}
\subfloat[\cHWBtil ]{
\includegraphics[width=0.47\textwidth]{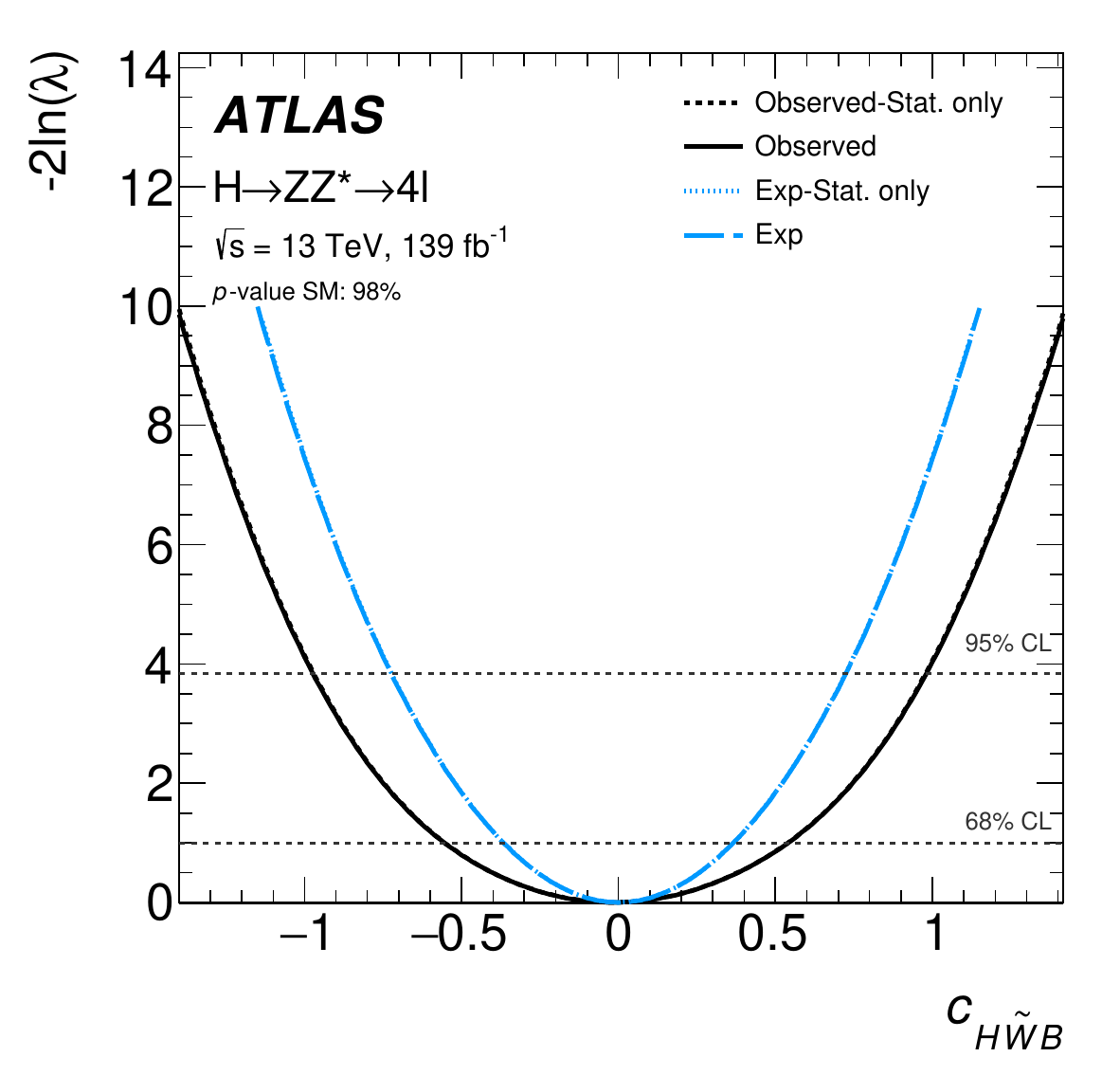}  \label{fig:cHWB_decay_scan}
}
\quad
\subfloat[\cHWtil ]{
\includegraphics[width=0.47\textwidth]{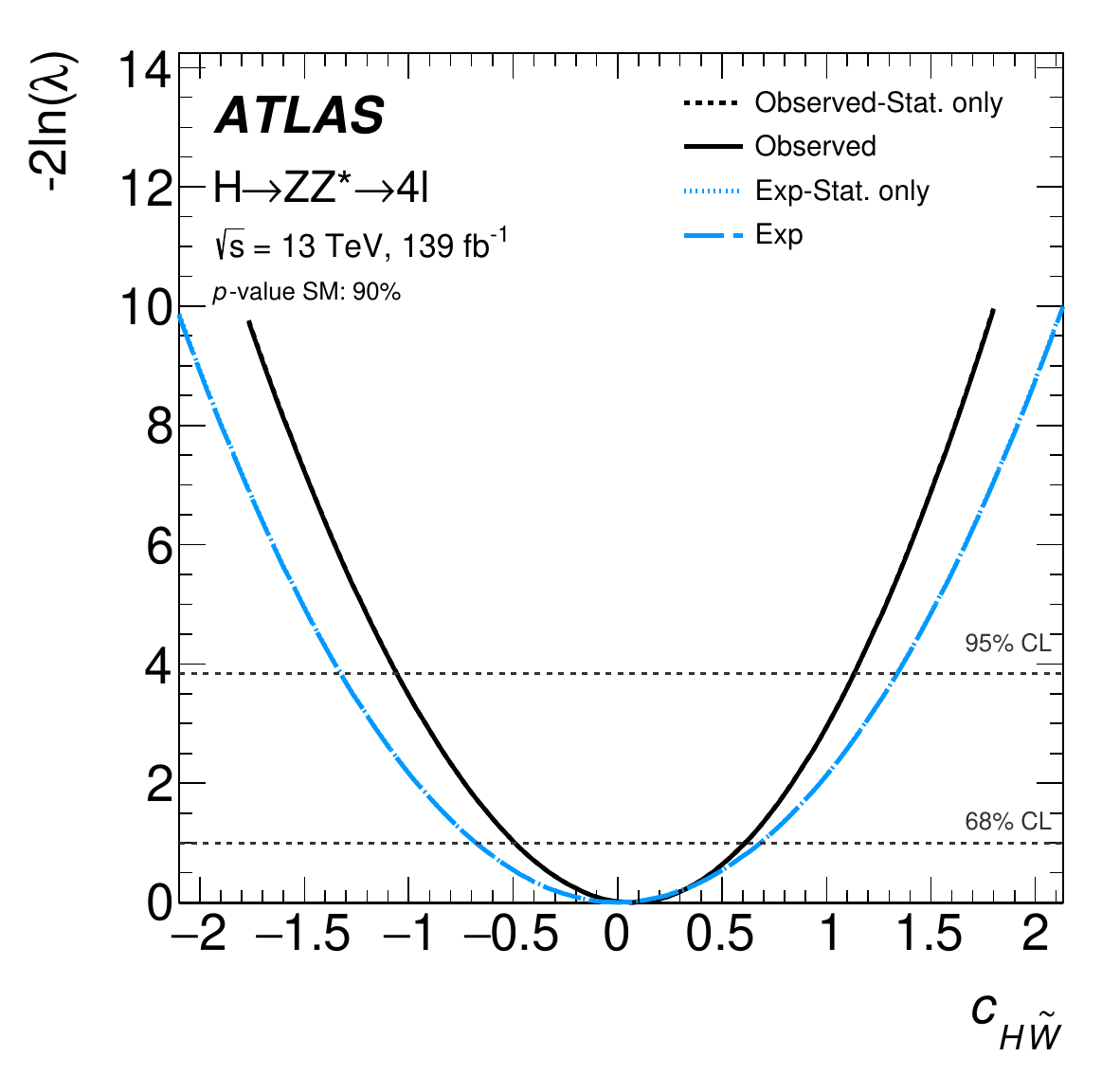}  \label{fig:cHW_decay_scan}
 
}
\subfloat[\dTilde ]{
\includegraphics[width=0.47\textwidth]{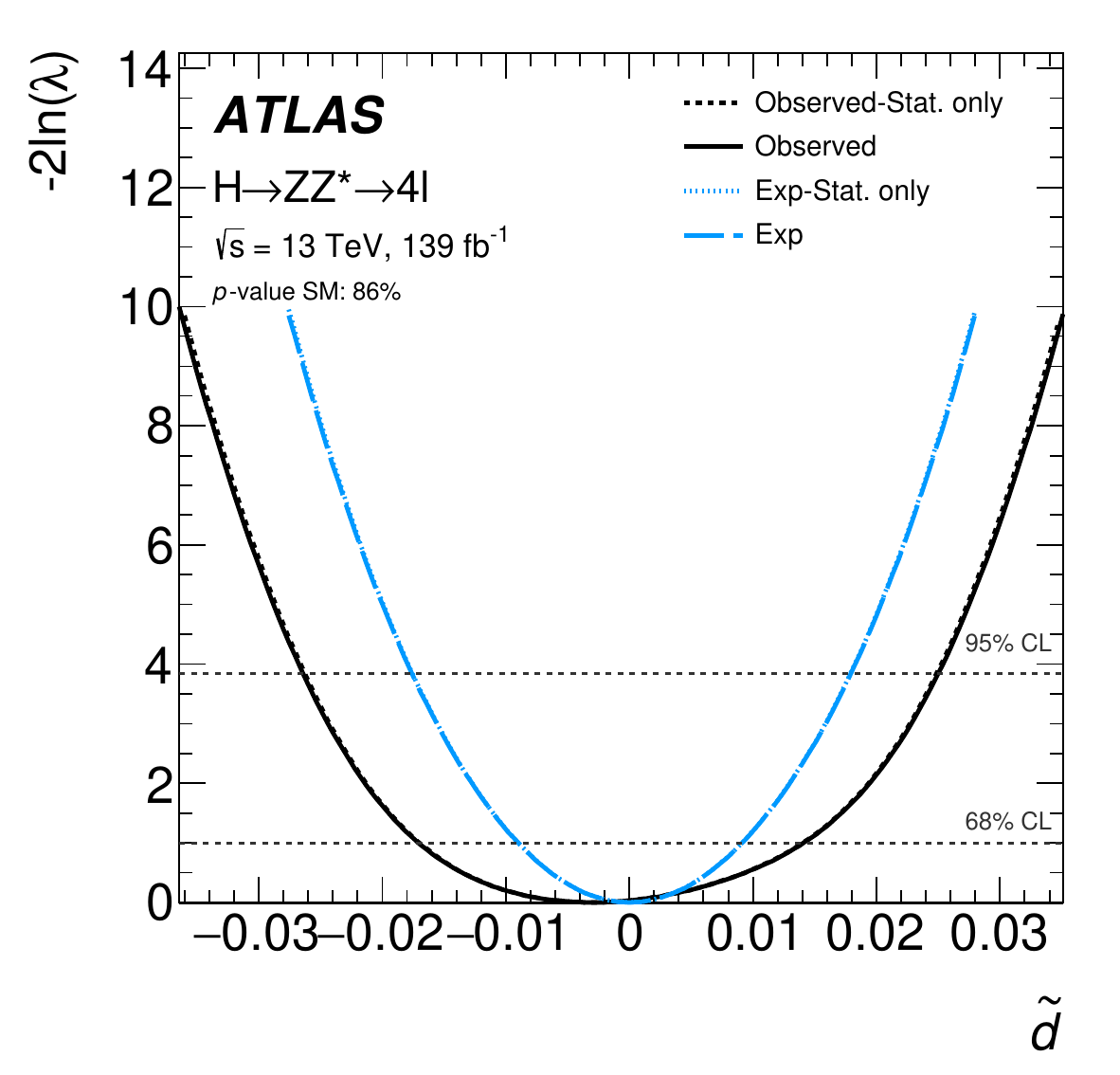}  \label{fig:dTilde_decay_scan}
}
\caption{
The NLL observable scans for the Warsaw-basis couplings for decay-only fits: \protect\subref{fig:cHB_decay_scan} \cHBtil (\OOllllcHBtil), \protect\subref{fig:cHWB_decay_scan} \cHWBtil (\OOllllcHWBtil), \protect\subref{fig:cHW_decay_scan} \cHWtil (\OOllllcHWtil), and \protect\subref{fig:dTilde_decay_scan} \dTilde (\OOlllldTilde).
The expected and observed distributions are shown with (long-dashed/solid lines) and without (dashed line) systematic uncertainties.
The effect of systematics is small for decay-only fits, so that the two sets of lines largely overlap.
The $p$-values for agreement with the SM are 0.86, 0.98, 0.90 and 0.86 for the \cHBtil, \cHWBtil, \cHWtil, and \dTilde coupling NLL scans, respectively.
All couplings scale as $1/\Lambda^2$ with the assumed value of $\Lambda =1$~\TeV.
}
\label{fig:wb_decay_scans}
\end{figure}
 
\begin{figure}[!htbp]
\centering
\subfloat[]{\includegraphics[width=0.45\linewidth]{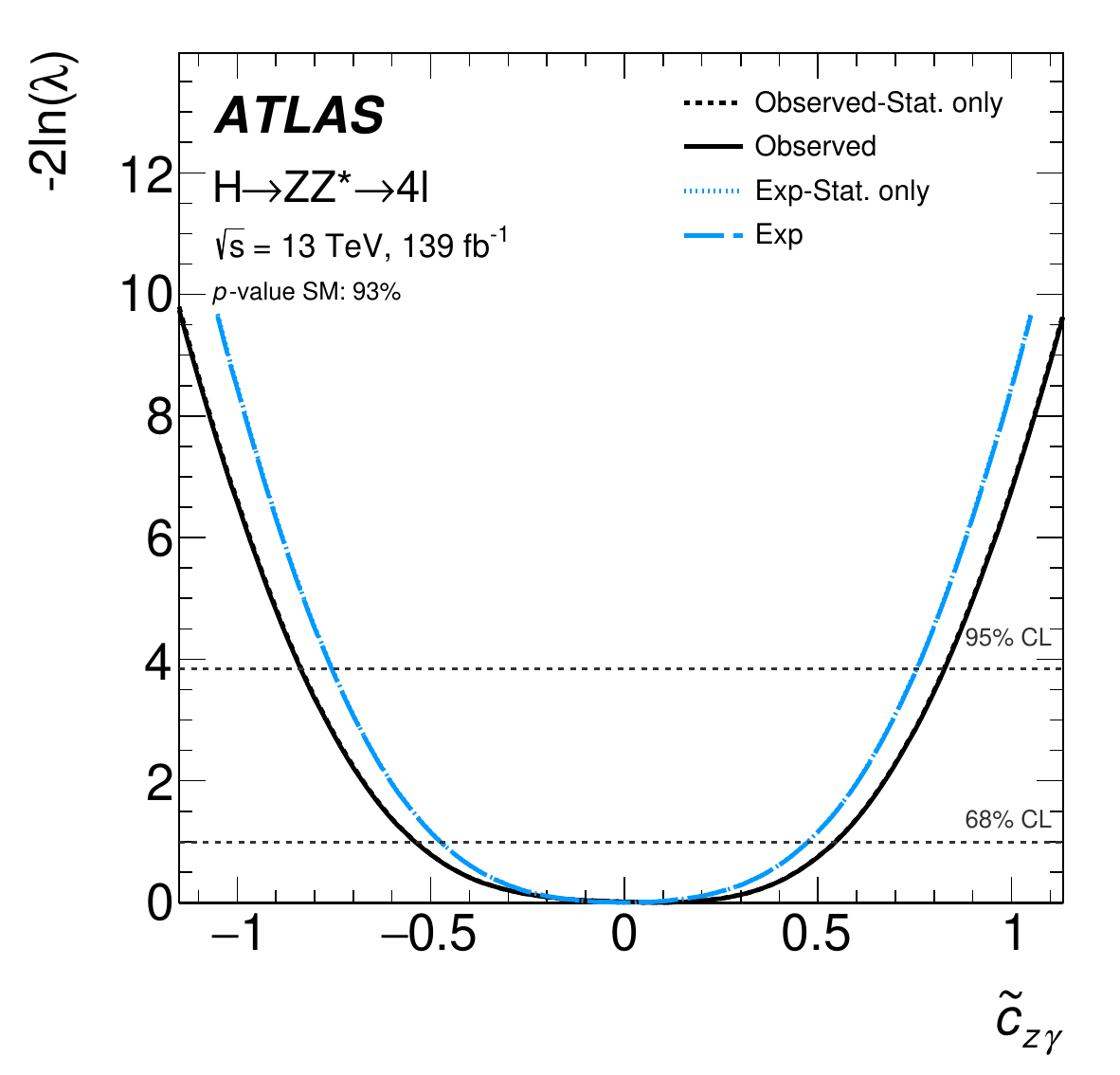}
\label{fig:tCza_prod_scan}
}
\subfloat[]{\includegraphics[width=0.45\linewidth]{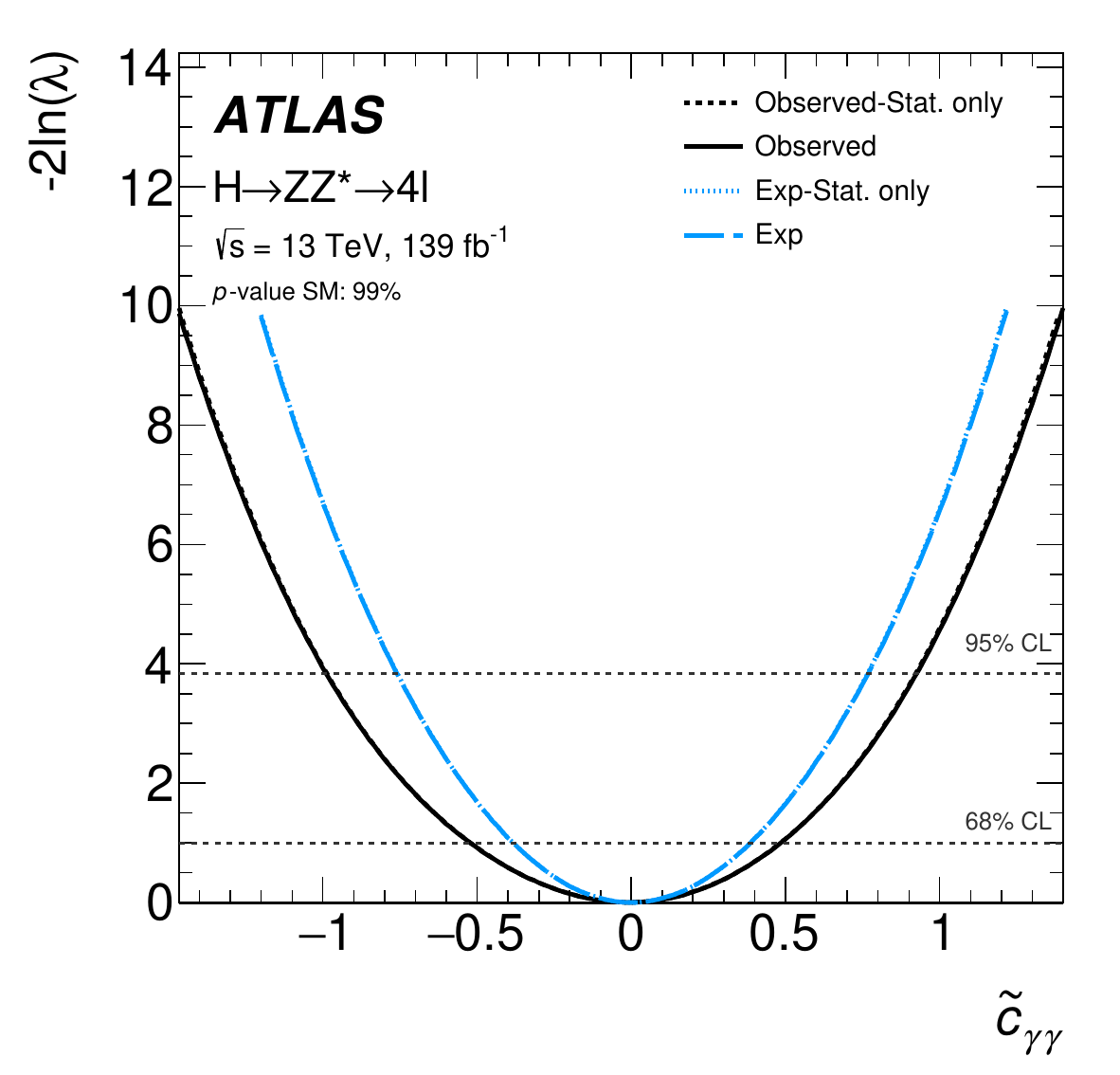}
\label{fig:tCaa_prod_scan}
}
\\
\caption{
The NLL observable fits for the Higgs-basis couplings for decay-only fits: \protect\subref{fig:tCza_prod_scan} \tCza (\OOlllltCza) and \protect\subref{fig:tCaa_prod_scan} \tCaa (\OOlllltCaa).
The expected and observed distributions are shown with (long-dashed/solid lines) and without (dashed line) systematic uncertainties.
The effect of systematics is small for decay-only fits, so that the two sets of lines largely overlap.
The $p$-value for agreement with the SM is 0.93 (0.99) for the \tCza (\tCaa) coupling NLL scan.
All couplings scale as $1/\Lambda^2$ with the assumed value of $\Lambda =1$~\TeV.
\label{fig:hb_decay_scans}}
\end{figure}
 
The observed and expected 68\% and 95\% CL two-dimensional contours for all three pairings of the Warsaw-basis couplings, \cHBtil, \cHWBtil and \cHWtil, are shown in \Figrange{\ref{fig:2D_cHB_cHWB}}{\ref{fig:2D_cHW_cHWB}}.
The two-dimensional observable \OOllllcHBtil vs \OOllllcHWtil is used for these decay-only contours.
This observable pair is found empirically to have the best expected sensitivity compared with any of the individual observables or other observable pairings.
The observed minima in \Figrange{\ref{fig:2D_cHB_cHWB}}{\ref{fig:2D_cHW_cHWB}} all lie along the diagonals of the confidence interval ellipses, well within the 68\% CL\@.
These diagonals are relatively flat in terms of the likelihood minima.
\Fig{\ref{fig:2D_tCzz_tCza}} shows the two-dimensional contours for scans along \tCzz versus \tCaa, where \tCzz is sensitive only to the production and \tCaa is sensitive only to the decay.
For \tCzz, the events in \VBF SR1 to SR4 are used with the observable \OOjjtCzz in 12 bins, as shown in \Fig{\ref{fig:tCzz_prod_scan}}, and for \tCza the events in the \VBF-depleted region are used with  \OOlllltCaa in 48 bins as shown in \Fig{\ref{fig:tCaa_prod_scan}}.
As shown in \Eqn{\ref{eqn:LSMEFT_leading}}, all couplings of \Figrange{\ref{fig:prod_scans}}{\ref{fig:2D_scans}} scale as $1/\Lambda^2$ with the assumed value of $\Lambda =1$~\TeV.

\begin{figure}[!htbp]
\centering
\subfloat[\cHWBtil vs \cHBtil  ]{
\includegraphics[width=0.47\textwidth]{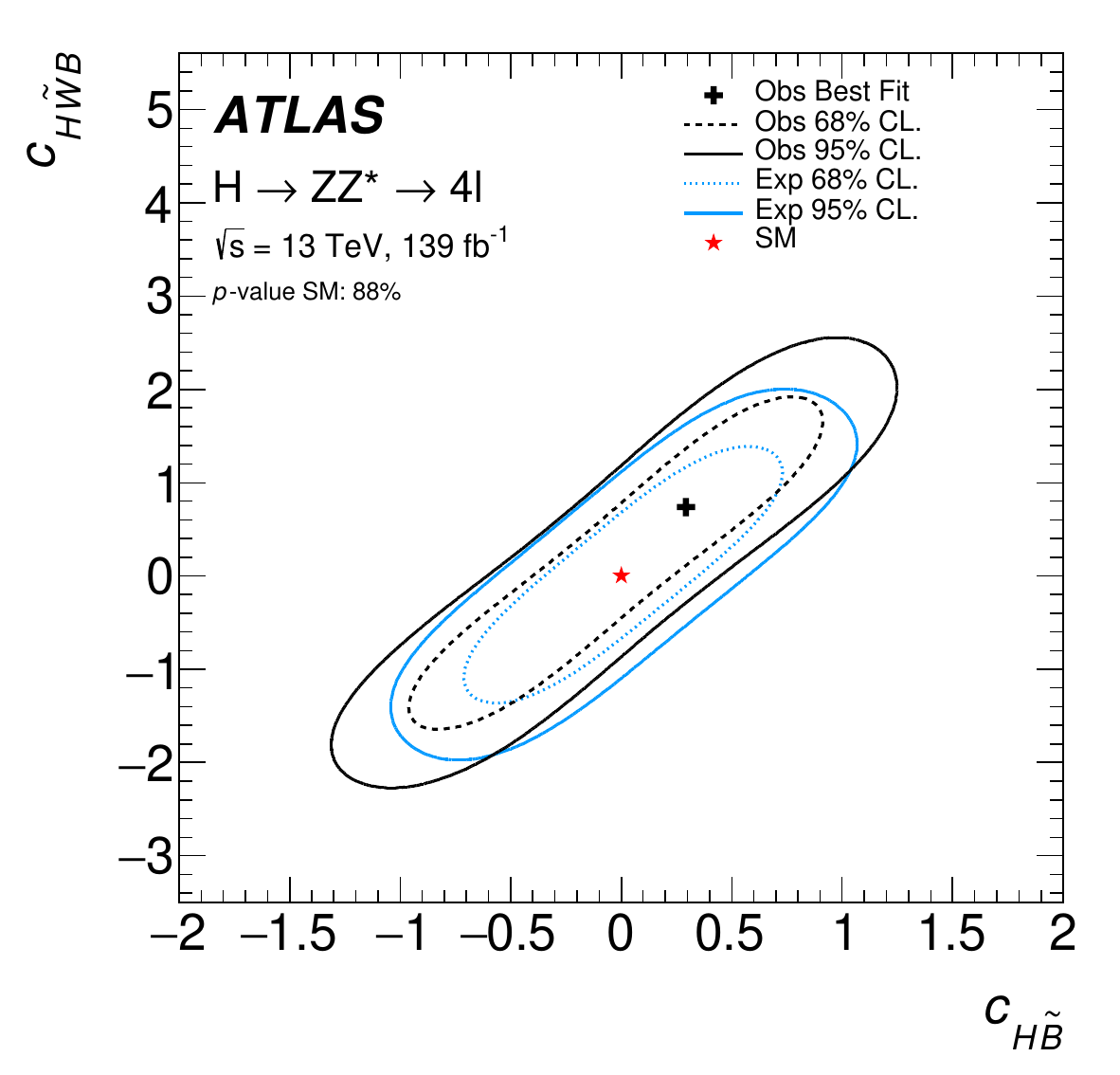}  \label{fig:2D_cHB_cHWB}
}
\subfloat[\cHWtil vs \cHBtil  ]{
\includegraphics[width=0.47\textwidth]{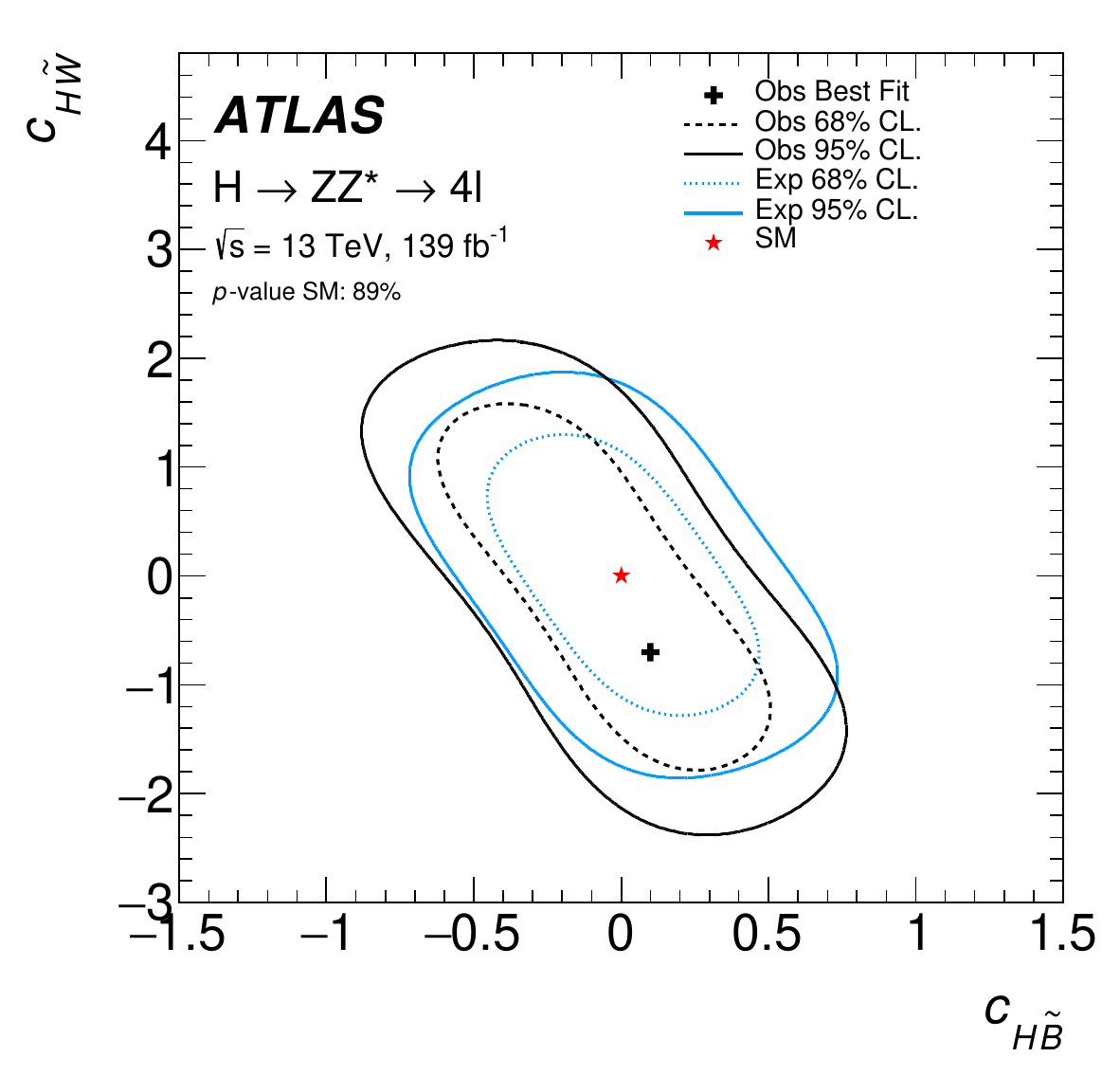}  \label{fig:2D_cHW_cHB}
}
\quad
\subfloat[\cHWtil vs \cHWBtil  ]{
\includegraphics[width=0.47\textwidth]{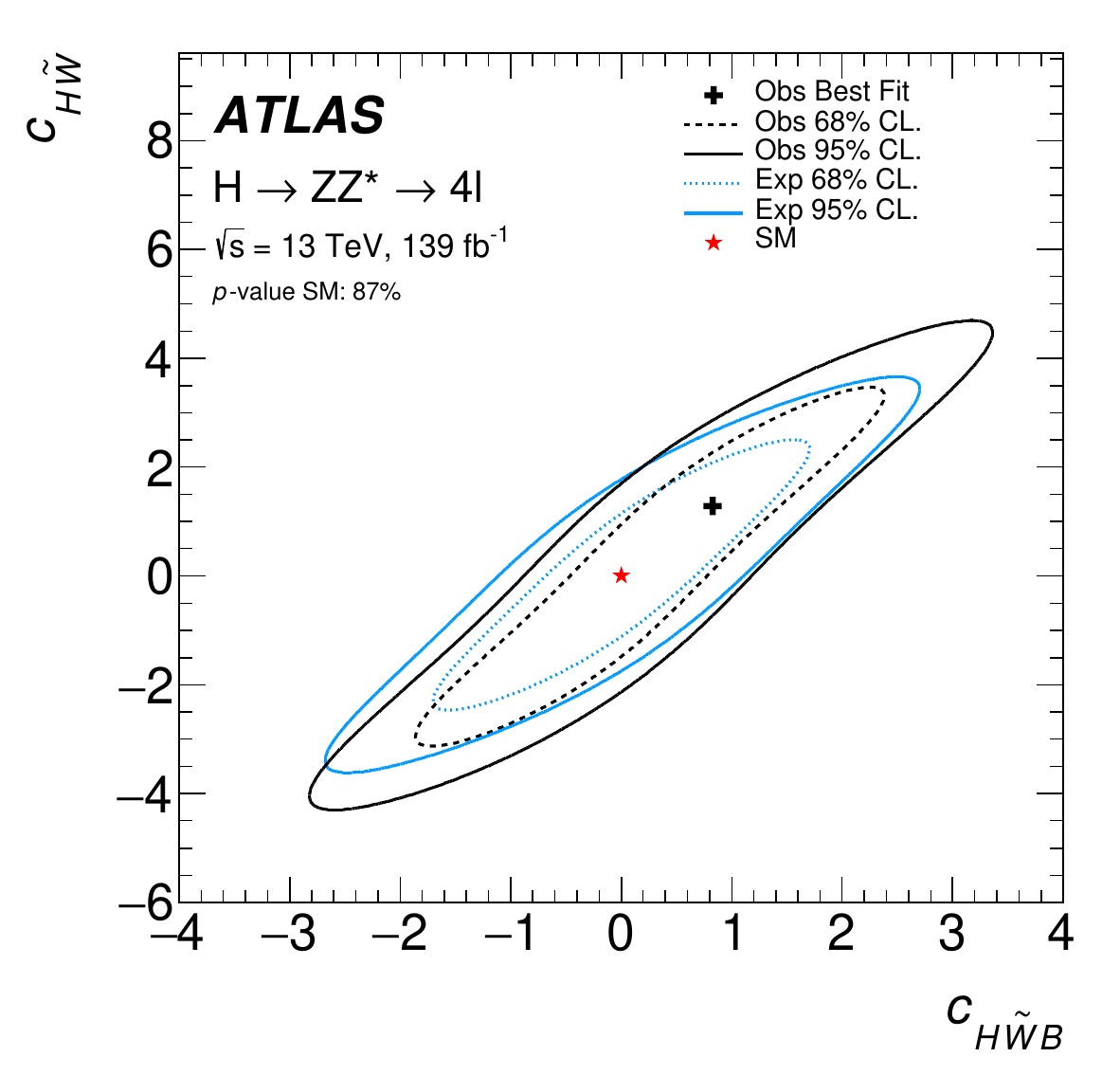}  \label{fig:2D_cHW_cHWB}
}
\subfloat[\tCaa vs \tCzz  ]{
\includegraphics[width=0.47\textwidth]{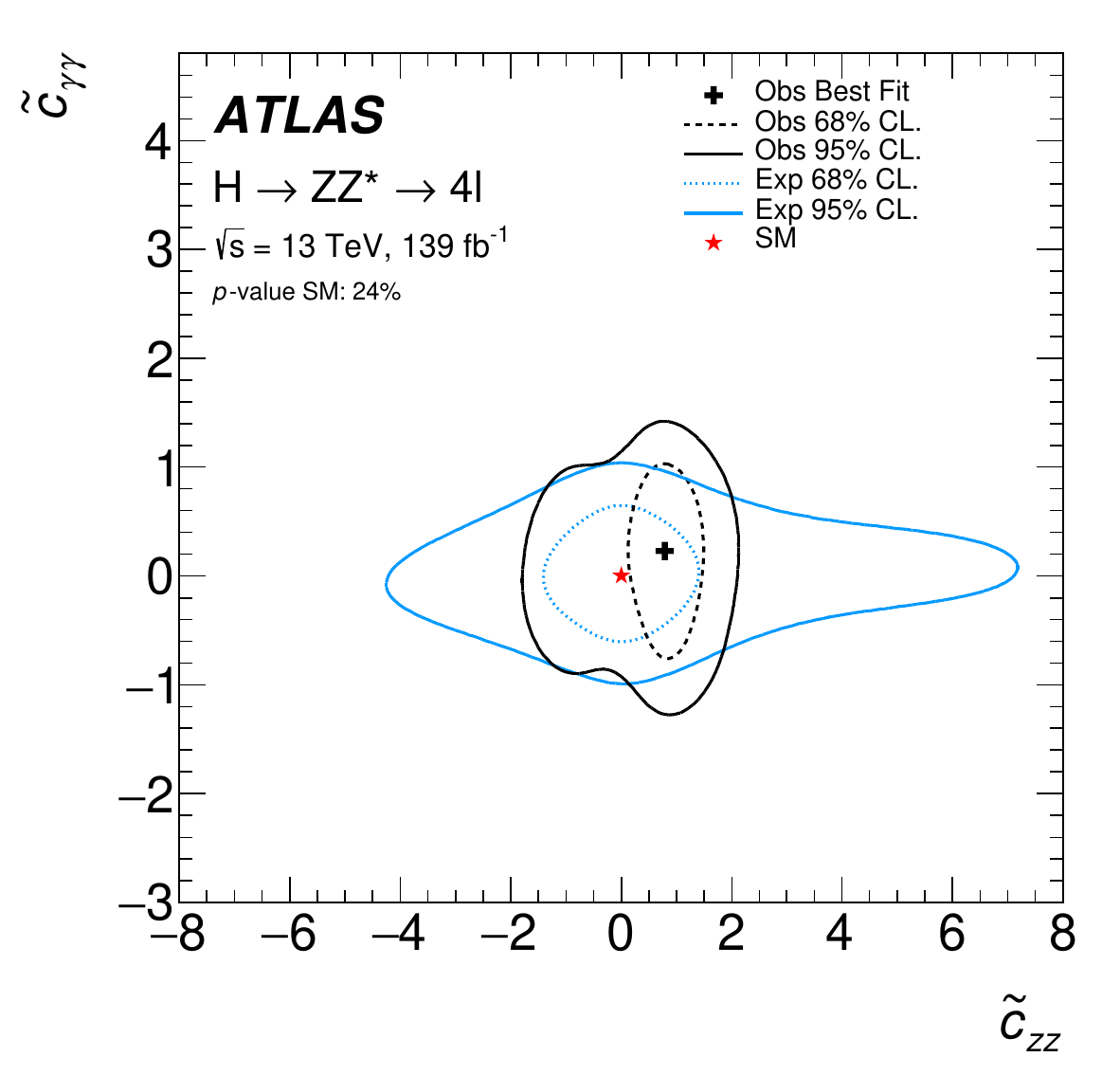}  \label{fig:2D_tCzz_tCza}
}
 
\caption{
The observed and expected 68\% and 95\% CL two-dimensional contours for all three pairings of the Warsaw-basis couplings: \protect\subref{fig:2D_cHB_cHWB} \cHWBtil versus \cHBtil, \protect\subref{fig:2D_cHW_cHB}  \cHWtil versus \cHBtil and \protect\subref{fig:2D_cHW_cHWB}  \cHWtil versus \cHWBtil.
These use the two-dimensional observable \OOllllcHBtil versus \OOllllcHWtil.
In \protect\subref{fig:2D_tCzz_tCza}, the two-dimensional contours for scans along \tCaa, sensitive only to the decay, and \tCzz, sensitive only to the production are shown, where the observable \OOlllltCaa is used for the decay and \OOjjtCzz for the production.
All couplings scale as $1/\Lambda^2$ with the assumed value of $\Lambda =1$~\TeV.
}
\label{fig:2D_scans}
\end{figure}

\Fig{\ref{fig:coupling_results}} and \Tab{\ref{tab:coupling_results}} summarize the expected and observed confidence intervals at 68\% and 95\% CL for the CP-odd Wilson coefficients.
The limits for the Higgs-basis Wilson coupling \tCzz are from a production-only observable fit.
Those for the coupling \cHWtil are from a combined-observable fit as this coupling has sensitivity in both production and decay.
The rest of the couplings are from a decay-only observable fit, having little sensitivity to the \VBF production, as previously shown in \Tab{\ref{tab:coupling_expected}}.

\begin{figure}[!htbp]
\centering
\includegraphics[width=0.7\textwidth]{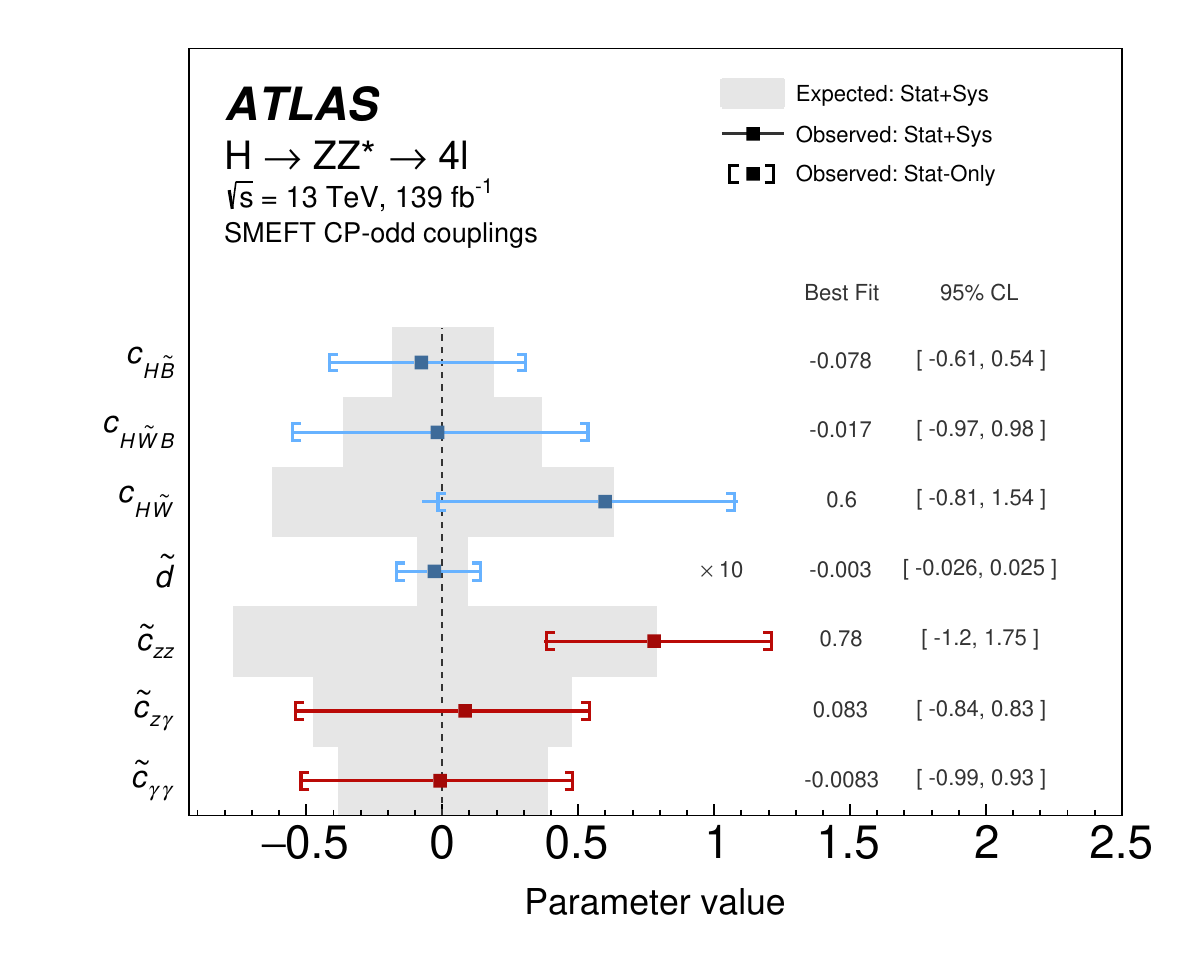}
\caption{
The expected and observed measurements
of the CP-odd Wilson coefficients for an integrated luminosity of $139$~fb$^{-1}$ at $\sqrt{s}=13~\TeV$.
The data points and 68\% CL uncertainty bars show the observed values with statistical and systematic uncertainties, and the statistical-only uncertainties are shown as square brackets.
The Warsaw-basis couplings and \dTilde are in blue and the Higgs-basis couplings are in red.
The \cHWtil measurement is from the combined production and decay fit, and the \tCzz measurement is from the production-only fit.
The other measurements are from decay-only fits, having little sensitivity to the \VBF production.
The \dTilde best fit point and uncertainty bars are scaled by a factor of ten.
The best fit values and 95\% CL limits are shown separately.
Only one Wilson coefficient is fitted at a time while all others are set to zero.
All couplings scale as $1/\Lambda^2$ with the assumed value of $\Lambda =1$~\TeV.
}
\label{fig:coupling_results}
\end{figure}

\begin{DIFnomarkup}
\begin{table}[!htbp]
\centering
\renewcommand{\arraystretch}{1.3}
\caption{The expected and observed confidence intervals at 68\% and 95\% CL for the CP-odd Wilson coefficients for an integrated luminosity of $139$~fb$^{-1}$ at $\sqrt{s}=13~\TeV$. Only one Wilson coefficient is fitted at a time while all others are set to zero.
The observed best fit value and $p$-value for agreement with the SM is provided.
The last column indicates whether the limits come from production (prod), decay or a combination of production and decay (comb).
All couplings scale as $1/\Lambda^2$ with the assumed value of $\Lambda =1$~\TeV.
}
\vspace{0.1cm}
\resizebox{\textwidth}{!}{
\begin{tabular}{*{5}{c}rcc}
\hline\hline
\noalign{\vspace{0.05cm}}
EFT coupling  &
\multicolumn{2}{c}{Expected} &
\multicolumn{2}{c}{Observed} &
\multicolumn{1}{c}{Best-fit} &
SM &
Fit type\\
parameter &
68\% CL &  95\% CL &
68\% CL &  95\% CL &
\multicolumn{1}{c}{value} &
$p$-value &
\\
\hline
$\cHBtil$  &
[$-0.18$, 0.19] & [$-0.37$, 0.37] &
[$-0.42$, 0.31] & [$-0.61$, 0.54] &
$-0.078$ &
$0.86$ &
decay\\
$\cHWBtil$ &
[$-0.36$, 0.36]  &  [$-0.72$, 0.72] &
[$-0.56$, 0.53]  &  [$-0.97$, 0.98] &
$-0.017$ &
$0.99$ &
decay\\
$\cHWtil$ &
[$-0.63$, 0.63] & [$-1.26$, 1.28] &
[$-0.07$, 1.09] & [$-0.81$, 1.54] &
$0.60$ &
$0.37$ &
comb\\
$\dTilde$ &
[$-0.009$, 0.009] & [$-0.018$, 0.018] &
[$-0.017$,  0.014] & [$-0.026$, 0.025]&
$-0.003$ &
$0.86$ &
decay\\
$\tCzz$ &
[$-0.77$, 0.79]  &  [$-2.4$, 2.4]&
[$ 0.37$, 1.21]  &  [$-1.20$, 1.75]&
$0.78$ &
$0.11$ &
prod\\
$\tCza$ &
[$-0.47$, 0.47]  &  [$-0.76$, 0.76]&
[$-0.54$, 0.54]  &  [$-0.84$, 0.83]&
$0.083$ &
$0.93$ &
decay\\
$\tCaa$ &
[$-0.38$, 0.38]  &  [$-0.76$, 0.77]&
[$-0.52$, 0.48]  &  [$-0.99$, 0.93]&
$-0.01$ &
$0.99$ &
decay\\
\noalign{\vspace{0.05cm}}
\hline\hline
\label{tab:coupling_results}
\end{tabular}
}
\end{table}
\end{DIFnomarkup}
 
The prediction of the optimal-observable distribution using the morphing method described in \Sect{\ref{sec:meth:morphing}} includes both a linear and quadratic dependence on the BSM matrix element (see \Eqn{\ref{eq:CouplingFormula}}).
Only the linear term generates an asymmetry of the optimal-observable distribution, while the quadratic term leads to a symmetric shape modification.
The normalization of the signal is left to float in the statistical model to allow a shape-only analysis, but since the quadratic terms contribute to the shapes in a symmetric way, they can still influence the measurements.
Their impact on the results is tested for \cHBtil decay-only and \cHWtil production and decay by repeating the analysis when including only linear terms in the morphing method. For the two couplings investigated, the 68\% CL (95\% CL) limits with linear-only terms change by $\sim$1\% ($\sim$3\%), indicating a negligible impact of the quadratic terms on the exclusion limits.

The effect of the choice to perform a shape-only analysis instead of additionally constraining the BSM couplings using the total cross-section predicted for each coupling value is also investigated using simulation. The measurement is repeated on simulated data, with the predicted signal event yield parametrised as a function of the coupling parameters instead of being left a free parameter. This parameterisation takes into account the effect of the BSM couplings on the production cross-section (for \VBF), the decay branching fraction through the modification of the Higgs boson partial and total decay widths resulting from the quadratic EFT terms, and the detector event acceptance. For example, the predicted yield decreases by less than 10\% relative to the SM yield for \cHWtil at its 95\% CL limit value of \Tab{\ref{tab:coupling_results}} from the combination of the three effects.
The cross-section change caused by the BSM couplings has a small effect on the decay-only limits as the predicted $68\%$ ($95\%$) CL exclusion limits would improve by less than $5$\% ($10\%$).
There would be an important effect for production-only measurements with the expected limits tightening by $10\%$ ($50\%$).
The cross-sections are CP-even quantities and are only affected by the quadratic term, as the linear term integrates to zero over the optimal-observable distribution.
Therefore the larger coupling values compatible with the data shape in the production scans result in a more pronounced effect on the limits from the quadratic term driving the rate change.

Finally, the present analysis assumes the presence of no other active CP-even BSM couplings in its signal model.
The effects of a potential CP-even BSM admixture are evaluated in simulation by injecting simulated signals with additional active \cHB, \cHWB and \cHW SMEFT couplings on the order of the current experimental limits~\cite{HIGG-2019-13}.
The effect on the production-only limits is negligible. For the decay analysis, a CP-even coupling of the magnitude investigated would weaken the limits and increase an observed non-zero CP-odd value by a few percent.

\Fig{\ref{fig:compare_results}} provides a comparison of the present measurements with those from~\cite{HIGG-2018-28, HIGG-2020-08, HIGG-2018-14}.
The 68\% confidence levels are compared for the Warsaw-basis couplings (\cHBtil, \cHWBtil and \cHWtil) and the \dTilde coupling.
The \htollllbrief simplified-template-cross-section (STXS) measurements~\cite{HIGG-2018-28} provide two degenerate positive/negative values as the measured cross-sections are only sensitive to the BSM quadratic terms ($\left|\mathcal{M}_{BSM}\right|^2$ of \Eqn{\ref{eq:CouplingFormula}}).
The STXS measurements are based on event rates in CP-insensitive categories and so cannot distinguish between a CP-even and CP-odd effect.
This also explains the higher expected sensitivity of the present measurement that includes the four-lepton decays.
The expected values for the present measurement are the most sensitive, except for \cHWtil of the \VBF \htoyy decay channel measurement~\cite{HIGG-2020-08} that has a larger number of \VBF produced events.
For \dTilde, the improved sensitivity of the present measurement over the ATLAS \htoyy + \htautau measurement arises from the contribution of the \htollllbrief decays.

Two CMS measurements, \htollllbrief~\cite{CMS-HIG-19-009} and \htautau~\cite{CMS:2022uox}, provide  results for CP-odd couplings with a similar approach as the present analysis. The CMS results, however, are not directly comparable to the present Warsaw basis results due to different assumptions applied on the CP-odd couplings.  The assumptions of~\cite{CMS-HIG-19-009} and~\cite{CMS:2022uox} are compatible with those of the present \tCzz measurement, however results in terms of the \tCzz coupling are not provided in~\cite{CMS:2022uox}.

\begin{figure}[!htbp]
\centering
\includegraphics[width=0.7\textwidth]{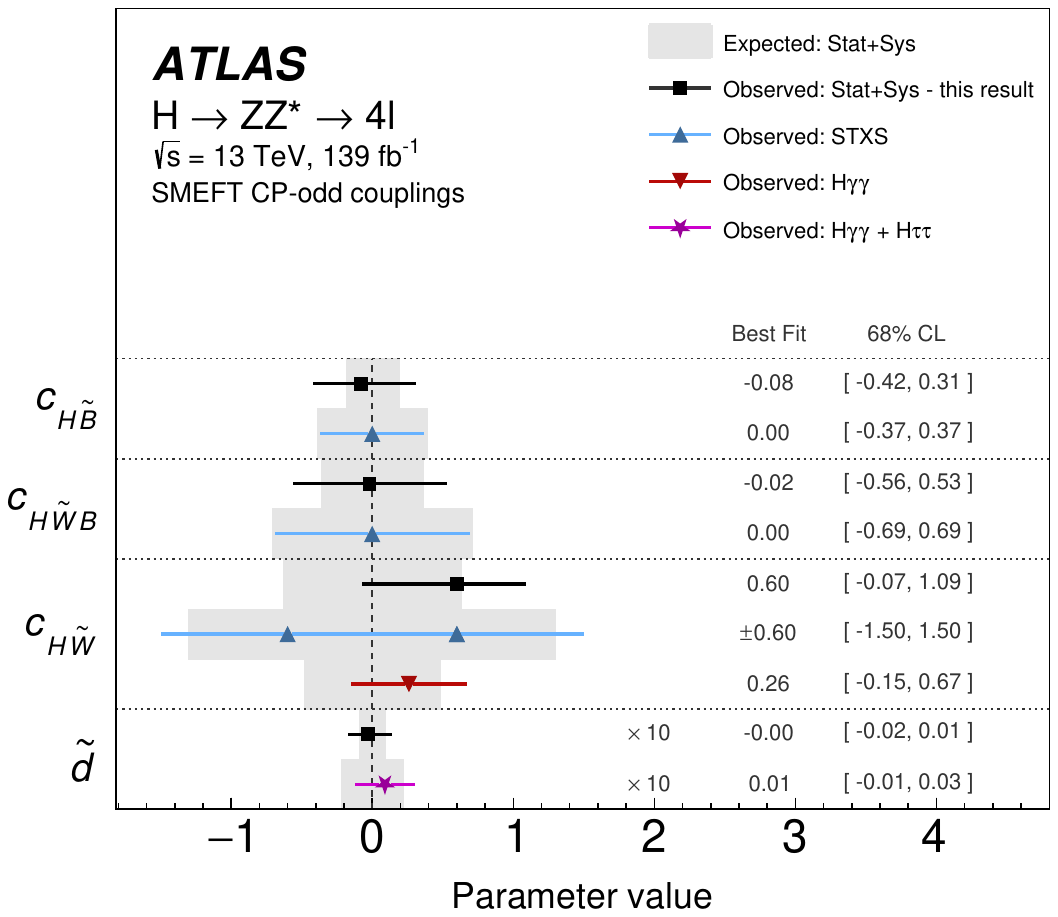}
\caption{
Comparison of results for the present measurement, the ATLAS \htollllbrief STXS cross-section measurement of Ref.~\cite{HIGG-2018-28},  the ATLAS \htoyy \VBF measurement~\cite{HIGG-2020-08} of \cHWtil, and the combined ATLAS \VBF measurement of \dTilde in \htoyy and \htautau decay channels~\cite{HIGG-2018-14, HIGG-2020-08}
The data points and 68\% CL uncertainty bars show the observed values with statistical and systematic uncertainties.
The \dTilde best fit points and uncertainty bars are scaled by a factor of ten.
}
\label{fig:compare_results}
\end{figure}

In conclusion, the results of the coupling measurement are all in agreement with the SM\@. A fluctuation in the optimal observables leads to a slight preference for a non-zero BSM coupling in the production-level analysis, which is compatible with the SM at better than two standard deviations and not confirmed by the decay-level analysis.

\subsection{Differential optimal-observable cross-section results}
\label{sec:differential_results}
The differential cross-sections for the optimal observables are measured using a binned profile likelihood ratio fit, fitting simultaneously the \mfourl\ distribution in each bin across all bins of a given distribution, as described in \Sect{\ref{sec:diff_xsec}}.
 
The measured differential cross-section for the production optimal observable \OOjjcHWtil and those for the decay optimal observables \OOllllcHWtil, \OOllllcHBtil and \OOllllcHWBtil are shown in \Fig{\ref{fig:OO_unfolded}}.
The couplings used for these optimal observables correspond to those with good expected sensitivity for \VBF production or four-lepton decay of the direct coupling measurement, as previously described.
The first bin of \Fig{\ref{fig:summary_OOcHWtil}} is not part of the \OOjjcHWtil differential distribution but rather contains events with fewer than two selected jets with $\pt >  30$ \GeV.
These events are included as part of the fiducial distribution unfolding as they have a correlation of up to 15\% with the cross-section measurements of the \OOjjcHWtil differential bins as can be seen in \Fig{\ref{fig:corrMatrix_OOcHWtil}} of \App{\ref{sec:appdiff}}.
This correlation arises from the jet energy resolution.
The larger number of decay events and the relatively better resolution for leptons allows a finer binning for the decay observable differential distributions.
The correlation between neighbouring bins is up to 30\% for \OOllllcHWtil with 12 bins, and up to 20\% for the six bin distributions of \OOllllcHBtil and \OOllllcHWBtil, as shown in \App{\ref{sec:appdiff}}.
 
These figures show the measured differential cross-section compared with SM predictions for ggF production provided by \progname{NNLOPS} and \progname{MadGraph5\_aMC@NLO}-FxFx (\progname{MG5} FxFx), which are both normalized to the N$^3$LO calculation~\cite{Anastasiou:2015ema,Anastasiou:2016cez}.
The other Higgs boson production modes, denoted by $XH$, are normalized to the most accurate predictions available, as given in~\cite{HIGG-2018-29}.
All of the figures include the $p$-values that quantify the probability of compatibility of the measurements and the corresponding SM prediction for \ggF + $XH$, including the theoretical uncertainties in the ggF predictions.
For the $p$-value estimations, QCD scale and PDF uncertainties for \progname{NNLOPS} are evaluated  individually for each bin; however for \progname{MG5} FxFx only total QCD scale and PDF uncertainties are available for all bins.
In both cases the theoretical uncertainties are treated as correlated across the bins.
All $p$-values reveal a good compatibility of the data with the SM predictions.
For \OOllllcHWtil in \Fig{\ref{fig:summary_OO4lcHWtil}}, the lower $p$-value is from a large fluctuation in one differential bin.

\begin{figure}[!htbp]
\centering
\subfloat[]{\includegraphics[width=0.45\linewidth]{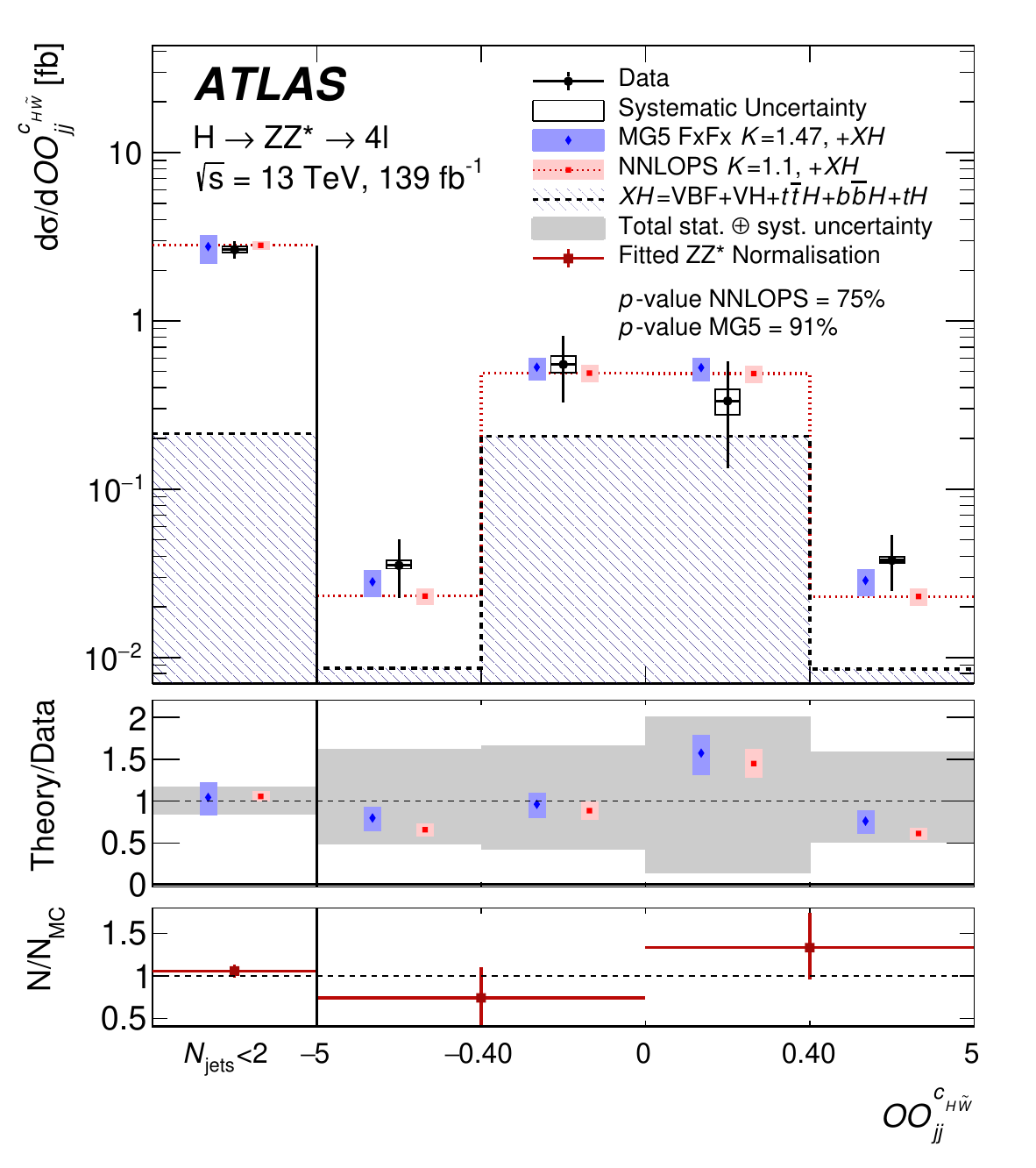}
\label{fig:summary_OOcHWtil}
}
\subfloat[]{\includegraphics[width=0.45\linewidth]{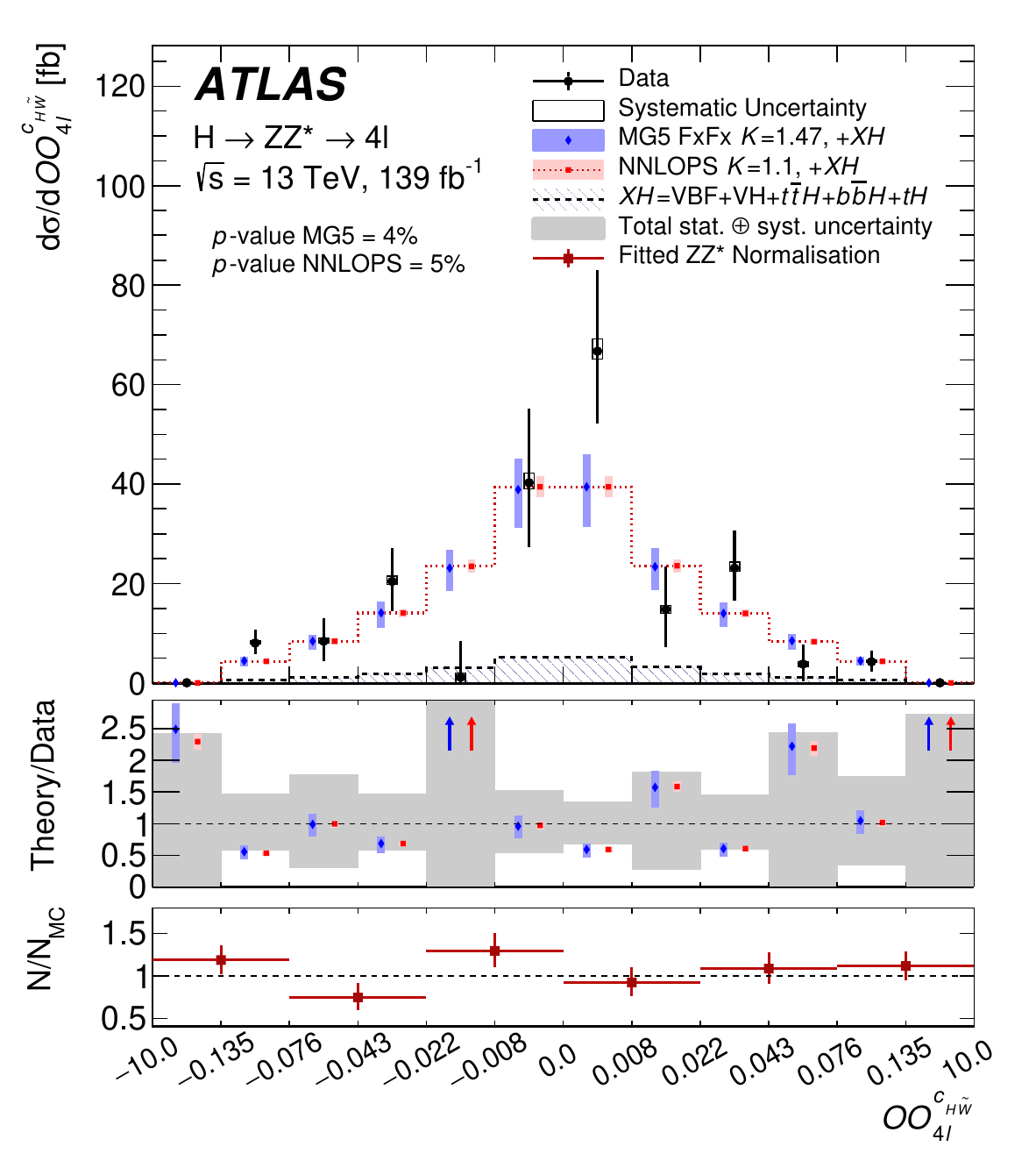}
\label{fig:summary_OO4lcHWtil}
}
\\
\subfloat[]{\includegraphics[width=0.45\linewidth]{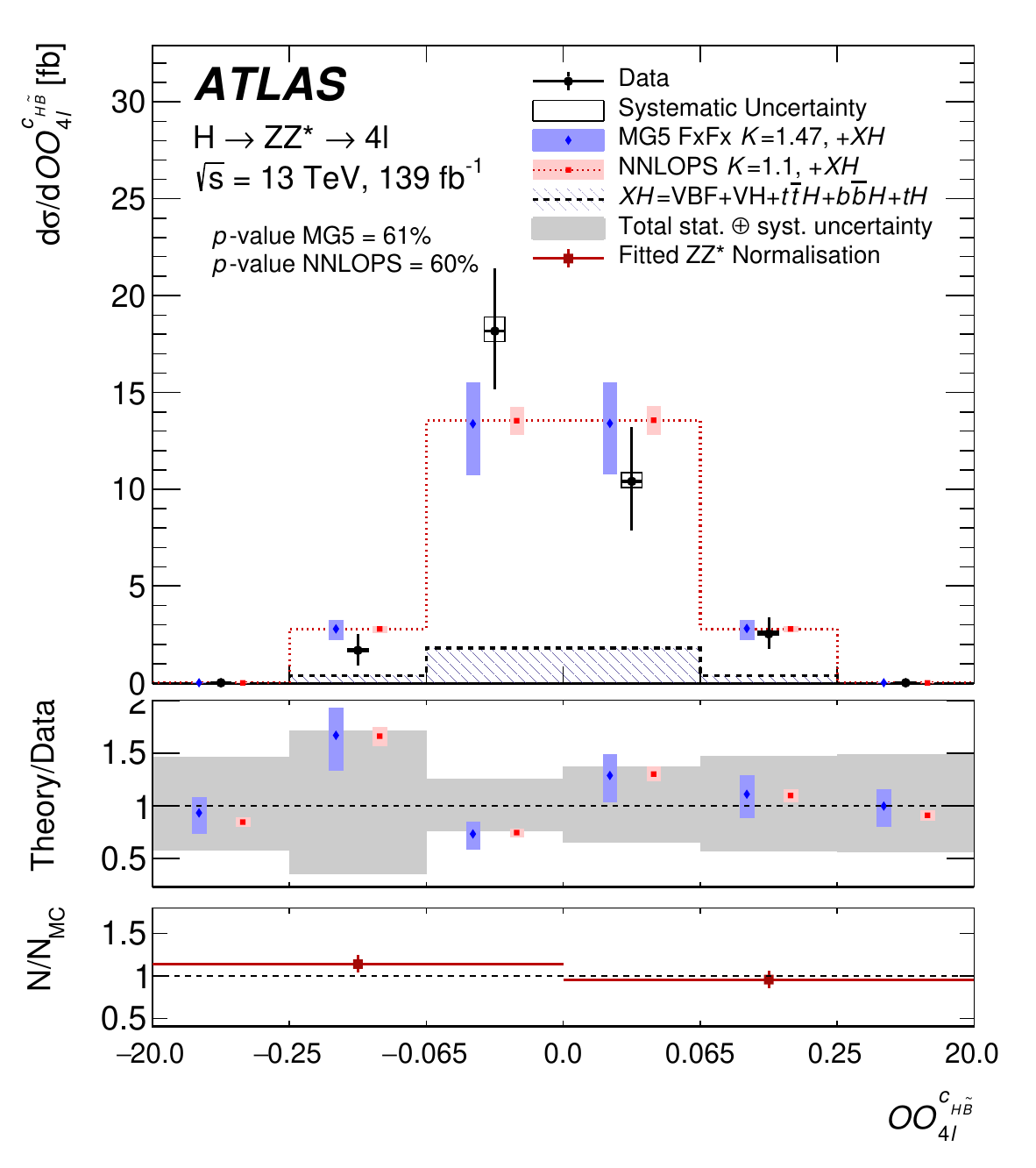}
\label{fig:summary_OO4lcHBtil}
}
\subfloat[]{\includegraphics[width=0.45\linewidth]{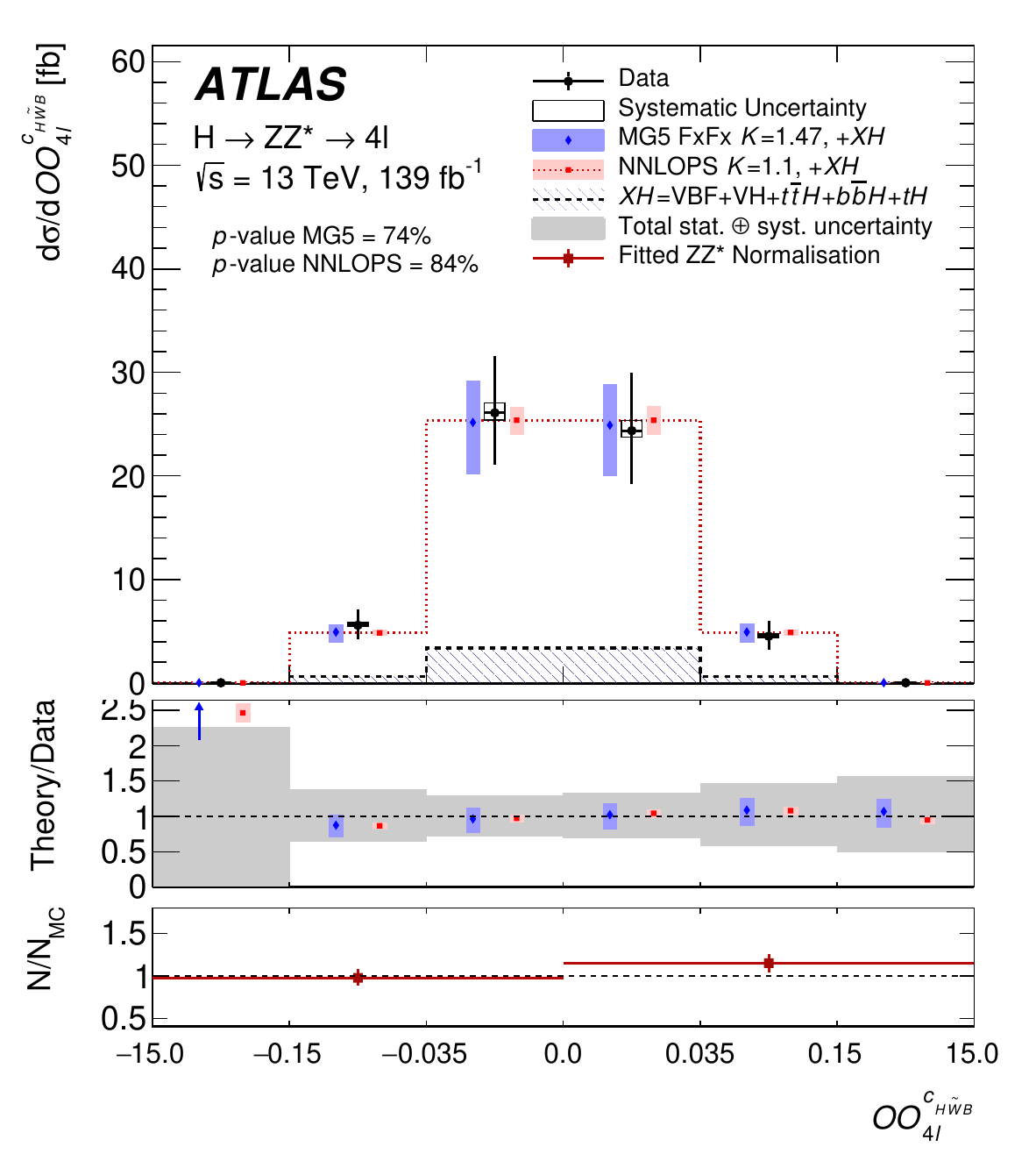}
\label{fig:summary_OO4lcHWBtil}
}
\\
\caption{Results for differential fiducial cross-section for the optimal observables \protect\subref{fig:summary_OOcHWtil} \OOjjcHWtil, \protect\subref{fig:summary_OO4lcHWtil} \OOllllcHWtil, \protect\subref{fig:summary_OO4lcHBtil} \OOllllcHBtil and \protect\subref{fig:summary_OO4lcHWBtil} \OOllllcHWBtil.
The first bin of \OOjjcHWtil distribution contains events with fewer than two jets that satisfy the jet selection requirements.
The error bars on the data points show the total uncertainties, while the systematic uncertainties are indicated by the boxes.
The shaded bands on the expected differential cross-sections indicate the PDF and scale systematic uncertainties.
The central panel shows the ratio of different predictions to the data, and the grey area represents the total uncertainty in the measurement. The bottom panel shows the ratios of the fitted values of the \zzstar normalization factors to the predictions from MC simulation.
As indicated by the horizontal error bars, the same \zzstar normalizations are estimated for different cross-section bins of the distributions.
\label{fig:OO_unfolded}}
\end{figure}

\FloatBarrier
 
\subsection{\VBF-enriched fiducial cross-section results}\label{sec:VBFenriched}
 
As described in \Sect{\ref{sec:diff_xsec}}, the \VBF-enriched fiducial cross-section measurement is performed by dividing the two-dimensional plane $m_{jj}$ versus $|\Delta\eta_{jj}|$ into three bins with a signal region defined by $m_{jj} \ge 400$ \GeV and $|\Delta \eta_{jj}| \ge 3.0$, and two control regions outside this selection.
The expected and observed events in this two-dimensional plane can be seen in \Fig{\ref{fig:VBF_mjj_detajj}}.
Events are unfolded to the corresponding fiducial phase space and two \VBF-enriched fiducial cross-sections are extracted: with the \ggF expectation treated as part of the signal region and with it considered as background at the detector-level.
The results for these two measurements are presented in Table~\ref{tab:FidVBF}.

\begin{figure}[!htbp]
\centering
\includegraphics[width = 0.7 \textwidth]{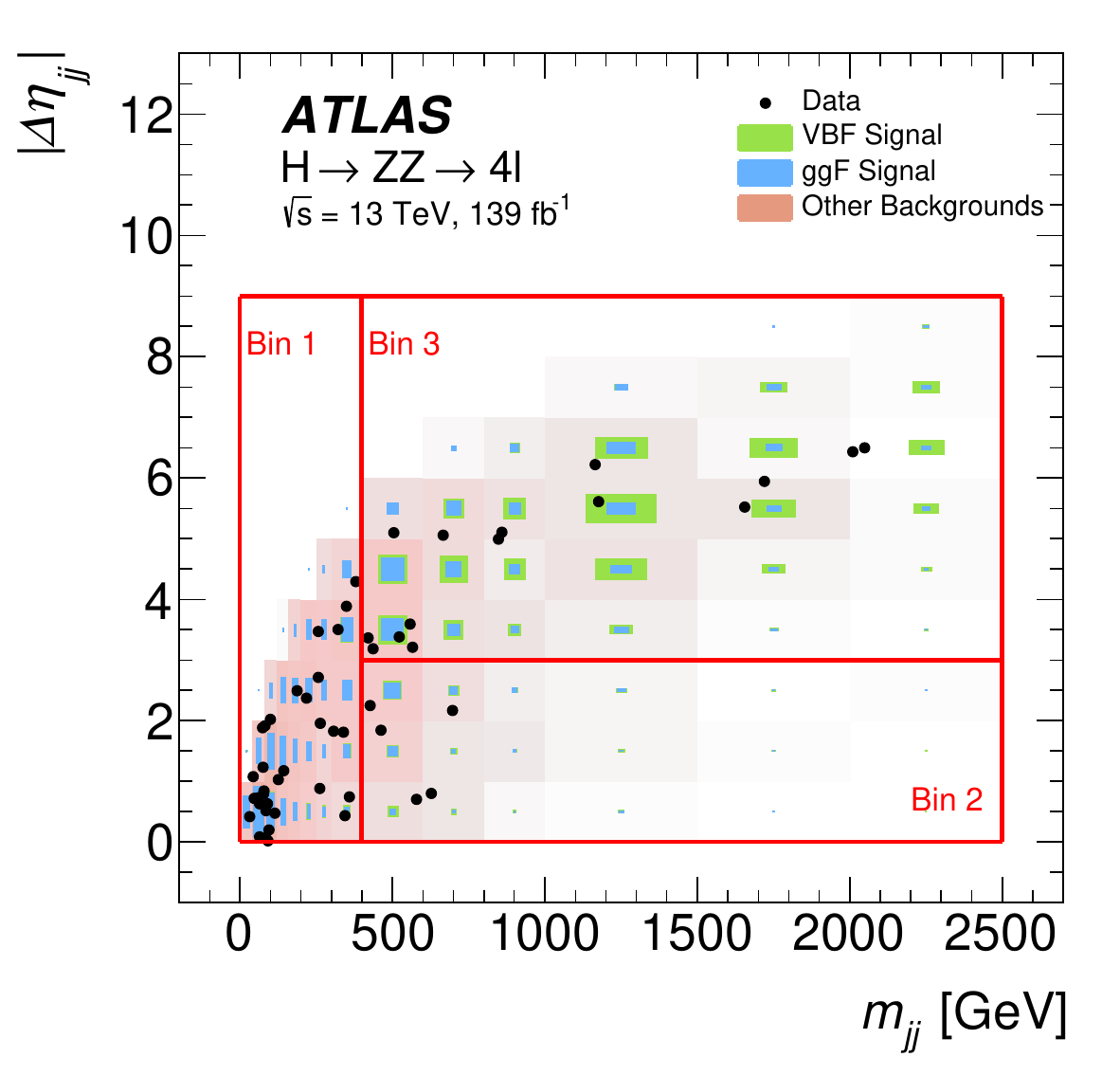}
\caption{
The expected and observed events in the $m_{jj}$ vs $|\Delta\eta_{jj}|$ two-dimensional plane for the \VBF-enriched fiducial cross-section.
The dots depict data and the shaded areas represent the simulated signal for \VBF, \ggF, and other backgrounds, respectively.
The lines depict the bin boundaries for unfolding into the corresponding fiducial phase space.
\label{fig:VBF_mjj_detajj}
}
\end{figure}
 
\begin{DIFnomarkup}
\begin{table}[!h]
\renewcommand\arraystretch{1.8}
\caption{Expected and observed \VBF fiducial cross-sections measured in a \VBF-enriched fiducial region with and without including the \ggF contribution as part of the signal region.}
\centering \small
\begin{tabular}{c|c|c|c|c}
\hline
\hline
\VBF--enriched & Signal for cross-& Purity of & Expected & Observed \\
region & section estimates& \VBF signal & cross-section [fb] & cross-section [fb] \\
\hline
$N_{\mathrm{jets}}\ge 2$,  $\ m_{jj} \ge 400 \mbox{ GeV}$& All production modes & 59 \% & \scalebox{1.}{$0.134^{+0.065}_{-0.053}$ $^{+0.014}_{-0.012}$}  & \scalebox{1.}{$0.215^{+0.075}_{-0.063}$ $^{+0.016}_{-0.013}$} \\
\cline{2-5}
$|\Delta \eta_{jj}| \ge 3.0$ & \VBF+\VH+\ttH & 95 \% & \scalebox{1.}{$0.088^{+0.063}_{-0.053}$ $^{+0.017}_{-0.020}$} & \scalebox{1.}{$0.172^{+0.072}_{-0.062}$ $^{+0.016}_{-0.018}$} \\
\hline
\hline
\end{tabular}
\label{tab:FidVBF}
\end{table}
 
\end{DIFnomarkup}

For both cases, the measured fiducial cross-section is larger than the expected, with a signal strength of about 1.6 and 2, but remain compatible with the SM as evaluated from $p$-values.
In the first scenario the $p$-value compatibility with \progname{NNLOPS} and \progname{MG5} FxFx is 82\% and 80\%,  respectively.
In the second scenario, where the \ggF is treated as background, the $p$-value compatibility with the SM predictions for the remaining three production modes is 19\%.

These measured fiducial cross-sections can be compared with the \VBF production cross-section of  \htollllbrief decay channel measured in the context of the STXS framework~\cite{HIGG-2018-28}.
The STXS measurement has a relative observed cross-section times branching ratio $(\sigma \cdot \cal{B})/(\sigma \cdot \cal{B})_\mathrm{SM}$ of $1.21\pm0.45$ for the \VBF production mode.
The STXS measurement extracts the \VBF signal using neural networks, similar to the direct measurement above, and combines \VBF signals from several kinematic bins separated according to the number of jets and \ptH. The present fiducial analysis instead applies explicit selections on $m_{jj}$ and $|\Delta\eta_{jj}|$ for events with two jets, which selects events in a different phase space.
All three measurements yield cross-sections that are compatible with the SM expectations.


\FloatBarrier

\section{Conclusion}

A measurement searching for CP violation in \VBF Higgs boson production and its decay into four leptons is presented based on proton--proton collision data produced at the LHC at a centre-of-mass energy of 13~\TeV\ and recorded by the ATLAS detector from 2015 to 2018, corresponding to an integrated luminosity of 139 $\mbox{fb\(^{-1}\)}$.
The measurement is performed using optimal observables defined as the ratio of the interference term between the SM and CP-odd BSM matrix elements to the SM matrix element squared.
Limits are obtained in both the Higgs and Warsaw bases using the SMEFT\@.
They are dominated by the interference term between the SM and BSM contributions linear in the Wilson coefficients, $\mathcal{O}\left(\Lambda^{-2}\right)$ in the cross-section, with only a small sensitivity to the pure BSM contributions quadratic in the Wilson coefficients, $\mathcal{O}\left(\Lambda^{-4}\right)$.
This qualitatively implies a low expected sensitivity to missing dimension-eight contributions that also enter at $\mathcal{O}\left(\Lambda^{-4}\right)$ in the cross-section, and it is therefore an improvement on analyses relying on rates rather than shapes.
Measurements of fiducial differential cross-sections of the optimal-observable distributions are also presented, completing the set of fiducial differential cross-section measurements in the Higgs boson to four-lepton channel.
In addition, fiducial cross-sections for \VBF production are measured for a \VBF-enriched fiducial region for two scenarios: with and without the explicit subtraction of the \ggF background.
All measurements are consistent with the SM expectation of a CP-even Higgs boson.
No significant CP-odd component is observed.
 
The data for all tables and the figures with differential distributions can be viewed at \href{https://www.hepdata.net/record/sandbox/1677693099}{HEPData}.


\section*{Acknowledgements}


We thank CERN for the very successful operation of the LHC and its injectors, as well as the support staff at
CERN and at our institutions worldwide without whom ATLAS could not be operated efficiently.
 
The crucial computing support from all WLCG partners is acknowledged gratefully, in particular from CERN, the ATLAS Tier-1 facilities at TRIUMF/SFU (Canada), NDGF (Denmark, Norway, Sweden), CC-IN2P3 (France), KIT/GridKA (Germany), INFN-CNAF (Italy), NL-T1 (Netherlands), PIC (Spain), RAL (UK) and BNL (USA), the Tier-2 facilities worldwide and large non-WLCG resource providers. Major contributors of computing resources are listed in Ref.~\cite{ATL-SOFT-PUB-2023-001}.
 
We gratefully acknowledge the support of ANPCyT, Argentina; YerPhI, Armenia; ARC, Australia; BMWFW and FWF, Austria; ANAS, Azerbaijan; CNPq and FAPESP, Brazil; NSERC, NRC and CFI, Canada; CERN; ANID, Chile; CAS, MOST and NSFC, China; Minciencias, Colombia; MEYS CR, Czech Republic; DNRF and DNSRC, Denmark; IN2P3-CNRS and CEA-DRF/IRFU, France; SRNSFG, Georgia; BMBF, HGF and MPG, Germany; GSRI, Greece; RGC and Hong Kong SAR, China; ISF and Benoziyo Center, Israel; INFN, Italy; MEXT and JSPS, Japan; CNRST, Morocco; NWO, Netherlands; RCN, Norway; MEiN, Poland; FCT, Portugal; MNE/IFA, Romania; MESTD, Serbia; MSSR, Slovakia; ARRS and MIZ\v{S}, Slovenia; DSI/NRF, South Africa; MICINN, Spain; SRC and Wallenberg Foundation, Sweden; SERI, SNSF and Cantons of Bern and Geneva, Switzerland; MOST, Taipei; TENMAK, T\"urkiye; STFC, United Kingdom; DOE and NSF, United States of America.
 
Individual groups and members have received support from BCKDF, CANARIE, CRC and DRAC, Canada; CERN-CZ, PRIMUS 21/SCI/017 and UNCE SCI/013, Czech Republic; COST, ERC, ERDF, Horizon 2020, ICSC-NextGenerationEU and Marie Sk{\l}odowska-Curie Actions, European Union; Investissements d'Avenir Labex, Investissements d'Avenir Idex and ANR, France; DFG and AvH Foundation, Germany; Herakleitos, Thales and Aristeia programmes co-financed by EU-ESF and the Greek NSRF, Greece; BSF-NSF and MINERVA, Israel; Norwegian Financial Mechanism 2014-2021, Norway; NCN and NAWA, Poland; La Caixa Banking Foundation, CERCA Programme Generalitat de Catalunya and PROMETEO and GenT Programmes Generalitat Valenciana, Spain; G\"{o}ran Gustafssons Stiftelse, Sweden; The Royal Society and Leverhulme Trust, United Kingdom.
 
In addition, individual members wish to acknowledge support from Chile: Agencia Nacional de Investigaci\'on y Desarrollo (FONDECYT 1190886, FONDECYT 1210400); China: National Natural Science Foundation of China (NSFC - 12175119, NSFC 12275265, NSFC-12075060); Czech Republic: PRIMUS Research Programme (PRIMUS/21/SCI/017); EU: H2020 European Research Council (H2020-MSCA-IF-2020: HPOFHIC - 10103); European Union: European Research Council (ERC - 948254), Horizon 2020 Framework Programme (MUCCA - CHIST-ERA-19-XAI-00), European Union, Future Artificial Intelligence Research (FAIR-NextGenerationEU PE00000013), Italian Center for High Performance Computing, Big Data and Quantum Computing (ICSC, NextGenerationEU), Marie Sklodowska-Curie Actions (EU H2020 MSC IF GRANT NO 101033496); France: Agence Nationale de la Recherche (ANR-20-CE31-0013, ANR-21-CE31-0013, ANR-21-CE31-0022), Investissements d'Avenir Idex (ANR-11-LABX-0012), Investissements d'Avenir Labex (ANR-11-LABX-0012); Germany: Baden-Württemberg Stiftung (BW Stiftung-Postdoc Eliteprogramme), Deutsche Forschungsgemeinschaft (DFG - CR 312/5-1); Italy: Istituto Nazionale di Fisica Nucleare (FELLINI G.A. n. 754496, ICSC, NextGenerationEU); Japan: Japan Society for the Promotion of Science (JSPS KAKENHI JP21H05085, JSPS KAKENHI JP22H01227, JSPS KAKENHI JP22H04944); Netherlands: Netherlands Organisation for Scientific Research (NWO Veni 2020 - VI.Veni.202.179); Norway: Research Council of Norway (RCN-314472); Poland: Polish National Agency for Academic Exchange (PPN/PPO/2020/1/00002/U/00001), Polish National Science Centre (NCN 2021/42/E/ST2/00350, NCN UMO-2019/34/E/ST2/00393, UMO-2020/37/B/ST2/01043, UMO-2021/40/C/ST2/00187); Slovenia: Slovenian Research Agency (ARIS grant J1-3010); Spain: BBVA Foundation (LEO22-1-603), Generalitat Valenciana (Artemisa, FEDER, IDIFEDER/2018/048), La Caixa Banking Foundation (LCF/BQ/PI20/11760025), Ministry of Science and Innovation (MCIN \& NextGenEU PCI2022-135018-2, MICIN \& FEDER PID2021-125273NB, RYC2019-028510-I, RYC2020-030254-I, RYC2021-031273-I, RYC2022-038164-I), PROMETEO and GenT Programmes Generalitat Valenciana (CIDEGENT/2019/023, CIDEGENT/2019/027); Sweden: Swedish Research Council (VR 2018-00482, VR 2022-03845, VR 2022-04683, VR grant 2021-03651), Knut and Alice Wallenberg Foundation (KAW 2017.0100, KAW 2018.0157, KAW 2018.0458, KAW 2019.0447); Switzerland: Swiss National Science Foundation (SNSF - PCEFP2\_194658); United Kingdom: Leverhulme Trust (Leverhulme Trust RPG-2020-004); United States of America: U.S. Department of Energy (ECA DE-AC02-76SF00515), Neubauer Family Foundation.


\clearpage
\appendix
\part*{Appendix}
\addcontentsline{toc}{part}{Appendix}
\section{Conversion between Warsaw and Higgs bases}
\label{sec:appconv}

The CP-odd operators studied in the present analysis are given in Table~\ref{tab:CPcoup}.
The couplings of the two bases
\begin{itemize}
\item Warsaw basis: \cHWtil, \cHBtil, \cHWtilB
\item Higgs basis: \tCzz, \tCza, \tCaa
\end{itemize}
can be expressed as linear combination of each other:
\begin{align}
\tilde{c}_{zz} &=
4 \left( \frac{ g^2 c_{H\widetilde{W}} + g^{\prime2} c_{H\widetilde{B}} + g g^{\prime} c_{H\widetilde{W}B} }{ (g^2 + g^{\prime2})^2 } \right) \frac{\varv^2}{\Lambda^2} \\
\tilde{c}_{z\gamma} &=
4  \left( \frac{ c_{H\widetilde{W}} - c_{H\widetilde{B}} - \frac{g^2 - g^{\prime2}}{2gg^{\prime}} c_{H\widetilde{W}B} }{ g^2 + g^{\prime2} } \right) \frac{\varv^2}{\Lambda^2} \\
\tilde{c}_{\gamma\gamma} &=
4 \left( \frac{1}{g^2} c_{H\widetilde{W}} + \frac{1}{g^{\prime2}} c_{H\widetilde{B}} - \frac{1}{g g^{\prime}} c_{H\widetilde{W}B} \right) \frac{\varv^2}{\Lambda^2}
\end{align}
and
\begin{align}
\frac{\varv^2}{\Lambda^2} c_{H\widetilde{W}} &= \frac{ g^2 }{ 4(g^2 + g^{\prime2})^2 }
\left( (g^2 + g^{\prime2})^2 \tilde{c}_{zz} + 2(g^2 + g^{\prime2}) g^{\prime2} \tilde{c}_{z\gamma} + g^{\prime4} \tilde{c}_{\gamma\gamma} \right) \\
\frac{\varv^2}{\Lambda^2} c_{H\widetilde{B}} &= \frac{ g^{\prime2} }{ 4(g^2 + g^{\prime2})^2 }
\left( (g^2 + g^{\prime2})^2 \tilde{c}_{zz} - 2(g^2 + g^{\prime2}) g^2 \tilde{c}_{z\gamma} + g^{\prime4} \tilde{c}_{\gamma\gamma} \right) \\
\frac{\varv^2}{\Lambda^2} c_{H\widetilde{W}B} &= \frac{ gg^{\prime} }{ 2(g^2 + g^{\prime2})^2 }
\left( (g^2 + g^{\prime2})^2 \tilde{c}_{zz} - (g^4 - g^{\prime4}) \tilde{c}_{z\gamma} - g^2g^{\prime2} \tilde{c}_{\gamma\gamma} \right)\!,
\end{align}
where $g$ and $g^\prime$ are the gauge couplings of the $SU(2)$ and $SU(1)$ local symmetries, respectively, $\varv$ is the Higgs boson vacuum expectation value and $\Lambda$ is is the energy cutoff scale of the effective field theory.


\section{Additional differential cross-section results}
\label{sec:appdiff}

This appendix shows additional results of the differential fiducial cross-section measurements presented in \Sect{\ref{sec:differential_results}}.
 
\Fig{\ref{fig:OO_corrMatrix}} shows the correlation matrices for the differential fiducial cross-sections presented in \Fig{\ref{fig:OO_unfolded}}. Non-negligible anti-correlations exist between the cross-sections in neighbouring bins due to detector resolution effects. In addition, there are anti-correlations between the cross-section for $\njets < 2$ and the four optimal observable bins in the $\OOjjcHWtil$ measurement shown in \Fig{\ref{fig:corrMatrix_OOcHWtil}}, driven by jet energy resolution. Finally, the normalization of the $\qqZZ$ background is anti-correlated with the signal cross-sections measured in the respective bins.
 
\begin{figure}[!htbp]
\centering
\subfloat[]{\includegraphics[width=0.5\linewidth]{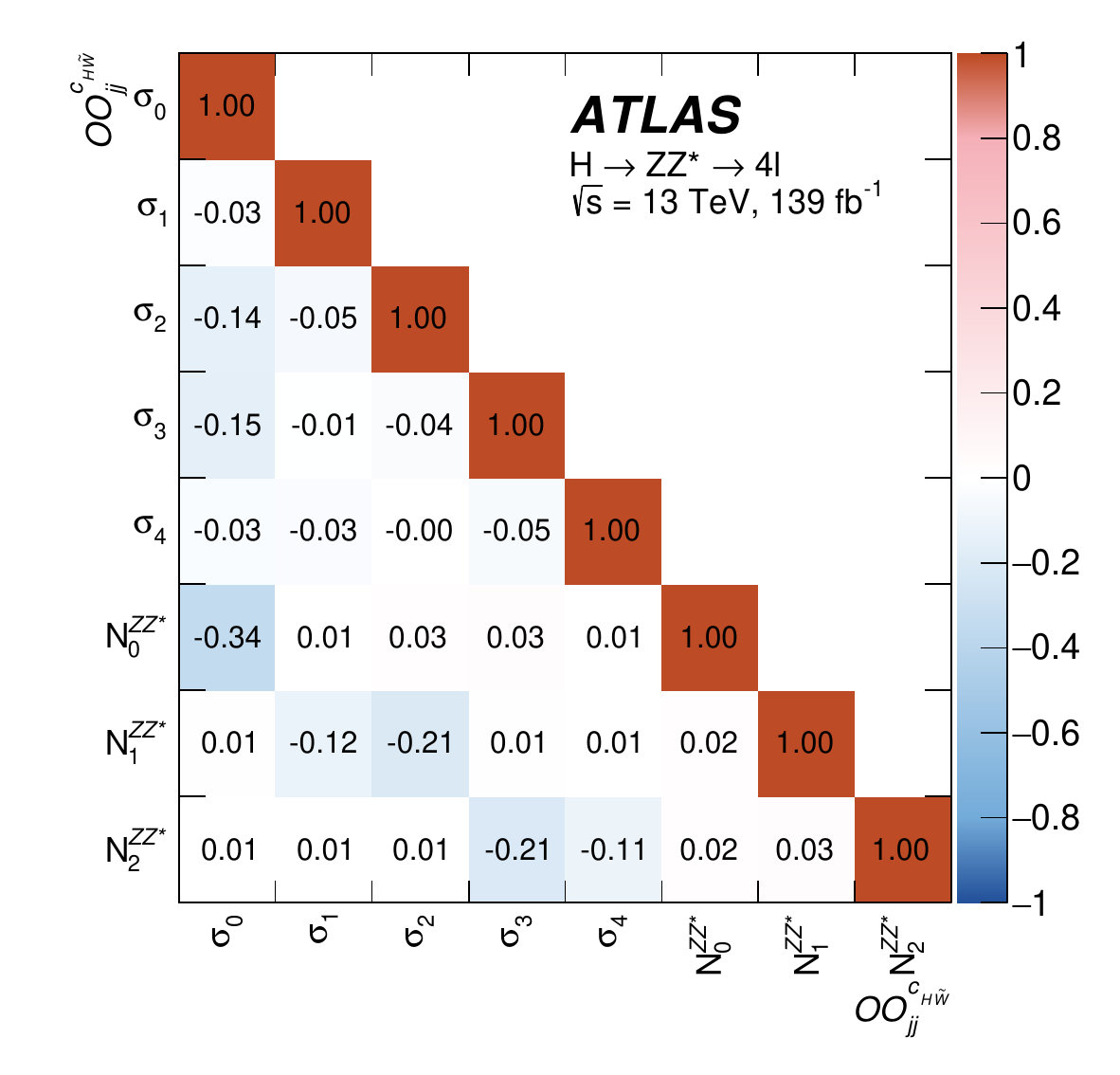}
\label{fig:corrMatrix_OOcHWtil}
}
\subfloat[]{\includegraphics[width=0.5\linewidth]{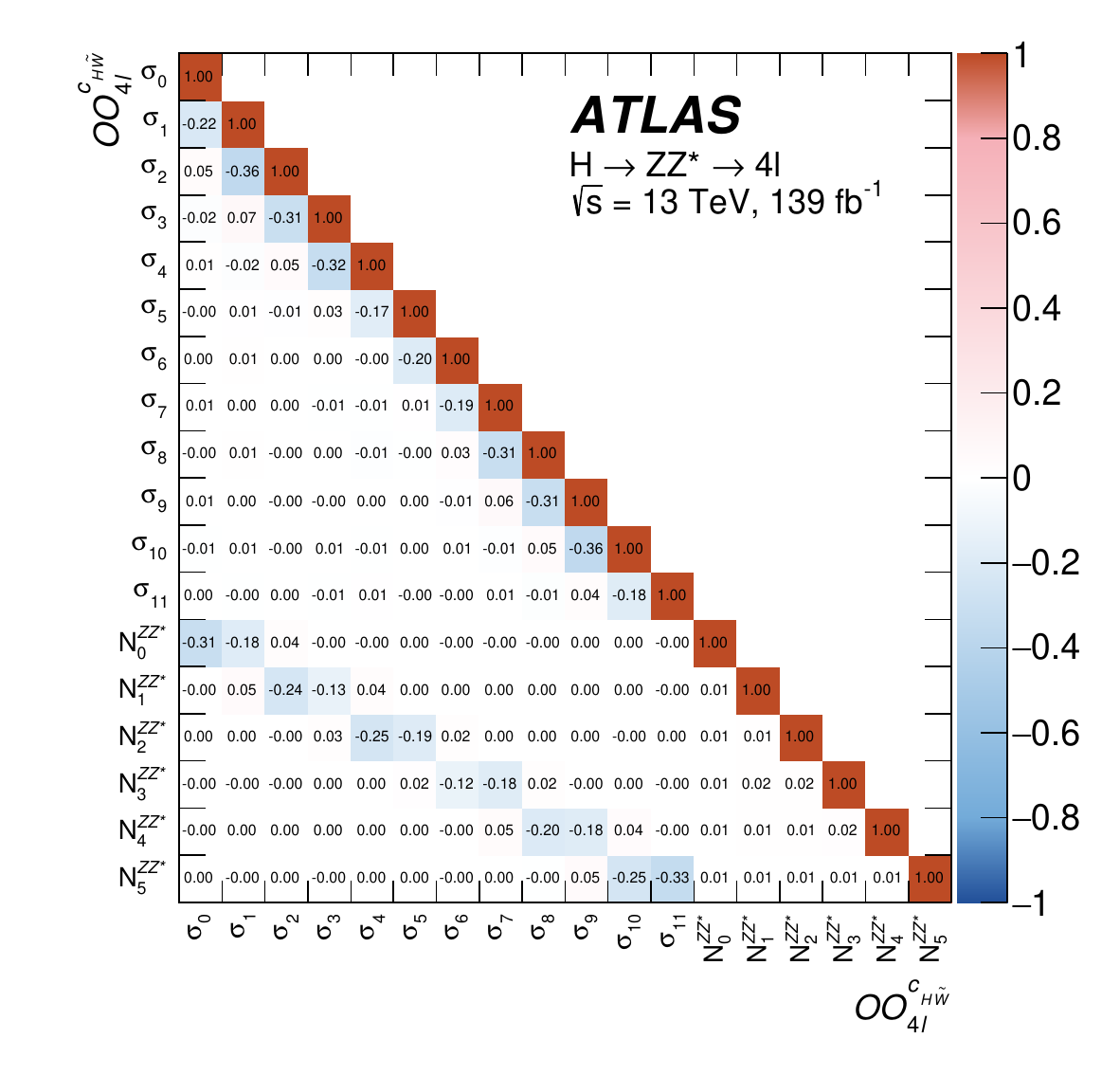}
\label{fig:corrMatrix_OO4lcHWtil}
}
\\
\subfloat[]{\includegraphics[width=0.5\linewidth]{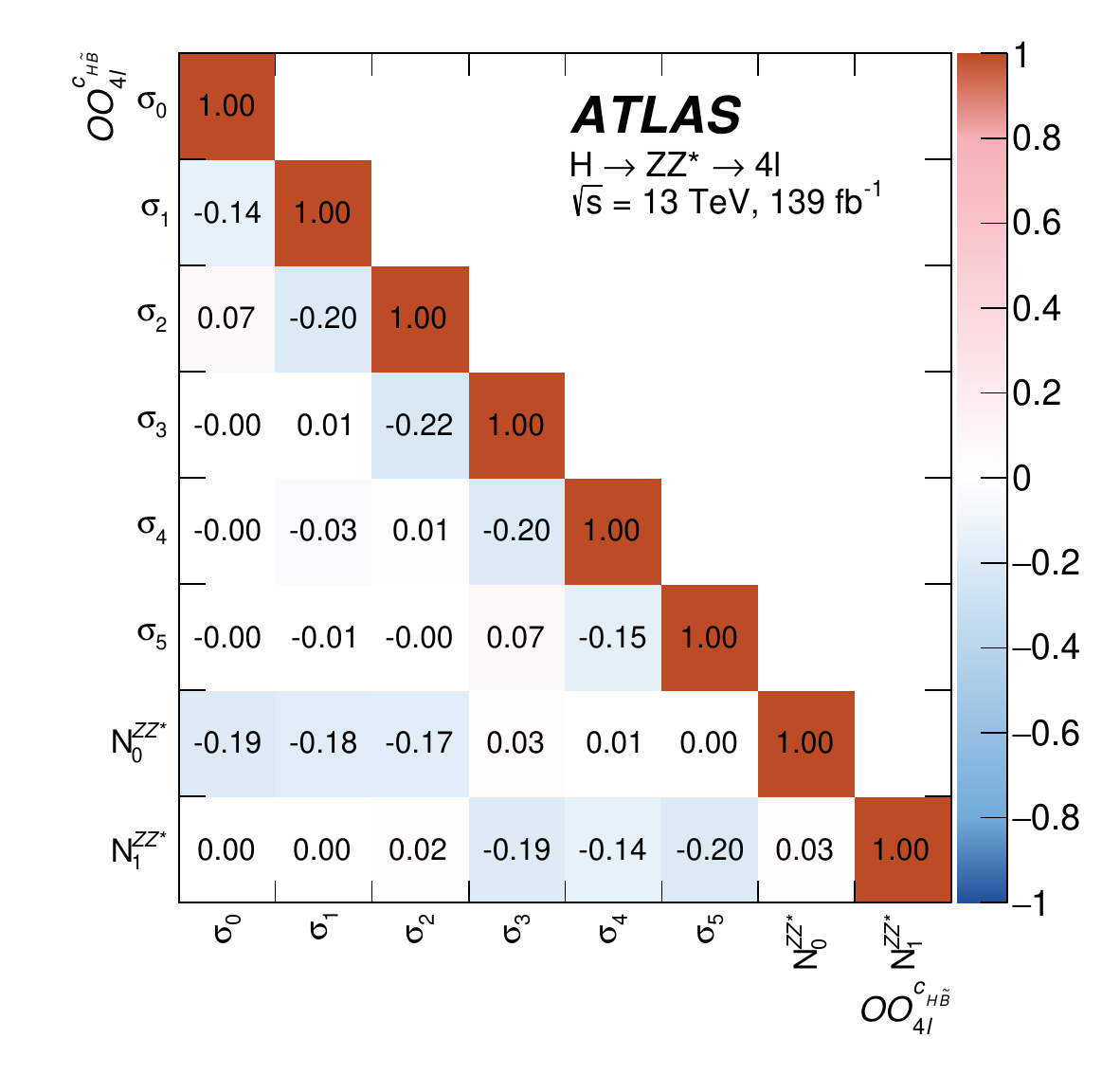}
\label{fig:corrMatrix_OO4lcHBtil}
}
\subfloat[]{\includegraphics[width=0.5\linewidth]{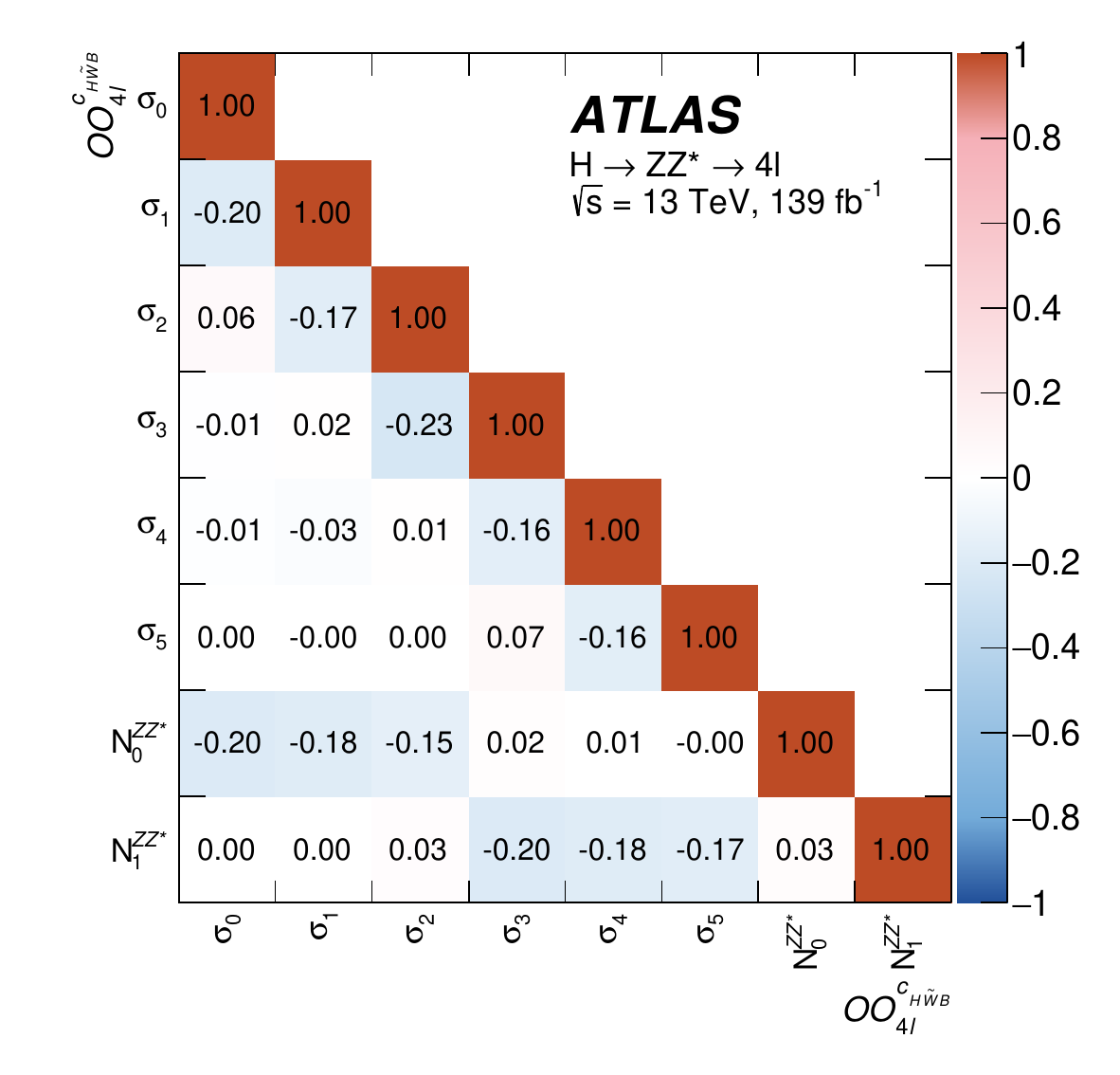}
\label{fig:corrMatrix_OO4lcHWBtil}
}
\caption{Correlation matrices between the measured differential fiducial cross-section and \zzstar normalization bins for the \protect\subref{fig:corrMatrix_OOcHWtil} \OOjjcHWtil, \protect\subref{fig:corrMatrix_OO4lcHWtil} \OOllllcHWtil, \protect\subref{fig:corrMatrix_OO4lcHBtil} \OOllllcHBtil and \protect\subref{fig:corrMatrix_OO4lcHWBtil} \OOllllcHWBtil optimal observables.
The bin labeled $\sigma_0$ of \protect\subref{fig:corrMatrix_OOcHWtil} is not part of the \OOjjcHWtil differential distribution, but contains events with fewer than two selected jets with $\pt >  30$ \GeV.
\label{fig:OO_corrMatrix}}
\end{figure}
 
\Fig{\ref{fig:OOtCzz_unfolded}} shows the differential fiducial cross-section for the $\OOjjtCzz$ observable and the associated correlation matrix.
This observable is strongly correlated with $\OOjjcHWtil$ and the two measurements thus share many features.
 
\begin{figure}[!htbp]
\centering
\subfloat[]{\includegraphics[width=0.45\linewidth]{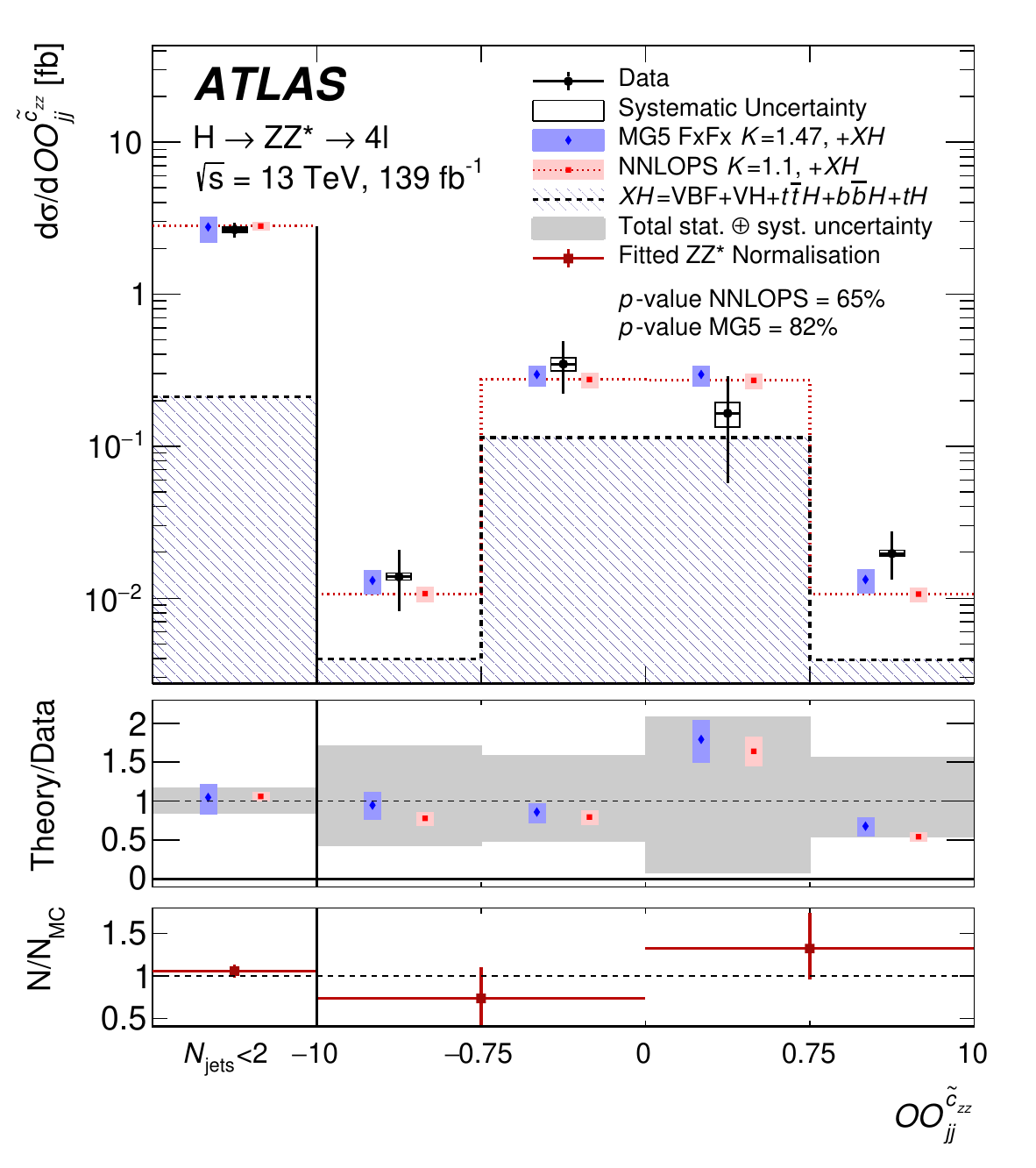}
\label{fig:summary_OOtCzz}
}
\subfloat[]{\includegraphics[width=0.45\linewidth]{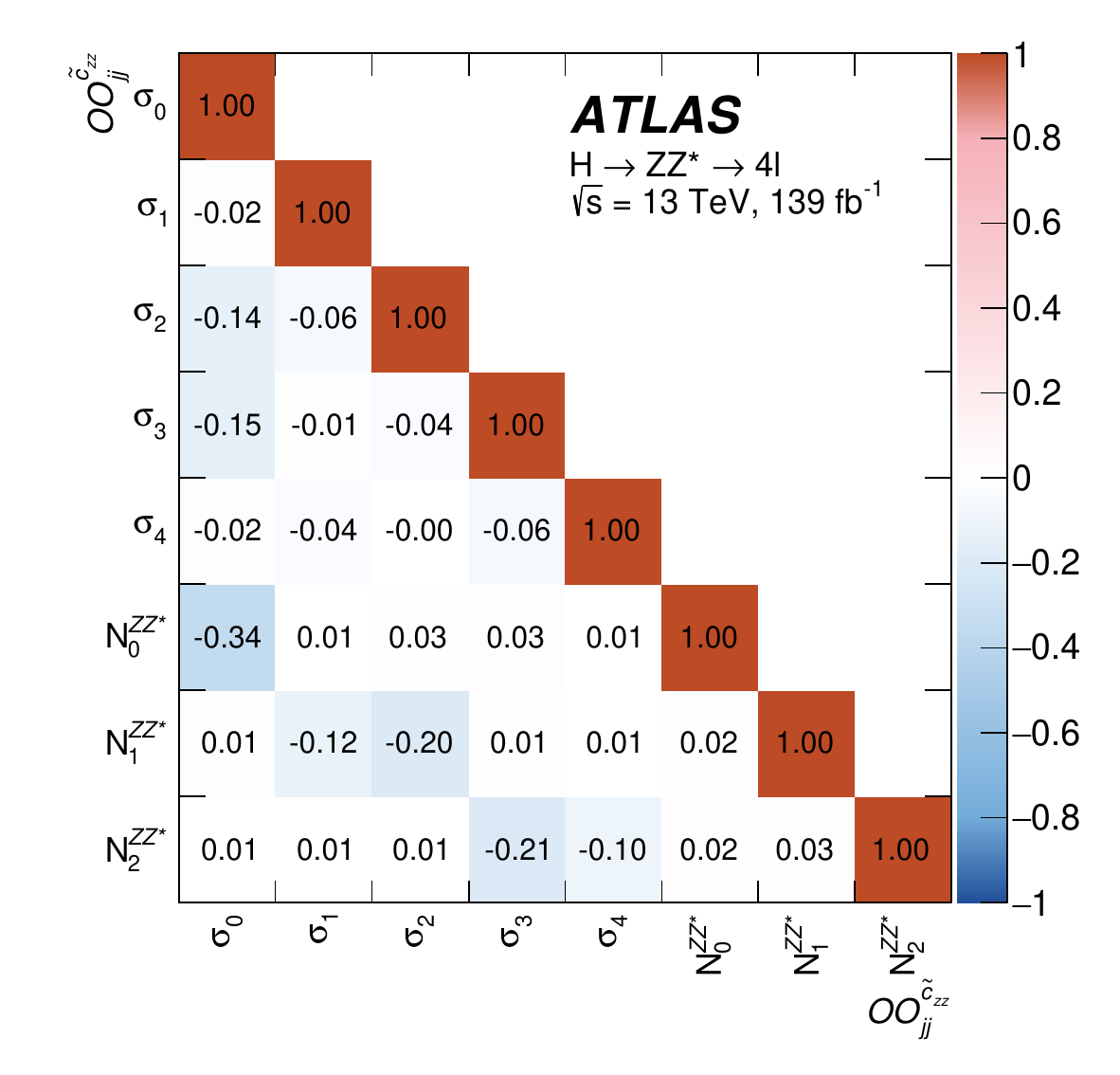}
\label{fig:corrMatrix_OOtCzz}
}
\\
\caption{\protect\subref{fig:summary_OOtCzz} Results for the differential fiducial cross-section for the \OOjjtCzz optimal observable, along with~\protect\subref{fig:corrMatrix_OOtCzz}  the corresponding correlation matrix between the measured cross-sections and the $ZZ^{\ast}$ background normalization factors.
The first bin of \protect\subref{fig:summary_OOtCzz} is not part of the \OOjjtCzz differential distribution, but contains events with fewer than two selected jets with $\pt >  30$ \GeV.
This bin corresponds to $\sigma_0$ of \protect\subref{fig:corrMatrix_OOtCzz}.
\label{fig:OOtCzz_unfolded}}
\end{figure}
 
\Fig{\ref{fig:summary_VBFfid}} shows the per-bin fiducial cross-section values and the correlation matrix for the extraction of the VBF-enriched fiducial cross-section described in \Sect{\ref{sec:VBFenriched}}.
 
\begin{figure}[!htbp]
\centering
\subfloat[]{\includegraphics[width=0.45\linewidth]{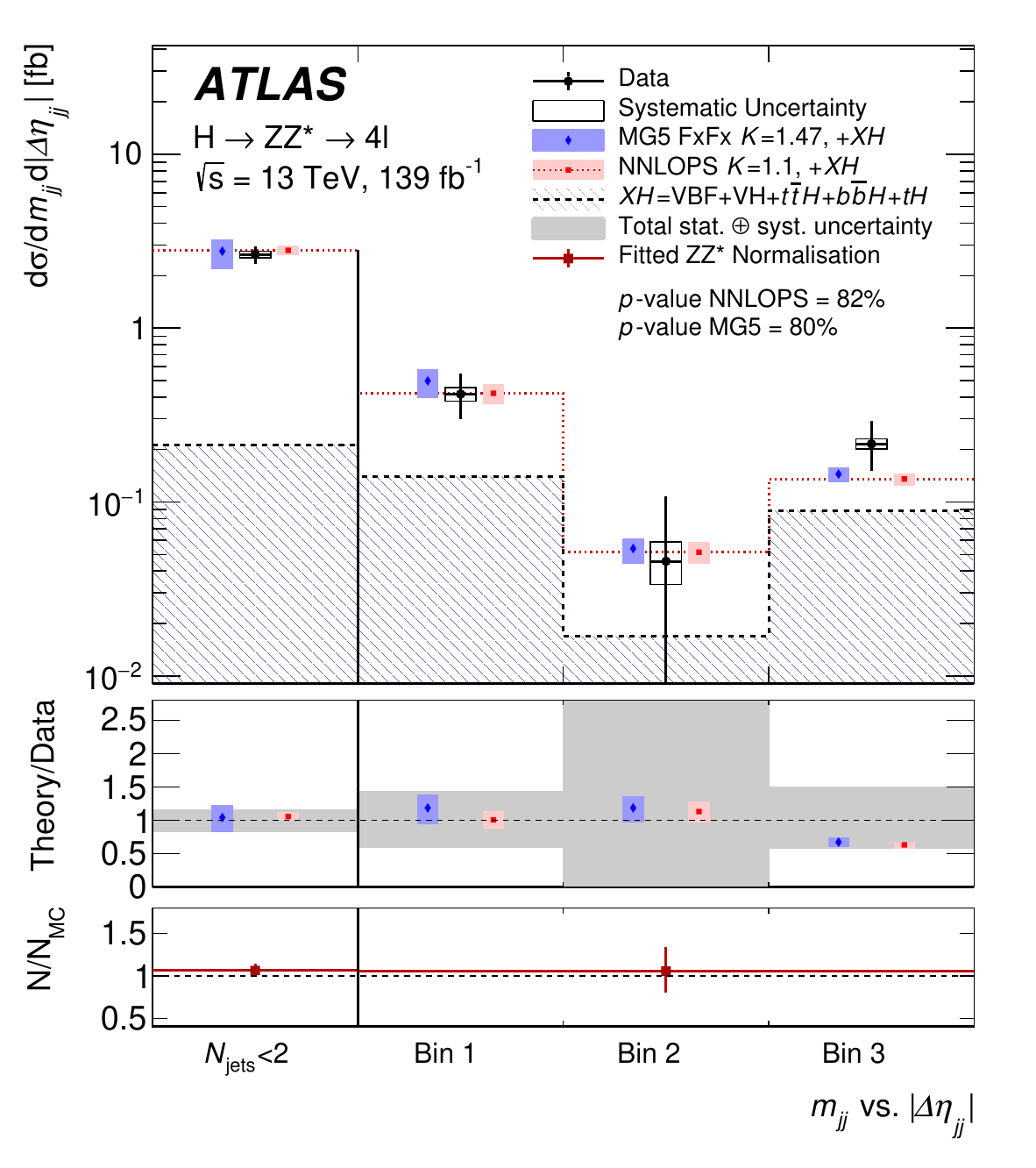}
\label{fig:summary_VBFfid}
}
\subfloat[]{\includegraphics[width=0.45\linewidth]{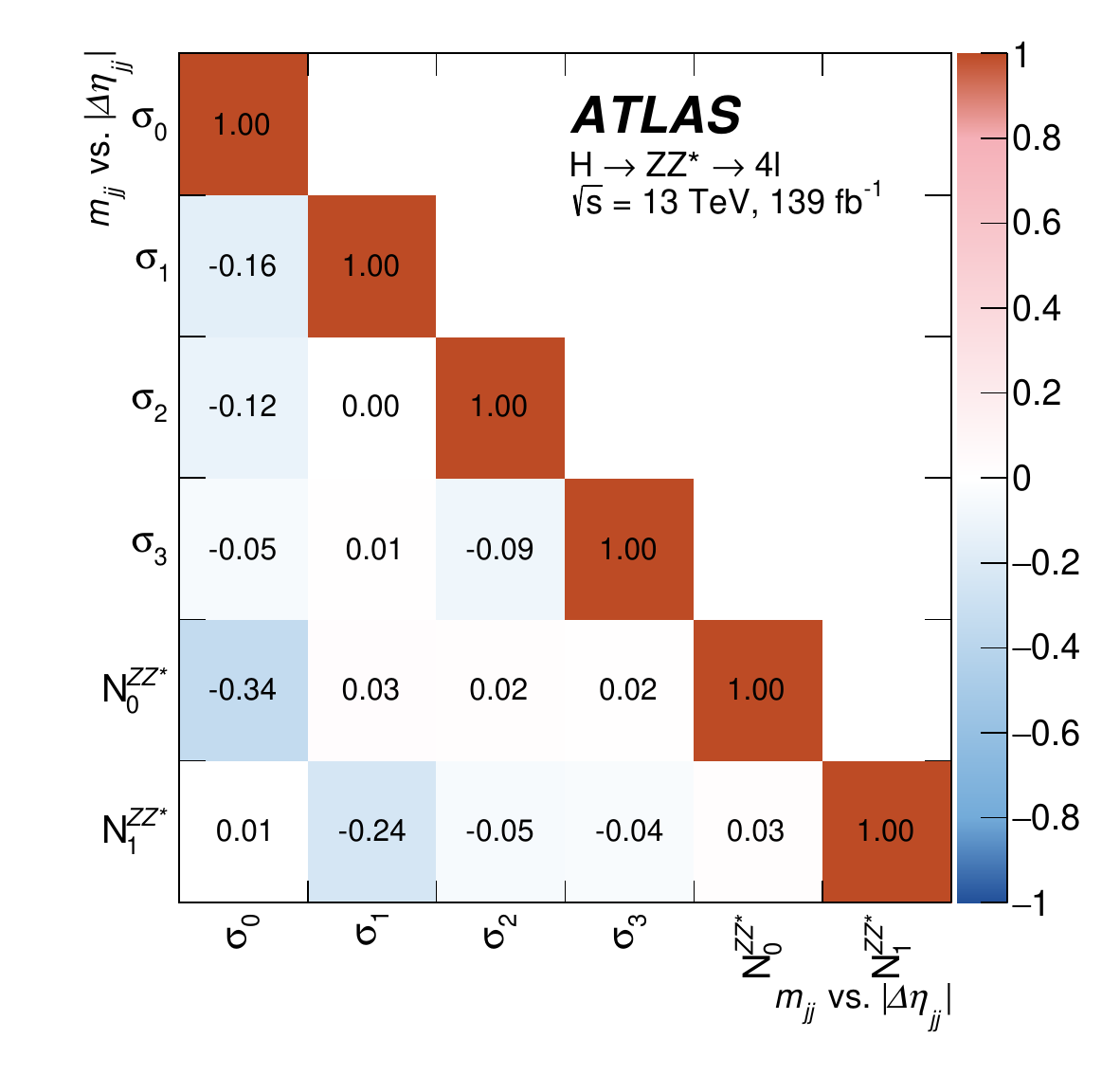}
\label{fig:corrMatrix_VBFfid}
}
\\
\caption{\protect\subref{fig:summary_VBFfid} Results for the \VBF-enriched fiducial cross-section measurement extracted from the double-differential fiducial cross-section for $m_{jj}$ versus $|\Delta\eta_{jj}|$, along with~\protect\subref{fig:corrMatrix_VBFfid}  the corresponding correlation matrix between the measured cross-sections and the $ZZ^{\ast}$ background normalization factors.
}
\end{figure}


\clearpage

\printbibliography
 
\clearpage
 
\begin{flushleft}
\hypersetup{urlcolor=black}
{\Large The ATLAS Collaboration}

\bigskip

\AtlasOrcid[0000-0002-6665-4934]{G.~Aad}$^\textrm{\scriptsize 102}$,
\AtlasOrcid[0000-0002-5888-2734]{B.~Abbott}$^\textrm{\scriptsize 120}$,
\AtlasOrcid[0000-0002-1002-1652]{K.~Abeling}$^\textrm{\scriptsize 55}$,
\AtlasOrcid[0000-0001-5763-2760]{N.J.~Abicht}$^\textrm{\scriptsize 49}$,
\AtlasOrcid[0000-0002-8496-9294]{S.H.~Abidi}$^\textrm{\scriptsize 29}$,
\AtlasOrcid[0000-0002-9987-2292]{A.~Aboulhorma}$^\textrm{\scriptsize 35e}$,
\AtlasOrcid[0000-0001-5329-6640]{H.~Abramowicz}$^\textrm{\scriptsize 151}$,
\AtlasOrcid[0000-0002-1599-2896]{H.~Abreu}$^\textrm{\scriptsize 150}$,
\AtlasOrcid[0000-0003-0403-3697]{Y.~Abulaiti}$^\textrm{\scriptsize 117}$,
\AtlasOrcid[0000-0003-0762-7204]{A.C.~Abusleme~Hoffman}$^\textrm{\scriptsize 137a}$
% ...
\bigskip
\\

% ...
$^{29}$Physics Department, Brookhaven National Laboratory, Upton NY; United States of America.\\
% ...
$^{35}$$^{(a)}$Facult\'e des Sciences Ain Chock, R\'eseau Universitaire de Physique des Hautes Energies - Universit\'e Hassan II, Casablanca;$^{(b)}$Facult\'{e} des Sciences, Universit\'{e} Ibn-Tofail, K\'{e}nitra;$^{(c)}$Facult\'e des Sciences Semlalia, Universit\'e Cadi Ayyad, LPHEA-Marrakech;$^{(d)}$LPMR, Facult\'e des Sciences, Universit\'e Mohamed Premier, Oujda;$^{(e)}$Facult\'e des sciences, Universit\'e Mohammed V, Rabat;$^{(f)}$Institute of Applied Physics, Mohammed VI Polytechnic University, Ben Guerir; Morocco.\\
% ...
$^{49}$Fakult\"{a}t Physik , Technische Universit{\"a}t Dortmund, Dortmund; Germany.\\
% ...
$^{55}$II. Physikalisches Institut, Georg-August-Universit\"{a}t G\"ottingen, G\"ottingen; Germany.\\
% ...
$^{102}$CPPM, Aix-Marseille Universit\'e, CNRS/IN2P3, Marseille; France.\\
% ...
$^{117}$Department of Physics, New York University, New York NY; United States of America.\\
% ...
$^{120}$Homer L. Dodge Department of Physics and Astronomy, University of Oklahoma, Norman OK; United States of America.\\
% ...
$^{137}$$^{(a)}$Departamento de F\'isica, Pontificia Universidad Cat\'olica de Chile, Santiago;$^{(b)}$Millennium Institute for Subatomic physics at high energy frontier (SAPHIR), Santiago;$^{(c)}$Instituto de Investigaci\'on Multidisciplinario en Ciencia y Tecnolog\'ia, y Departamento de F\'isica, Universidad de La Serena;$^{(d)}$Universidad Andres Bello, Department of Physics, Santiago;$^{(e)}$Instituto de Alta Investigaci\'on, Universidad de Tarapac\'a, Arica;$^{(f)}$Departamento de F\'isica, Universidad T\'ecnica Federico Santa Mar\'ia, Valpara\'iso; Chile.\\
% ...
$^{150}$Department of Physics, Technion, Israel Institute of Technology, Haifa; Israel.\\
$^{151}$Raymond and Beverly Sackler School of Physics and Astronomy, Tel Aviv University, Tel Aviv; Israel.\\
% ...
$^{a}$ Also Affiliated with an institute covered by a cooperation agreement with CERN.\\
$^{b}$ Also at An-Najah National University, Nablus; Palestine.\\
$^{c}$ Also at Borough of Manhattan Community College, City University of New York, New York NY; United States of America.\\
$^{d}$ Also at Center for High Energy Physics, Peking University; China.\\
$^{e}$ Also at Center for Interdisciplinary Research and Innovation (CIRI-AUTH), Thessaloniki; Greece.\\
$^{f}$ Also at Centro Studi e Ricerche Enrico Fermi; Italy.\\
% ...

\end{flushleft}


\end{document}